# Synthetic Biology in Leishmaniasis: Design, simulation and validation of constructed Genetic circuit

**Seat No.: 06**

**Dixita Limbachiya**

**OF**

**G H Patel Post Graduate Department of Computer Science and Technology (GDCST), Sardar Patel University, V.V. Nagar**

# Synthetic Biology in Leishmaniasis: Design, simulation and validation of constructed Genetic circuit

**Project Report**

Submitted to the
**G H Patel Post Graduate Department of Computer Science and Technology (GDCST)**
**Sardar Patel University, V. V. Nagar**
In Partial Fulfillment of the award for the
Degree of Master of Science
By

**Dixita Limbachiya**

Under the guidance of

**Dr. Shailza Singh**
Scientist 'C'

**National Centre for Cell Science Pune**

June 2012

# Synthetic Biology in Leishmaniasis: Design, simulation and validation of constructed Genetic circuit

**Project Report**

Submitted to the
**National Centre for Cell Sciences**

In Partial Fulfillment of the award for the
Degree of Master of Science
By

Dixita Limbachiya

Under the guidance of

**Dr. Shailza Singh**
Scientist 'C'
**National Centre for Cell sciences Pune**

June 2012

**G H PATEL POST GRADUATE DEPARTMENT OF COMPUTER SCIENCE AND TECHNOLOGY**
**SARDAR PATEL UNIVERSITY**
**VALLABH VIDYANAGAR**

## ***CERTIFICATE***

*This to certify that* ***Ms. Dixita Limbachiya*** *has worked on the project entitled* ***"Synthetic Biology in Leishmaniasis: Design Simulation and Validation of constructed Genetic Circuit"*** *towards the partial fulfillment of the degree of* ***M. Sc. (Bioinformatics)*** *during the final semester at* ***National Centre for Cell Sciences Pune*** *from 21-12-2011 to 39-5-2012.*

.

**Date of Submission:-2$^{nd}$ JUNE 2012**

| | |
|---|---|
| **Dr. Hetal J Panchal** | **Dr. D. B. Choksi** |
| **Project Coordinator** | **Head of Department** |
| **Lecturer** | **Director and professor** |
| **GDCST** | **GDCST** |

**Date: - 02/06/2012**

# UNDERTAKING

I, **Dixita Limbachiya**, student of the G.H. Patel Post Graduate Department of Computer Science and Technology, VallabhVidyanagar hereby undertake that the work presented in the dissertation project report entitled "**Synthetic Biology In Leishmaniasis:Design,Simulation and Validation Of Constructed Genetic Circuit**" comprises the results of independent and original work carried out by me under the supervision of **Dr.Shailza Singh** for the partial fulfillment of the award of the degree of M.Sc. Bioinformatics of the Sardar Patel University, VallabhVidyanagar.

I further declare that this work did not form a part of any other work published or unpublished.

**Dixita Limbachiya**

**G.H. Patel Post Graduate Department of**
**Computer Science and Technology,**
**VallabhVidyanagar, Gujarat**.

# ACKNOWLEDGEMENT

*The writing of this thesis has been one of the most significant academic challenges I have ever had to face. Without the support, patience and guidance of following people, this study would not have been completed. It is to them that I owe my deepest gratitude.*

*I would like to sincerely thank my guide* ***Dr. Shailza Singh*** *for her guidance, understanding, patience, and most importantly, her friendly gesture during the dissertation period. Her mentorship was paramount in providing a well rounded experience to pursue my long-term career goals. She encouraged me not only to grow as a researcher but also as an independent thinker, not many post graduate students are given the opportunity to develop their own individuality and self-sufficiency by being allowed to work with such independence. Words are inadequate in the available lexicon to avouch the guidance given by her. Her dedication to research, meticulous planning, consecutive counsel and unreserved help served as a beacon light throughout the course of study, research work and completion of this manuscript. For everything you have done for me, madam I thank you.*

*I am grateful to my Head of the Department* ***Dr. Hetal Panchal*** *of* ***G.H. Patel Department of Computer science and Technology*** *for their cooperation throughout the study and for allowing me to pursue my dissertation program at NCCS.*

*I would like to thank all the members of the Lab, especially Sonali Shinde, Vineetha Mandlik and Ashish Chaudhary for supporting me and encouraging me. They also provided for some much needed humor and entertainment in what could have otherwise been a somewhat stressful laboratory environment. I am thankful to my colleagues Radhika and Jyoti with whom I worked closely and made my journey of dissertation most memorable and enthusiastic. I am grateful to National Centre for Cell sciences for giving me opportunity to carry out my dissertation successfully.*

*Finally, and most importantly, I thank All Mighty for his blessing. It was under his watchful eyes that I gained so much drive and an ability to tackle challenges head on. I would like to thank my parents, brother and family members for their support, encouragement, patience and unwavering love which was always with me. I thank them for allowing me to be as ambitious as I wanted.*

*Dixita Limbachiya*

# CONTENTS

# LIST OF FIGURES

# LIST OF TABLES

## ABBREVIATIONS:

| | |
|---|---|
| WHO | World health Organization |
| AMA | Amastigotes |
| PRO | Promastigote |
| GSL | Glycosphingolipids; |
| IPC | Inositol phosphorylceramide |
| LPG | Lipophosphoglycan |
| PL | Phospholipids |
| GIPL | Glycoinositolphospholipid. |
| SL | Sphingolipids |
| IPCS | Inositol Phosphoryl ceramide synthase |
| DAG | Diacylglycerol |
| PC | Phosphorylceramide |
| SM | Sphingomyelin |
| PI | Phosphoryl Inositol |
| MIPC | Mannose inositol Phosphoryl ceramide |
| DNA | Deoxyribonucleic acid |
| mRNA | Messenger ribonucleic acis |
| GRN | Gene Regulatory Network |
| ODE | Ordinary Differential Equation |
| SBML | Systems Biology Markup Language |
| COPASI | Complex Pathway Simulator |
| LacI | Lactose repressor |
| Tetr | Tetracycline repressor |
| GRENITS | Gene Regulatory Network using Time Series Data |
| BoolNet | Boolean Network |
| MCMC | Monte Carlo Markov Chain |
| URL | Uniform Resource Locator |

In recent years, the disciplines of systems biology and synthetic biology have gained prominence as the embodiments of the future of biological science. For biological circuits, we need to produce quantitative predictions of cell behavior for a given genotype as consequence of the different molecular interactions. There is a great synergy between the fields of systems and synthetic biology such that methodologies from one can help make significant advances in the other. Recent systems biology paradigms such as computational systems analysis, methods for quantifying time-dependent gene expression and bioinformatics cataloging of cellular parts can help enable synthetic biology. Greatest advances in biology and biotechnology are arising at the intersection of the top-down systems approach and the bottom-up synthetic approach. Building circuits and studying their behavior in cells is a major goal of synthetic biology in order to evolve a deeper understanding of biological design principles from the bottom up. Collectively, these developments enable the precise control of cellular state for systems studies and the discovery of novel parts, control strategies, and interactions for the design of robust synthetic function (Isaacs, Dwyer, and Collins 2006).

Synthetic biology aims to design novel biological circuits for desired applications implemented through the assembly of biological parts including natural components of cells and artificial molecules that emulate biological behavior. Because of its parts-to-whole approach, synthetic biology has a significant engineering component(Heinemann and Panke 2006). It may appear that it should be possible to apply strategies such as standardization which ensures that components of a system are compatible and exchangeable toward constructing a synthetic biological circuit in a manner similar to constructing an electric or electronic circuit (Andrianantoandro et al., 2006). The attainment of this ideal goal is, however, impeded by the overwhelming complexity of biological systems with their myriad biomolecules and interconnections as well as sparse databases of gene function(Haseltine and Arnold 2007). Consequently it is challenging to convert design concepts to predicted results for which mathematical modeling serves as a bridge.

Mathematical modeling plays an important and often indispensable role in synthetic biology because it serves as a crucial link between the concept and realization of a biological circuit. Types of modeling frameworks such as deterministic and stochastic, the importance of parameter estimation and optimization in modeling plays an important role. Additionally mathematical techniques used to analyze a model such as sensitivity analysis and bifurcation analysis is also dealt with, which enable the identification of the conditions that cause a synthetic circuit to behave in a desired manner. In the project, mathematical modeling is incorporated as a central component of synthetic circuit design in *Leishmania.* Both deterministic approaches and stochastic approaches are used.

Stochastic models are used to deduce the effects of noise within a synthetic network, potentially leading to the manipulation of the network itself in order to improve the signal-to-noise ratio within these networks. A stochastic approach regards changes over time as random- walk processes, with no set of differential equations defined, and takes into account inherent fluctuations that are not considered in the deterministic kinetic approach. Stochastic effects may be particularly significant in some biological systems with small molecular populations involved. Although this stochastic basis is more accurate in modeling, it is more difficult to solve mathematically. However, numerical simulations are possible using Monte Carlo principles. Simulations using stochastic considerations have been reported for biological systems involving genetic and enzymatic reactions between molecular populations that was relatively small, including synthetic oscillatory networks, transcriptional regulation and circadian rhythms. For large populations of molecular species, the predictions obtained from stochastic approaches match with deterministic ones. However, at smaller population sizes, stochastic effects become more dominant, in which case, deterministic approaches become insufficient. Unfortunately, for many biological networks, stochastic simulations are still computationally expensive due to the huge differences in timescales of biological interactions and population sizes. Various improvements, approximations, and hybrid approaches have been presented. In one such study, stochastic simulations were done on multi-scaled systems to study reactions occurring in three different regimes (slow, medium and fast) as well as coupled reactions.

The presented approach in this project showed substantial improvement over using the basic stochastic simulation approach when applied to the study of expression and activity of IPCS in *L.major*. The constructed genetic circuit was modeled and simulated using diverse set of algorithms. This assumption was shown to greatly simplify the stochastic model and to significantly reduce the computational complexity. Despite providing a more complete representation of biological networks, stochastic approaches still face the challenge of dealing with several orders of magnitude in terms of scale and properties including binding affinities, specificities, and kinetic rates. Therefore, even statistics based theories have limitations. Although they provide insights into macroscopic properties of a network, they may have inaccurate predictions about specific interactions. These limitations can be addressed with new developments in integrative modeling.

Once the circuit is constructed, validation is performed in two stages. Observation of the qualitative behavior of the circuit can be very informative in models having multiple steady states and showing switch-like or oscillatory behaviors'. The qualitative behavior can be studied either over small parts of the parameter space by simply scanning over defined ranges of parameters and initial conditions or by doing global bifurcation analyses.Qualitative analysis gives hints as to which parameters offer the best success in achieving a desired behavior or whether a certain design can exhibit the wanted function at all. Identifying the most promising parameters to change, of course, depends not only on the mathematical analysis but also on the biological feasibility. While some characteristics such as promoter strength, transcript and protein stability are quite variable, enzymatic activities, for example, might be harder to tweak. Also as changes in the characteristics of biological components can at best be qualitative, it is important to find parameter ranges that show behavior robust to variations. In the repressilator for example, the qualitative analysis may lead to the identification of a few key properties important for obtaining stable oscillations—strong promoters with tight cooperative repression. Apart from helping to choose the right biological components, these criteria may help researchers to introduce tags into the repressor sequences. Model validity can also be checked by comparison of the results of simulation runs with quantitative experimental data such as time courses or steady-state concentrations and fluxes. These can sometimes be derived from the literature or retrieved from databases.

Based on the current approaches dealt, herein, we describe how different approaches of bioinformatics could enable novel synthetic biology applications in *Leishmania.*

**Aims and Objectives**

- To characterize dynamical properties of small gene networks displaying oscillations or bistability.

- To construct artificial genetic systems displaying oscillations and bistability through repressilator and the Toggle switch models.
- To investigate the various properties of built in genetic switch including the robustness to noise, signal amplifications and tuneable frequency which ultimately deals the synchronization ability of the constructed genetic switch/circuit.

## Systems and Synthetic Biology

Cellular complexity stems from the interactions among thousands of different molecular species. Thanks to the emerging fields of systems and synthetic biology (Hasty *et a*l., 2002; Hayete *et al*., 2007; Kaern *et al*., 2003; Sprinzak and Elowitz, 2005), scientists are beginning to unravel these regulatory, signaling, and metabolic interactions and to understand their coordinated action. A system is a network of mutually dependent and thus interconnected components comprising a unified whole. Every system exhibits emergent behavior, a unique property possessed only by the whole system and not shared to any great degree by the individual components on its own. Fields of systems and synthetic biology are important for accelerating both our understanding of biological systems and our ability to quantitatively engineer cells. Synthetic biology is the engineering of biology: the synthesis of complex, biologically based (or inspired) systems which display functions that do not exist in nature .At the nexus of these two fields is a unique synergy that can help attain these goals. Thus, the next greatest advances in biology and biotechnology are arising at the intersection of the top-down systems approach and the bottom-up synthetic approaches. Collectively, these developments enable the precise control of cellular state for systems studies and the discovery of novel parts, control strategies, and interactions for the design of robust synthetic function. Combining these efforts can provide novel insights into cellular function and lead to robust, novel synthetic design (Lanza M *et al*., 2012). Likewise, we can design orthogonal synthetic systems that have predictable behavior in a complex and noisy environment. However, our methods are inverted: the most sophisticated methods for understanding cellular complexity at the gene level utilize top-down systems approaches whereas the most promising avenue for creating novel macroscopic function relies on bottom-up synthetic design principles. These disciplines aims to develop Predictive, Preventive, Personalized, Participative (P4) medicine that has potential to transform medicine by decreasing morbidity and mortality.

### Top-down systems biology

The advent of high-throughput biology has led to a rapid acceleration in obtaining systems-level information about living organisms including genomes, transcriptomes, proteomes, metabolomes, epigenetic states, and transcription factor binding profiles. This global information, integrated with computational approaches for analysis and model-based prediction, has led to an enormous understanding of bio molecular networks in a field termed 'systems biology'. Combining these global measurements lead to robust, high resolution information about the cellular responses and metabolism.

**Bottom-up synthetic biology**

Synthetic biology attempts the reverse bottom up approach in which the discrete bimolecular processes are organized into larger construct to generate complex behavior at cellular level. This engineering perspective may be applied at all levels of the hierarchy of biological structures – from individual molecules to whole cells, tissues and organisms. In essence, synthetic biology will enable the design of 'biological systems' in a rational and systematic way (*Lunes*, 2007). These constructs can be simple, such a promoter driving the expression of a single gene, or much more complex modules such as different promoters combination producing different behavior.

Systems biology aims to model and understand an entire organism by characterizing dynamic environment-dependent functional interrelationships between its constituent parts (for example, genes, RNAs, proteins and metabolites). Synthetic biology, however, uses well characterized parts that are shaped by natural evolution to construct artificial systems that perform new tasks. These fields are on trajectories that are bound to cross paths and even merge as they begin to inform one another. We envision that systems biology will provide both the parts and wiring diagrams for entire cells to guide complex circuit design for synthetic biology approaches.

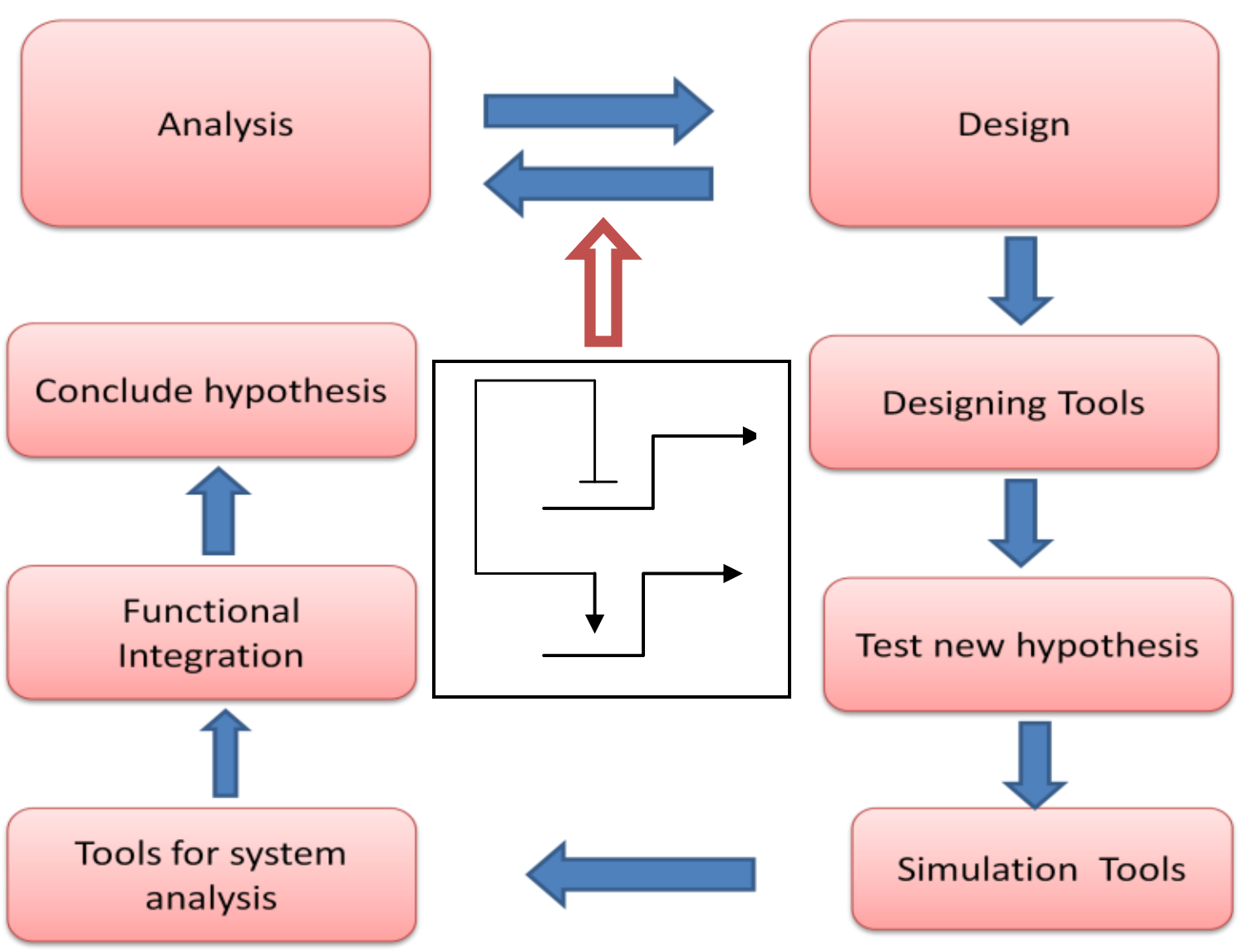

**Fig (5.1)** Synergy between System and synthetic biology

**Bioinformatics for Synthetic and Systems biology**

In the era of post-genomic research two new disciplines, Systems and Synthetic biology, act in a complementary way to shed light on the ever-increasing amount of data produced by tools and databases of bioinformatics .Bioinformatics cataloging of cellular parts can help in construction of synthetic modules. The ease of collecting genome-scale measurements available at Bioinformatics database and data repository has created a deluge of information that must be parsed in order to provide systems insight as well as utility for synthetic biology applications. Optimal design certainly requires these bioinformatics approaches to parse the rapid growth in available genome sequences. These fast, accurate, and efficient methods for predicting desired behavior from easily collected expression or genomic data is a major systems biology advancement that has direct implications for synthetic biology applications. In this regard, these approaches have been used to enable automated synthetic circuit design. It is clear that there is a great synergy between global cellular modeling efforts and synthetic circuit design. A final area of synergy between systems and synthetic biology lies in the efforts to expand and automate the process of parts identification. The rapid capacity to sequence new genomes has sparked interest in equally rapid annotation capacities. As a result, databases of prokaryotic (S. Gama-Castro *et al*, 2008) and eukaryotic (R. Gupta., 2011) motifs have been curated, enabling automatic annotation of promoters (Y.Y. Yamamoto, 2008.,R. Gupta., 2011) transcription factor binding sites (J. Zhu, M.Q. Zhang, 1999) and terminators for these organisms. In addition to providing a repository of natural biological parts for synthetic biology applications, these tools allow synthetic biologists to design novel regulatory elements by combining motifs in interesting ways (A. Mitra, A.K. Kesarwani, 2011) In turn, the increased understanding afforded by manipulating regulatory motifs in this fashion can serve to illuminate the complex systems biology underlying regulatory element performance. There are many software packages of bioinformatics for the analysis of the models generated by approaches of systems and synthetic biology. Bioconductor package uses the R statistical programming language, and is open source and open development, provides tools for the analysis and comprehension of high-throughput genomic data. It includes sequence analysis package, network analyses package using different reverse engineering methods like Bayesian and Boolean methods.

## Synthetic Circuit

An important aim of synthetic biology is to uncover the design principles of natural biological systems through the rational design of gene and protein circuits (M Shankar *et al*., 2009). The synthetic gene circuits discipline can be described succinctly as novel regulation of pre-existing or engineered cellular functions. Synthetic circuits generally consist of components optimized to function in their natural context, not in the context of the synthetic circuit. The emergence of synthetic gene circuits as an engineering goal has re-emphasized the importance of considering the continuous nature of gene regulation. **"Synthetic circuit"** describes a system that is designed to execute a useful function (a bistable state, oscillation, pulse etc) in a predictable and reliable manner. This highlights the process of engineering biological systems from genetic circuits to the control of cell–cell interactions, has contributed to our understanding of how endogenous systems are put together and function. Synthetic biological devices allow us to grasp intuitively the ranges of behavior generated by simple biological circuits, such as linear cascades and interlocking feedback loops, as well as to exert control over natural processes, such as gene expression and population dynamics (M Shankar *et al*., 2009)

## Design principles

Design principles helps in deciphering the quantitative laws that governs the behavior of biological systems. These principles will be able to design biological systems having desired properties by reconstructing and thus facilitates the better understanding of naturally occurring functions. Different laws are used for specific reactions e.g. **Hill-Hinze** equation is applied for the reactions for the formation of gene products.

<u>Modules of Gene Components</u>

- Its behavior must be predictable when it is integrated in the system.
- It must function more or less independently of the host organism.
- It must utilize few resources from host cell as much as possible to minimize the perturbations in the host.
- It should be portable and should function predictable in variety of host systems.

**Hill-Hinze equation**

Hill equation as kinetic law for gene product is:

$$V_{re2} = V_{max_re2}.\ \frac{[R1]n/c\ _re2_B2}{[R1]^{hic_re2_B2} + ksp_re2_s2^{hic_re2_s2}}$$

Since we had taken the transition from gene to protein as a “known transition omitted”, and ignored the RNA, the Hill equation is modified as per the reaction’s convenience and the equation was given as:

**Vmax_re**

Where, Vmax is maximum velocity attained.

According to Quassi steady state rule, d/dt[mRNA] = 0, i.e. for the intermediate compound .Moreover the concentration of the mRNA reaches steady state very quickly compared to protein concentration, thus we considered that [mRNA] concentration is always considered at steady state.

## Construction of genetic circuit

The first step in assembling a biological circuit is to gather the component parts. Equations and computer simulations that model relationships between biological processes (e.g., gene activation, protein production, cell division, etc.) guide the design of synthetic systems by determining the vital parameters and parameter spaces in which the system will function as intended.

Steps for construction of genetic circuit are as follows:

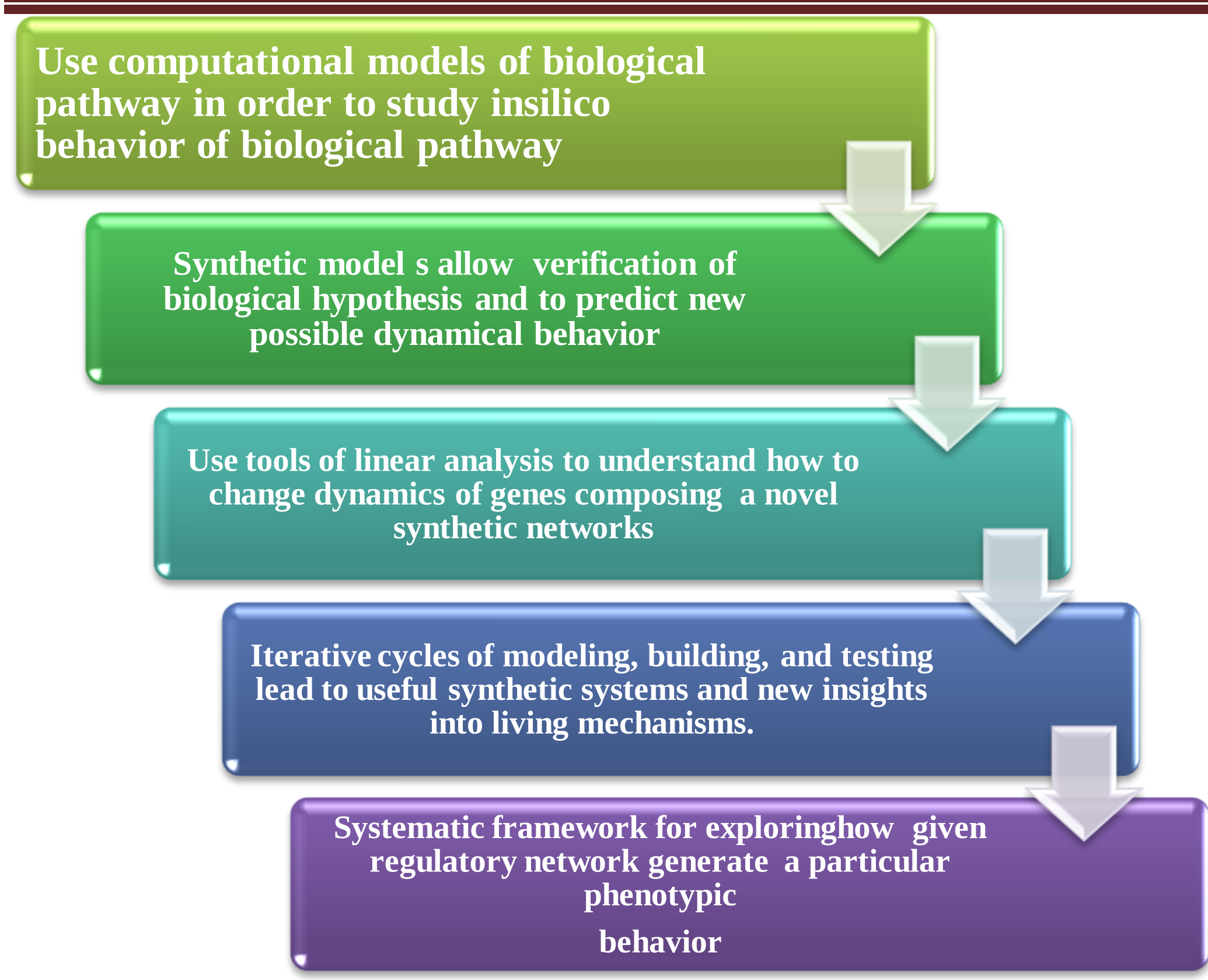

**Fig 5.2** Steps for construction of genetic circuit

**Synthetic biology approaches to the design of circuits**

<u>Parts</u>

Parts are the individual components that make up gene expression machinery which are specific DNA sequences that code for gene promoters and upstream regulatory sites, ribosome binding sites, gene or protein coding regions, and mRNA translation termination sites. Sequences for RNA-only machinery, such as small interfering RNA and ribozyme coding sequences, are also used.

<u>Devices</u>

Assemblage of parts (promoters and genes) that carry out specific functions. There are several basic devices, including reporters, inverters and devices that carry out signaling and protein generation. Each device includes one or two promoter–ribosome binding site–gene terminator constructs. Subsequently, these basic devices can be combined into 'composite devices' to achieve more complex behavior.

Systems

Systems are collections of specific devices that enable individual parts to be quantified. Generally, a subset of the composite devices is used to characterize the strength and efficiency of promoters under non-repressive conditions.

Chassis

The host cell in which the systems and devices described above are assembled and used. To date, these cells have been primarily from *Escherichia coli* strains, although a few devices and systems have been tested in yeast and mammalian cells. A more exotic chassis includes cell-free systems that use DNA transcription and mRNA translation chemistry derived from whole-cell extracts and encapsulated in artificial membranes.

## Logic Circuits or Regulatory Modules

Electrical circuits are based upon mathematical models and so are genetic circuits, and many of the techniques in predicting outcomes of genetic circuits are directly derived from electric circuits. Electric circuits often contain modular parts such as switches and oscillators, which have strong resemblance to the two genetic circuits, a genetic toggle switch and the repressilator respectively. Logic gates are the basic building blocks in electronic circuits that perform logical operations (G Agarwal., 2007). These have input and output signals in the form of 0's and 1's; '0' signifies the absence of signal while '1' signifies its presence (Fig-2.12). Similar to the electronic logic gates, cellular components can serve as logic gates. A typical biological circuit consists of i) a coding region, ii) its promoter, iii) RNA polymerase and the iv) regulatory proteins with their v) DNA binding elements, and vi) small signaling molecules that interact with the regulatory proteins. Messenger RNA or their translation products can serve as input and output signals to the logic gates formed by genes with which these gene products interact. The concentration of the gene product determines the strength of the signal. High concentration indicates the presence of signal (=1) whereas low concentration indicates its absence (=0).

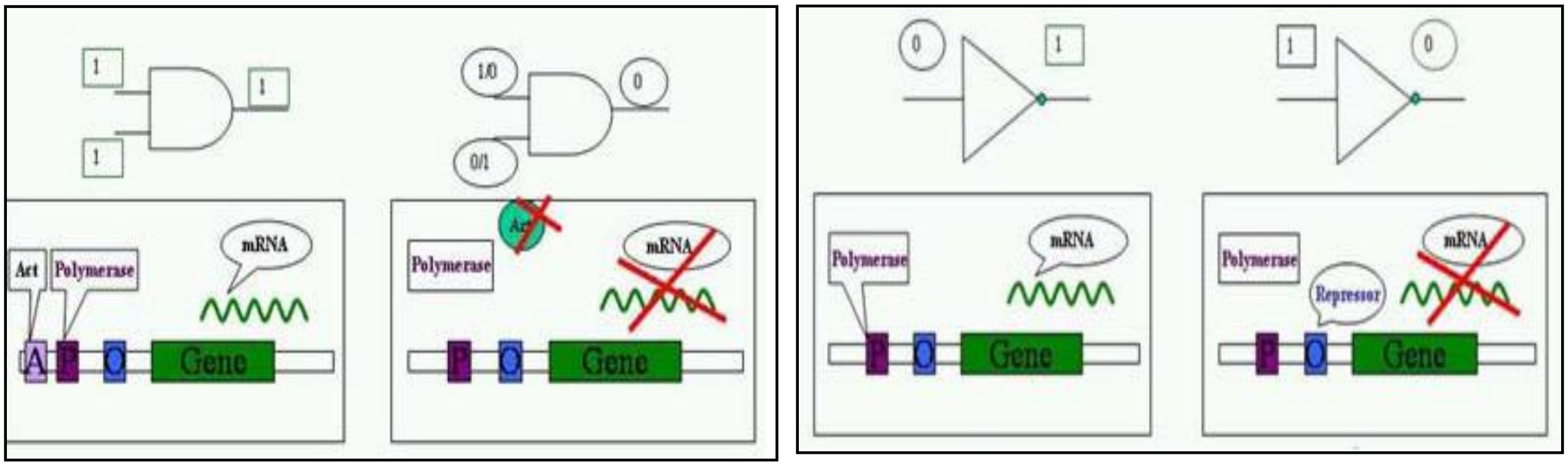

a.) NOT gate b.) AND gate

**Fig (5.3)** Logic gates of Biochemical processes

| Logic circuit | No of Input | No of Output | Logic | Application |
|---|---|---|---|---|
| **NOT** | 1 | 0 | Inverts the input signal | Determine intracellular state of the cell |
| **NAND** | 2 | 1 | If both input signal is present only then gives no output | Presence or absence of polymerase and repressor present gene is not transcribed |
| **AND** | 2 | 1 | If both input signal is present only then gives output | The activator and inducer together result in turning on a gene .It is used in cell to cell communication |
| **Implies** | 2 | 1 | If both input signal are present output signal is present, If one input signal present it behaves like NOT gate | When repressor and inducer are require for gene expression |

**Table 5.1** Logic gates in biological systems

## Synthetic circuit in protozoan parasite

There are reports and evidences for the synthetic circuit in prokaryotes and eukaryotes like *E.coli* and yeast systems but to the best of my knowledge there are no reports in the literature suggesting the synthetic circuit construction for protozoan parasites. The project serves to the first attempt made to use synthetic biology approach for the construction of genetic circuit for the protozoan parasite *Leishmania*.

## Leishmaniasis

Among all the emergent diseases, the ones caused by protozoan have great importance. Leishmaniasis is a disease caused by a parasite member of the Leishmania genus and presents high morbidity and mortality levels. The annual incidence of approximately two million new cases and around 350 million people that are living in endemic areas reveals the importance of this neglected disease (Campos *et al.*, 2004). Leishmaniasis, widespread parasitic disease caused by a heterogeneous group of protozoan parasite belonging to the genus *Leishmania*, spread by the bite of female sand fly of the genera: ***Phlebotomus*** (Old world) and ***Lutzomyia*** (New world).20 species of *Leishmania* was found to be pathogenic for humans [WHO].

### *LEISHMANIA*

This parasite was first time reported by W.B. Leishman and C. Donovan in 1903. Species of *Leishmania* are in **Old World** they are *L. tropica*, *L. aethiopica*, *L. major*, *L. infantum* and *L. donovani* and in new world ***New World*** are *L. mexicana*, *L. amazonensis*, *L. venezuelensis*, *L. pifanoi.*

<u>MORPHOLOGY</u>

*Leishmania* is a dimorphic protozoan parasite that resides as an extracellular flagellate-**Promastigotes** in the sand fly vector and as an obligate intracellular aflagellate-amastigotes within macrophages of vertebrate hosts (Liew and O'Donnell, 1993). The various species of *Leishmania* are not distinguishable from one another.

**Amastigotes** appear as round or oval bodies ranging from 2 - 3μm in diameter with a well-defined nucleus and kinetoplast, a rod shaped specialized mitochondrial structure that contains extra nuclear DNA. **Promastigote** form is spindle shaped, longer than amastigotes, measuring 10

- 20 µm in length with a central nucleus and anterior kinetoplast and a well-developed flagellum, which is used either for propulsion or for attachment. (http://www.cvbd.org/en/sand-fly-borne-diseases/leishmaniosis/pathogens)

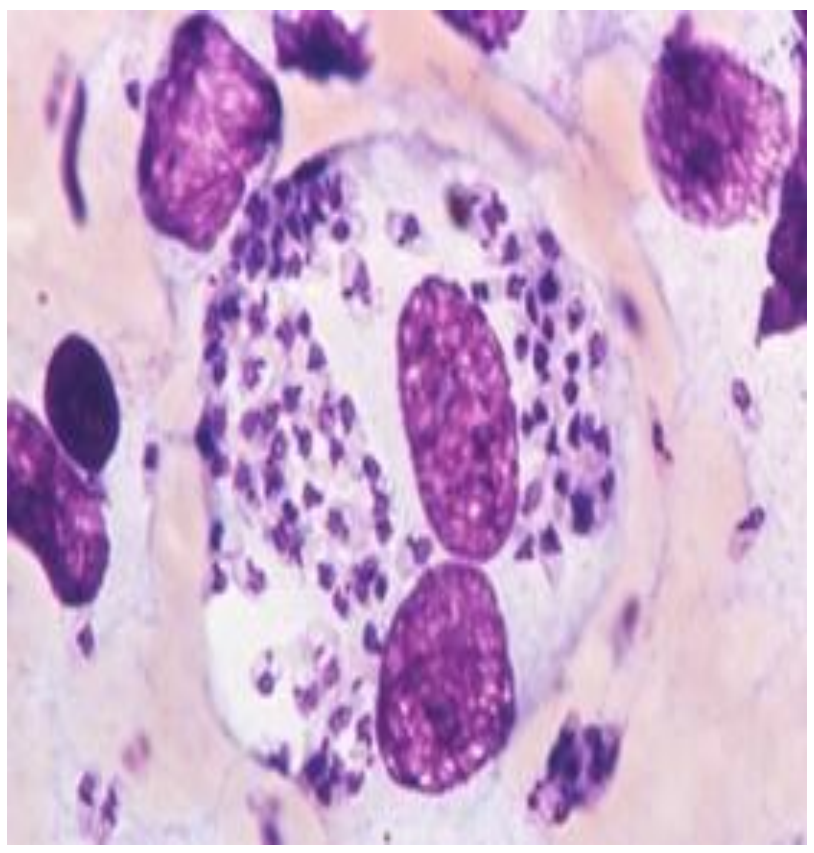

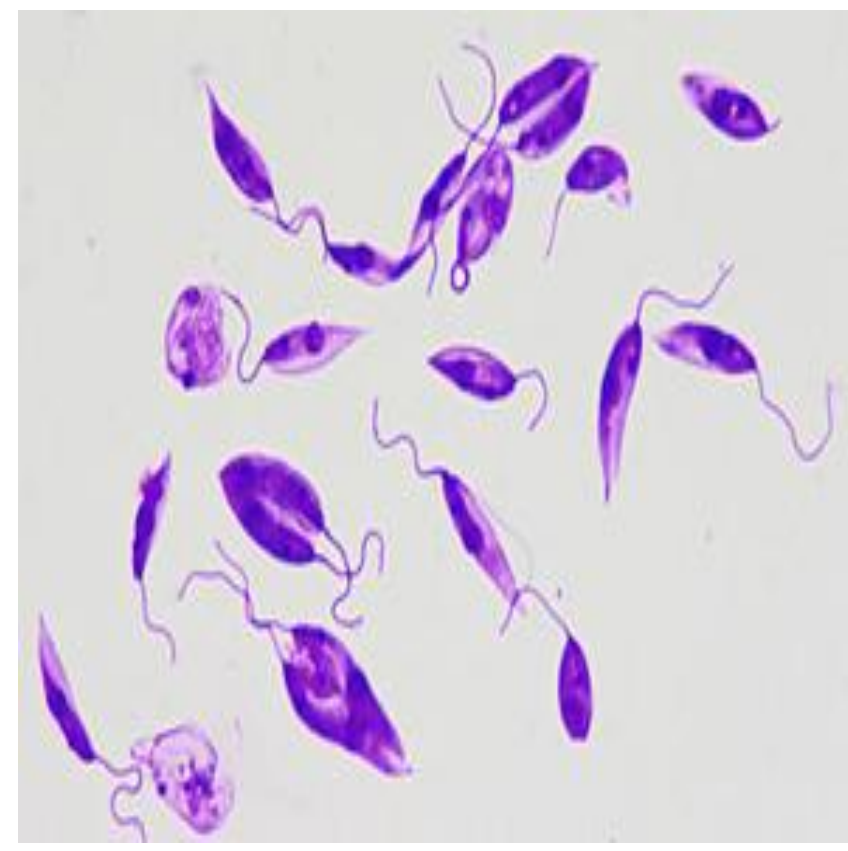

**Fig 5.4*:* A)** Amastigotes **B)** Promastigotes

***Picture Credit: Science photo library (Leishmaniasis)***

**Genome**

Mapping and sequencing of the *L. major* genome allowed the definition of 36 discrete chromosomes composing the genome of this protozoan parasite. Different *Leishmania* species have 34 to 36 chromosomes, varying in size from 268 to 2,680 kb. The arrangement of genes in trypanosomes and *Leishmania* (and in other related parasites from the same order, Kinetoplastida) is reminiscent of that in bacterial operons, especially as protein coding regions are almost never interrupted by introns; the single exception so far is the gene encoding poly(A) polymerase (Mair *et al*., 2000). So far nothing is known about the sequences involved in transcription initiation of protein-coding genes in the parasite *Leishmania.* Most genes in these organisms are transcribed polycistronically, and the mature mRNAs are generated from primary transcripts by *trans*-splicing (M Santiago *et* al., 2003). Promoters for *rRNA*, *VSG*, and *PARP* genes, which are transcribed by RNA polymerase I (Pol I), have been characterized in trypanosomatids (Zomerdijk *et al*. 1991) and (Chung *et al*. 1993), as has the *SL* gene promoter, which is transcribed by Pol II. The nature of the polymerase II complex is still one of the major mysteries of kinetoplast molecular biology.

**Leishmania Life cycle:**

Leishmania procyclic promastigote differentiate in sandflies into infective, non-dividing metacyclic Promastigotes, which are located ready for transmission at the stomodeal valve (an invagination of the foregut into the midgut). During blood feeding, the sand fly regurgitates metacyclic promastigote, together with immunomodulatory parasite-derived proteophosphoglycans and various salivary components. The metacyclic promastigote are then phagocytized by one of several possible cell types that are found in the local environment. After establishing an intracellular residence, metacyclic promastigote transform into aflagellate amastigotes. Amastigotes undergo replication within host cells, which rupture when too many amastigotes are present, allowing reinfection of local phagocytes. The transmission cycle is complete when infected phagocytes are taken up by another sand fly with the blood meal, and amastigotes then convert into Promastigotes in the sand fly midgut.

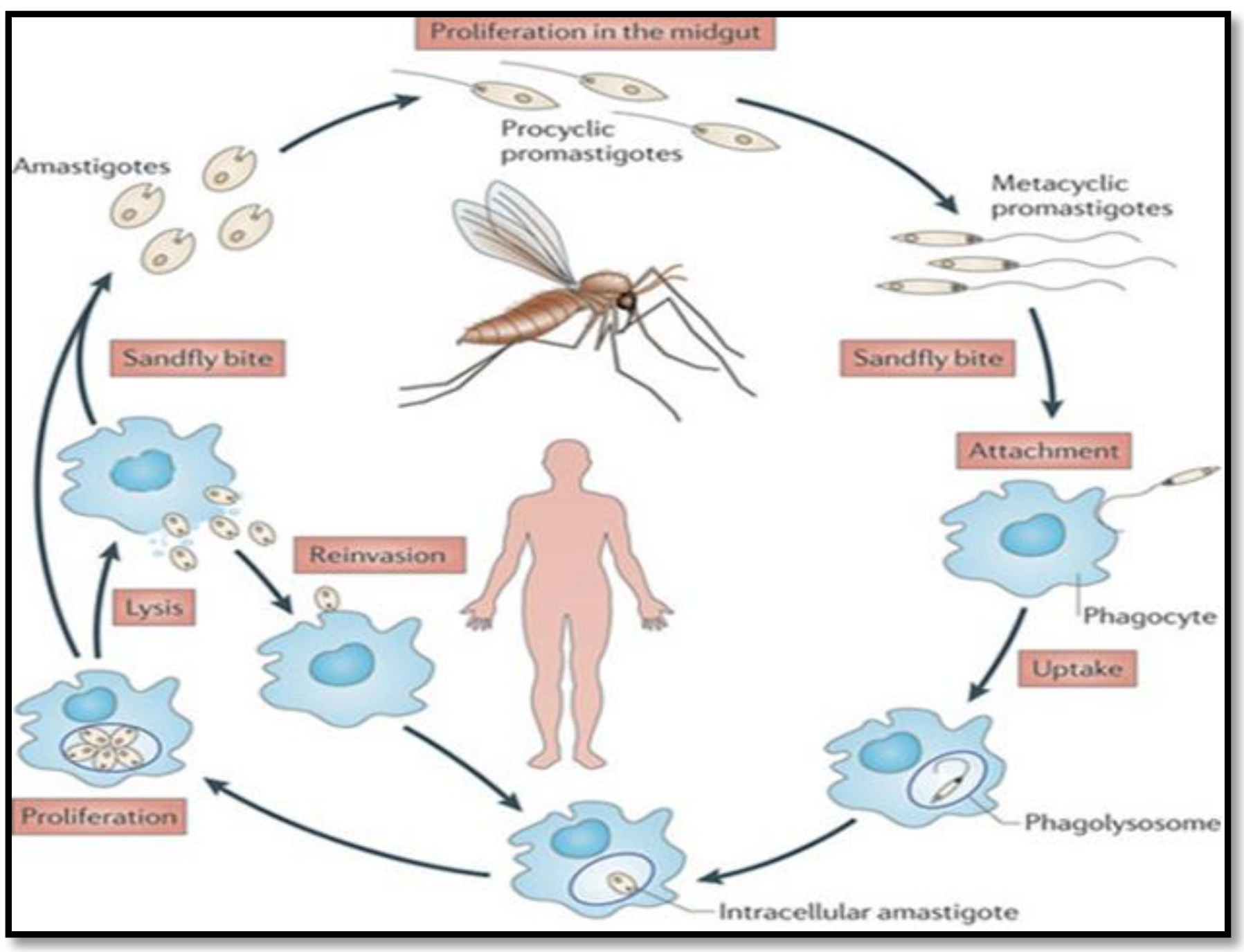

*Fig (5.5)* **Life cycle of Leishmania**

**Picture credit:** Kaye P, Scott P (2011).Leishmaniasis: complexity at the host–pathogen interface

***TYPES OF LEISHMANIASIS***

*There are 3 forms of Leishmaniasis:*

| Types | Cutaneous | Mucocutaneous | Visceral |
|---|---|---|---|
| **Pathogen** | *L.major*<br>*L.tropicana and*<br>*L.aethiopica* | *L.braziliensis* | *L.donovani,*<br>*L.infantum* |
| **Location** | India, Afghanistan, Brazil, Iran, Peru, Saudi Arabia and Syria | Bolivia, Brazil, Peru | Bangladesh, Brazil, India, Nepal and Sudan<br><br>Semi deserts in Middle East, North India, Pakistan, North, Africa Central Asia |
| **Symptom** | ➢ Skin sores, which may become a skin ulcer wearing away (erosion) in the mouth, tongue, gums, lips, nose, and inner nose that heals very slowly<br>➢ The skin lesions take on a variety plaques | ➢ Perforation of the nasal septum, and enlargement of the nose or lips<br>➢ They erode underlying tissue and cartilage separating two nostrils.<br>➢ Prevent speech.<br>➢ If the larynx is involved, the voice changes as well. Ulcerated lesions may lead to scarring and tissue destruction that can be disfiguring | ➢ Weight loss, which may be severe.<br>➢ Low blood counts (pancytopenia).<br>➢ Enlargement of the liver and spleen (hepatosplenomegaly).<br>➢ Fever, which is usually intermittent.<br>➢ High levels of immune globulin in the blood (hypergammaglobulinemia).<br>➢ The skin may turn dark, causing VL to be called "**kala-zar**," which means "black sickness." Some people who recover will have a persistent rash or pigment changes in the skin. The kidney is also affected, which may lead to renal failure. Other organs, including the bowel and the lung, may be affected. |

**Table 5.2** Type of leishmaniasis

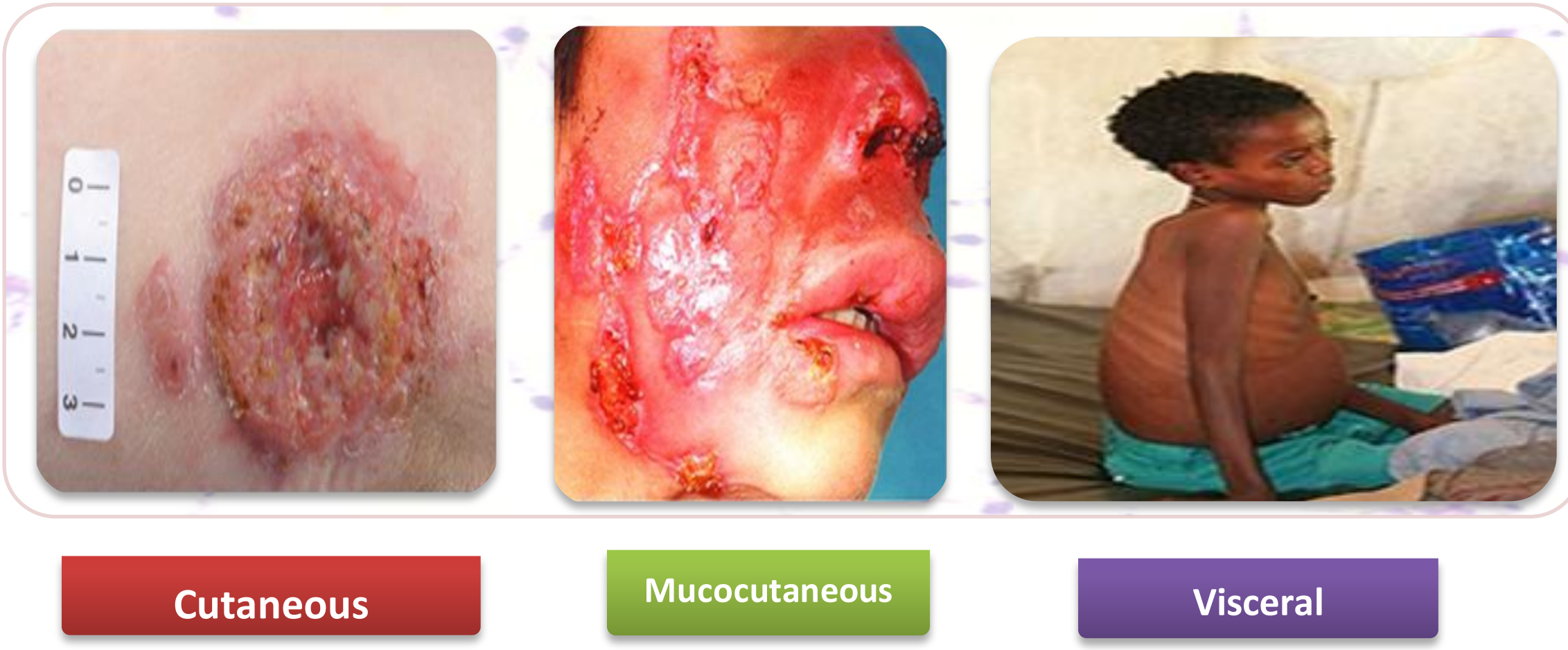

**Fig (5.6) Severity of lesions**

***Picture credit CGM Research Project Telemedicine*** ***Picture credit: Human Leishmaniasis***

***Fig (5.6) Visceral Leishmaniasis***

***Picture credit*** Jorge et *al*, WHO (2012)

**Lipid Metabolism**

In *Leishmania*, lipid play important role in formation of micro domains which are major virulence factors for the parasite. Lipid metabolism includes sphingolipid metabolism, Lipopolysaccharide metabolism, Glycan metabolism etc. The plasma membranes of the divergent eukaryotic parasites, *Leishmania* and *Trypanosoma*, are highly specialized, with a thick coat of glycoconjugates and glycoproteins playing a central role in virulence. Unusually, the majority of these surface macro-molecules are attached to the plasma membrane via a glycosylphosphatidylinositol (GPI) anchor. In mammalian cells and yeast, many GPI-anchored molecules associate with sphingolipid and cholesterol-rich detergent-resistant membranes, known as lipid rafts (Figure 2.6). Lipid micro domains, including rafts and caveolae, have an important role in the organization of membrane proteins, in cell–cell contact and in numerous signaling processes. Lipid rafts in *Leishmania major* as being enriched in sphingolipid (inositol phosphosphingolipid) and sterol is very essential for the survival of the parasite.

Schematic representation of membrane rafts components in *L.major* .It shows association between sphingolipid and sterols in the outer and inner leaflets of the membrane bilayer results in the formation of a liquid-ordered raft phase (red and blue regions) that is structurally distinct from the surrounding liquid-disordered phase (orange region). Lipid rafts from amastigotes are mainly constituted by GSLs and sterols. In procyclic promastigote forms, IPC and sterols are the major lipid components of the rafts.

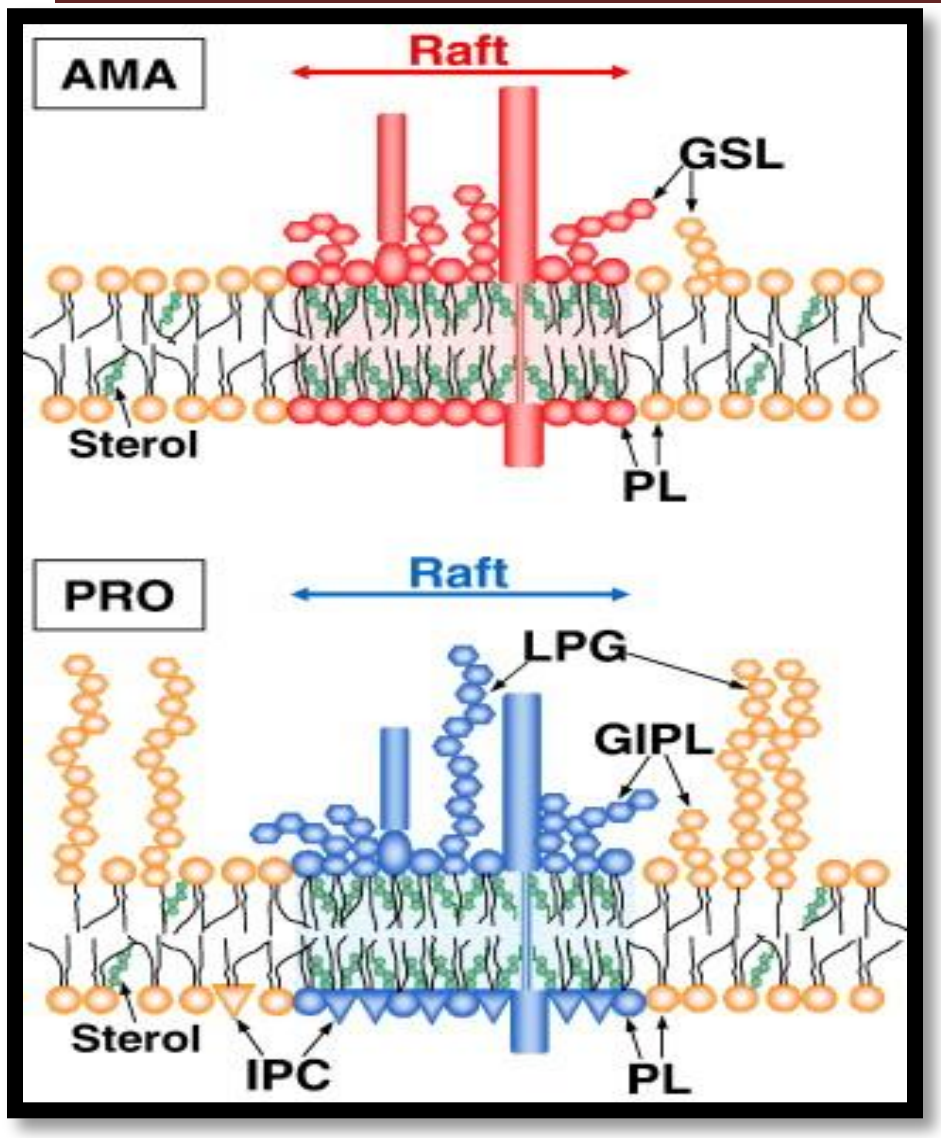

**Fig (5.7)** Lipid rafts ***Picture credit:*** Erika Suzuk**i,** Ameria K. *et al.,* 2008)

**Sphingolipid Metabolism**

Sphingolipids (SLs) are ubiquitous membrane components in pathogenic protozoan. The surface of most protozoan parasites relies heavily upon lipid-anchored molecules, to form protective barriers and play critical functions required for infectivity. Sphingolipids (SLs) play important roles through their abundance and involvement in membrane micro domain formation, as well as serving as the lipid anchor for many of these molecules. Interactions of parasite sphingolipid metabolism with that of the host may potentially contribute to parasite survival and/or host defense (W Paul Denny., 2001). The basic structure of a sphingolipid consists of a long-chain sphingoid base backbone (distinguishing it from glycerolipids which have a glycerol backbone) linked to a fatty acid via an amide bond with the 2-amino group and to a polar head group at the C-1 position via an ester bond. Trypanosomatids such as *Trypanosoma* and *Leishmania spp*. synthesize large amounts of unglycosylated inositol phosphorylceramide (IPC), a lipid found widely among fungi and plants but absent in mammals (Figure 2.7 ***Image Credit***: Lena J. Heung (2006). After the synthesis of dihydroceramide, the pathway can be generally divided into mammalian- and fungal/plant-specific branches. Synthesis of IPC occurs by the transfer of inositol phosphate from PI to the C-1 hydroxyl group of ceramide or phytoceramide. This reaction is catalyzed by IPC synthase, which is localized to the Golgi of yeasts.

**Ceramide + L-1-phosphatidyl-inositol IPC ⇄ S inositol phosphorylceramide + a 1, 2-diacylglycerol**

IPC is present together with other SLs and sterols in organized lipid rafts but it is never found attached to any GPI-anchored protein or GIPL. As IPCS is parasite specific and has no functional equivalent in mammals, it can serve as putative drug target for leishmaniasis

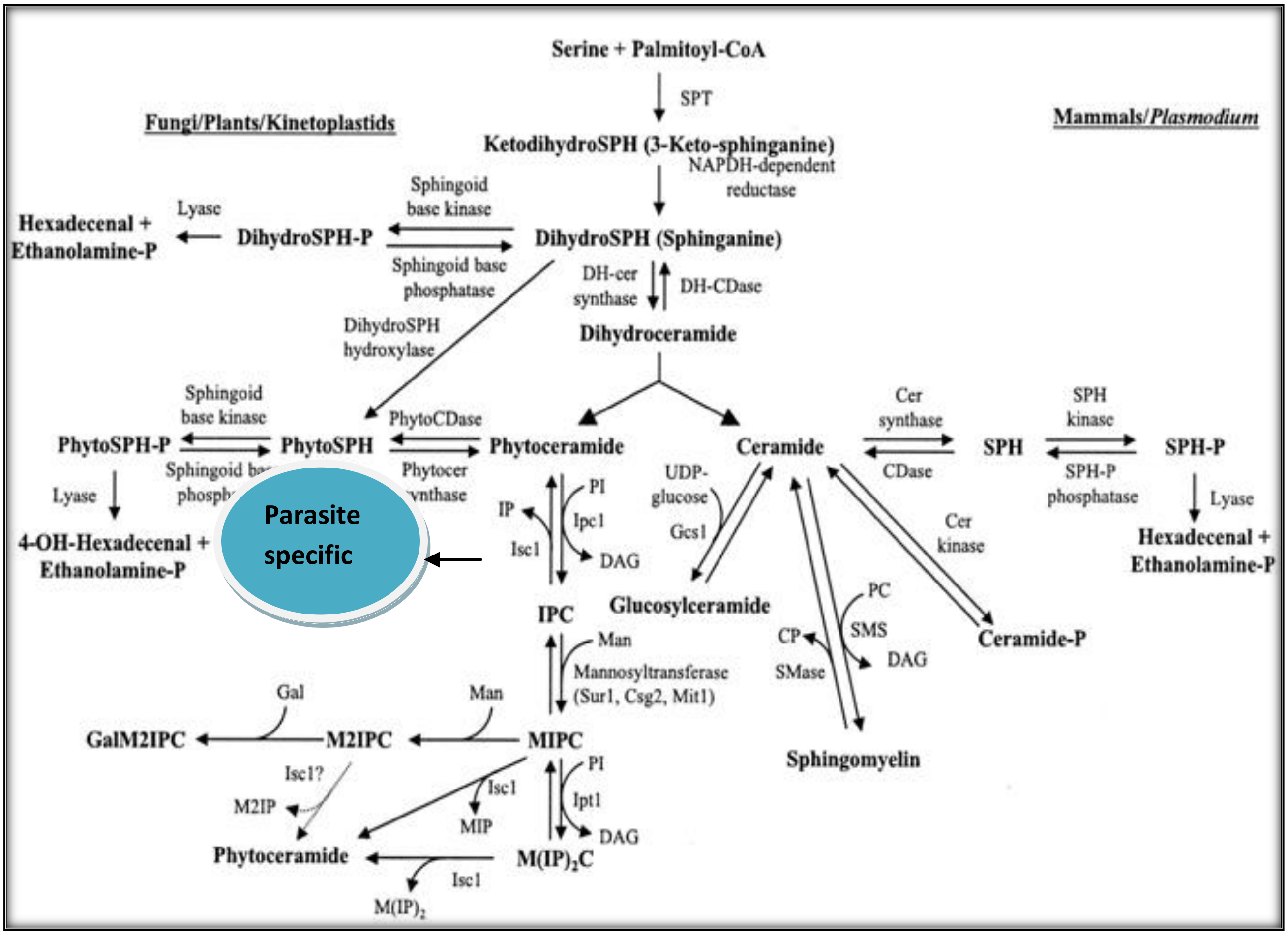

**Fig 5.8** Sphingolipid metabolism

## Existing Drugs:

Medicines called antimony-containing compounds are the main drugs used to treat leishmaniasis .There are several drugs available for the treatment of leishmaniasis, but many of the newer medications are not yet available in all endemic areas. Drug resistance is a concern in regions using monotherapies for treatment (. Lucio H., *et al*, 2012)

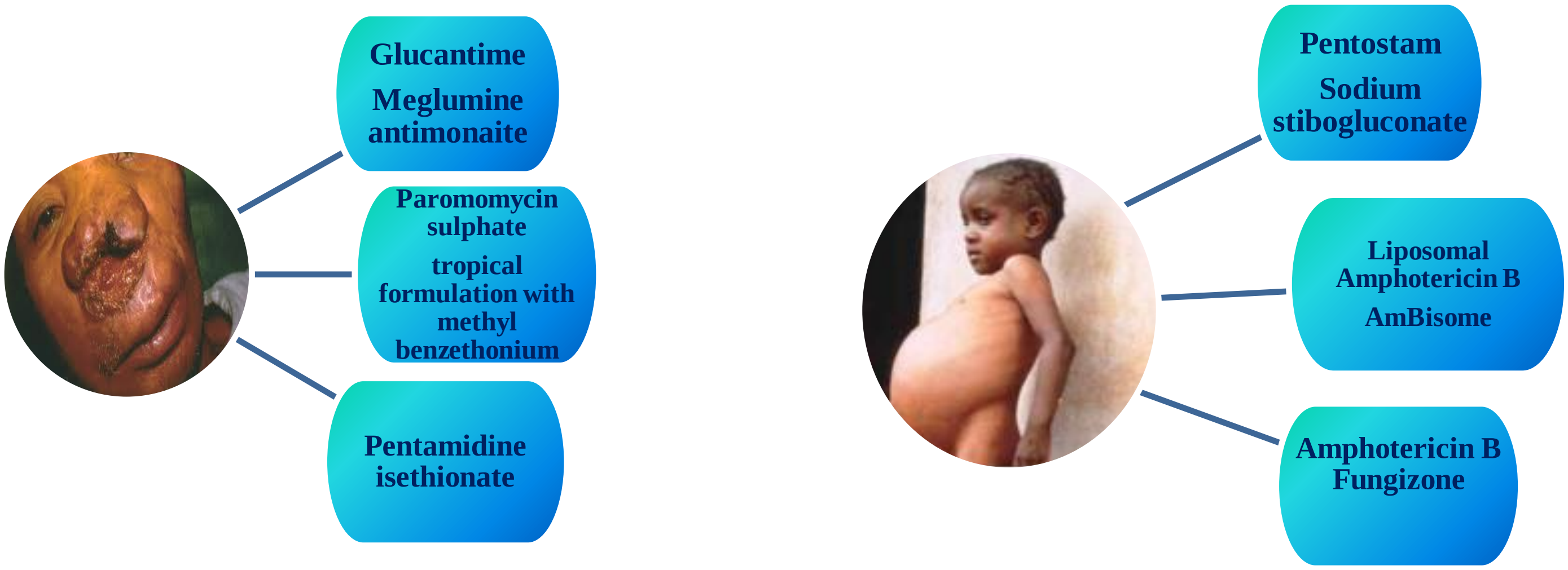

**Fig (5.9)** Drugs for Cutaneous and visceral leishmaniasis

**Limitations of available Drugs:**

- Due to side effects such as high cardiotoxicity pancreatitis and and nephrotoxicity patients should be hospitalized and monitored, as treatment may need to be suspended.
- Lack of efficacy
- Requirement for hospitalization and/or cost.
- Due to drug resistance
- Improper and inaccurate dosage of the drugs
- Several changes in the physiological parameters in the parasites.

Nevertheless, the lack of adequacy for administration in the field, toxicity and resistance issues of the current therapies highlights the need of new drugs for Leishmaniasis. To overcome the limitations of conventional drug discovery and delivery systems, new approaches must be followed using fascinating science like system and synthetic biology.

## Genetic toggle switch

Switch is important part of DNA that makes how and when to use the gene. E.g. Fruit fly sparks in the wings can be switched in and off. Switching of gene is controlled by gene regulatory network proteins. Genetic toggle switch—**a synthetic, bistable gene-regulatory network** (S. Gardner *et al., 2000)* is composed of two repressors and two constitutive repressible promoters (figure 2.8). Each promoter is inhibited by the repressor that is transcribed by the opposing promoter. When each gene encodes a transcriptional repressor of the other and each repressor is blocked by a chemical input, the system can be switched between two stable states (i.e., "gene 1 on, gene 2 off" vs. "gene 1 off, gene 2 on".

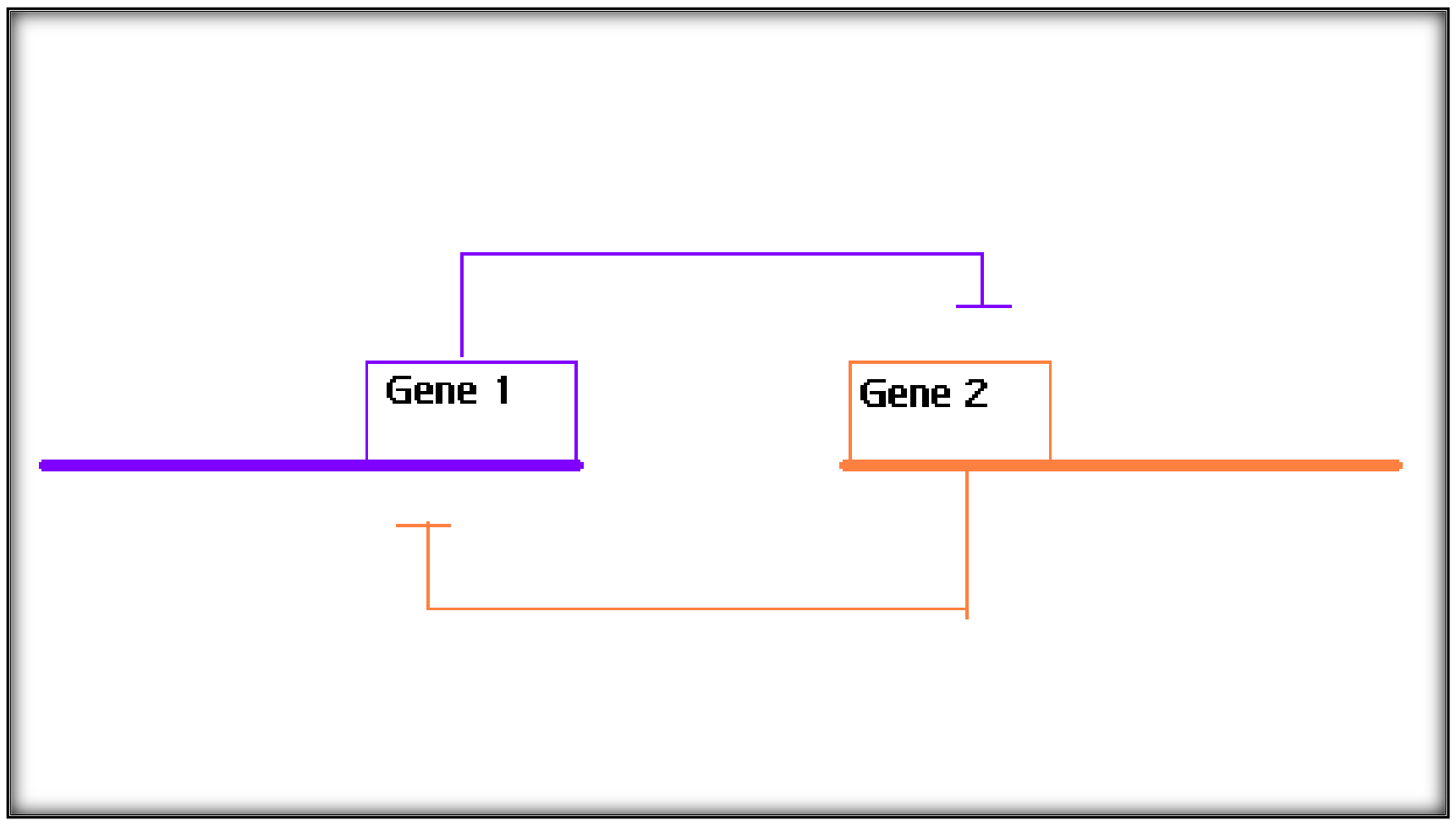

**Fig (5.10):** Schematic representation of Genetic toggle switch

**Bistability**

Bistability is a dynamic feature of a particular gene to preferentially toggle between two steady-states. Bistability arises within range of biological systems from lamda phage bacterial switch to signal transduction pathways. The state of gene switches plays pivotal roles in cell fate decision. The expression level of a gene switch does not change gradually but rather has two distinct steady-states: HIGH or LOW, ON or OFF, ALL or NONE. The ability of switches to convert a graded signal into a binary response ensures that a cell responds in a decisive manner or unambiguously commit to a specific program. There are generally two means of achieving bistability in a network: a positive feedback loop or mutual repression (i.e., double negative feedback). Bistable behavior of gene switches have been reported to play pivotal roles in many important aspects of cell physiology, including cell fate decisions cell cycle control cellular responses to environmental stimulation. Changes in regulatory mechanism may results in genetic switching in the bistable system.

**Steady states**

Attractors or steady states are cycles of states and are assumed to be associated with the stable states of cell function (F. Li, T. Long *et al*, 2003). The states in which the network resides most of the time, attractors in models of gene-regulatory networks are expected to be linked to phenotypes (S. A. Kauman *et al.*, 1993). Hence attractors represent long term behavior of genes or protein in regulatory network.

**Bifurcations**

The number and stability of the steady states may change as the value of some control parameter changes value. The critical value at which the qualitative change of the steady states occurs is called a *bifurcation point.* The saddle-node bifurcations are the bifurcations where by changing the control parameter two steady states, one stable and one unstable, will coalesce and disappear. Different bifurcation methods give different steady states but saddle point is most appropriate for the bistable genetic switches as it gives two states.

**Repressilator**

Three transcriptional repressor systems can be used to build an oscillating network. The repressilator consists of three genes connected in a feedback loop, such that each gene represses the next gene in the loop and is repressed by the previous gene (Fig 5.11). In addition, green fluorescent protein is used as a reporter so that the behavior of the network can be observed using fluorescence microscopy (M Elowitz *et al.*, 2009).

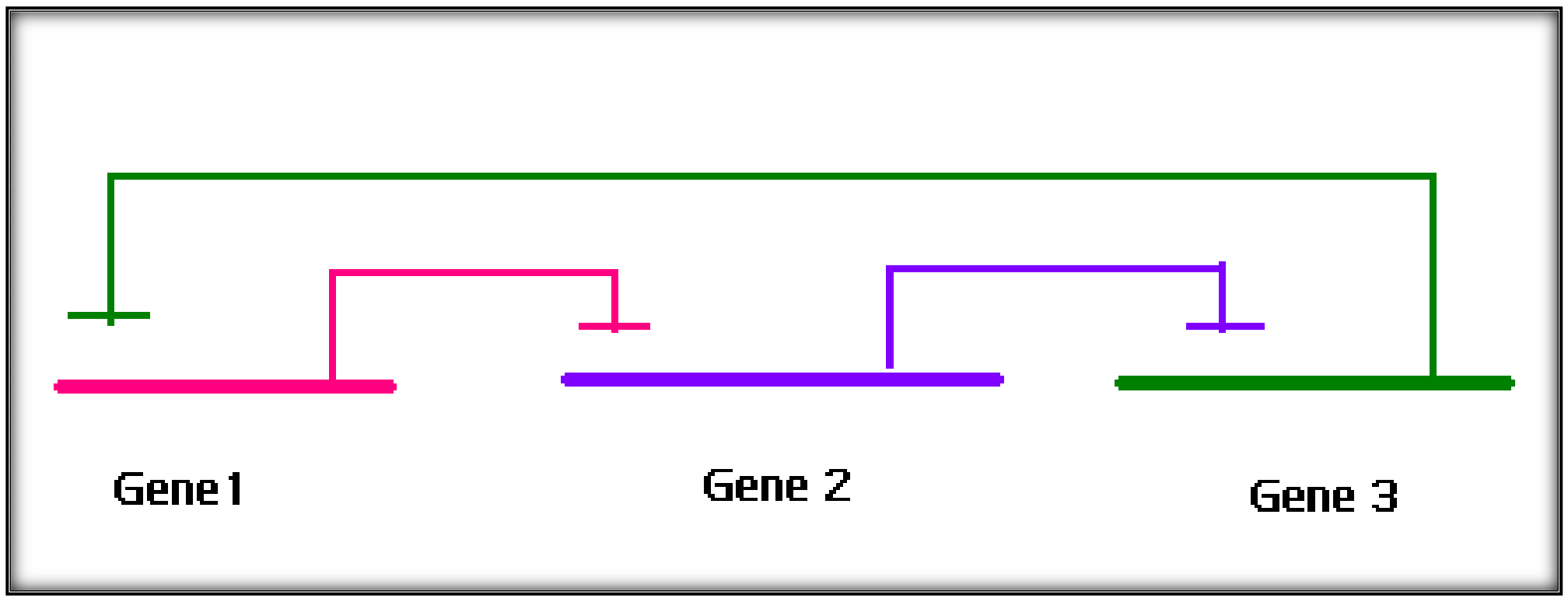

**Fig (5.11)** Gene 1 represses Gene2, Gene 2 represses Gene 3 and Gene 3 represses Gene 1 in cyclic manner

.

**Toggle switch coupled with Repressilator**

Coupling is obtained by common variable between two circuits. There are two types of coupling where in former type the expression of one gene of toggle switch is under control of one protein of the repressilator. In later type expression of one gene of repressilator is controlled by toggle switch, these are present at multi stages of Gene regulatory network supposedly at hierarchial one. Former type is present at Bistable stages of Gene Regulatory Network (GRN). In nutshell model should be able to connect toggle switch and oscillator in such way that transition from one steady state to another steady state of toggle switch via the coupling induce oscillation in repressilator. The design of the repressilator was guided by two simple mathematical models, one continuous and deterministic and the other discrete and stochastic.

**Stochastic Simulation**

Modeling approach for genetic circuit includes common approaches like Logical, continuous and stochastic modeling (G Karlebach., 2008) .Stochastic models are also called single-molecule level models as they take the fluctuating concentrations of single molecules into account when describing a circuit. The stochastic models are built up much like the ODEs but instead of a reaction rate they make use of a reaction probability. The system can then be run with a stochastic simulator, like Dizzy (S Ramsey., 2005), using algorithms made for stochastic simulation of coupled chemical reactions like the Gillespie Direct (Gillespie D., 1997) or Gibson-Bruck algorithms (M Gibson *et al*., 2000).Deterministic approach can used for analyzing bistability in terms of kinetic parameters but it can describe only average behavior of system based on large fluctuations of system behavior in different cells. Moreover it cannot realize experimental results with different genetic switching in different cells under same condition. For the biological systems when detailed information of biochemical reaction is not available, stochastic approach is boon for the system to study which is in our case. The circuits may be highly affected by stochastic mechanisms and the determinism produced by the continuous models may not be sufficiently descriptive. Stochastic analysis gives narrower bistable area than deterministic approach. Stochastic fluctuations push the system towards the other stable steady state (S Hagen Johansen 2011).

There are three types of stochastic simulations:

- Stochastic exact (Master Equation) where reactions are represented by variables like protein concentration or gene product.
- Stochastic (tau-leap): Solves master equation based on controllable dimensionless error parameters. (Poison distribution per unit time)
- Hybrid: Combines stochastic simulation algorithm of Gibson –Bruck with different algorithm for numerical integration of ODEs

From the above mentioned methods tau-leap type of stochastic simulation is most applicable for genetic circuit because it links to biochemical reaction system to Euler method which is useful to describe the reconstruct DNA sequence from its fragments .It gives more quantative description of stochastic dynamics of biological system.

**Stochasticity in genetic switches**

Gene expression is exposed to stochasticity caused by fluctuations in transcription and translation, despite constant environmental conditions giving rise to diversity and differentiation of cell types (M Kærn *et al*., 2005). Stochasticity in gene expression of genetic circuits can potentially predict transition among two states of genetic circuits. Gene expression is vulnerable to fluctuations that are caused by noise in the system. The total noise in a cellular environment can be divided in the noise arising from the gene expression itself, intrinsic noise, and that of the fluctuations in all the other components of a cell, extrinsic noise, like transcription factors and RNA polymerase abundance. In stochastic model intrinsic noise can switch the system from intermediate to steady state. Extrinsic noise can affect key regulatory processes. Also one can investigate switching behavior by the measurement of noise. The cellular processes have to be robust in order to cope up with the noise .Moreover this approach gives an appropriate technique for introducing noise into models to study robustness of genetic circuits.

## Robustness of synthetic circuit

Robustness is a property that allows a system to maintain its functions despite external and internal perturbations. Robustness is a feature often observed when studying biological systems.

Observable phenomena that characterize such robustness are adaptability, insensitivity to parameter changes and resistance to structural damages. Adaptive biological systems have the ability to change mode in a changing environment but still maintain the same phenotype. When designing and studying biological systems these properties will be able to ensure or explain the robustness of the system (H Kitano., 2004). In genetic regulatory network robustness of steady state can be defined as probability of a steady state reverting back to itself when the expression of one or more nodes is perturbed from its original expression value. It is imperative that robustness of cellular steady state is reflected by robustness of attractors under stochastic simulation of gene regulatory network .Hence biologically motivated stochastic models for quantifying the robustness properties of genetic circuits is essential to compare multiple network configuration for same biological synthetic model.

**Statistical Inference of genetic Circuit**

Mathematical and computational modeling of genetic regulatory networks promises to uncover the fundamental principles governing biological systems in an integrative and holistic manner. It also paves the way toward the development of systematic approaches for effective therapeutic intervention in disease**.** The difference between inference and design of synthetic circuits is that in design we try to reconstruct the system from data that we observe; we seek to construct the system that produces the data we would like to observe i.e. desired behavior. Hence after construction of circuit and its simulation, it is mandatory to check for the circuit performance. If it performs well, it is applicable but if not then check for sensitivity and robustness analysis. When desired robust solution is obtained after perturbations and optimization, we can implement the design in wetlab. Need for the inference of gene regulatory network due to protein expression when needed in spite of infrequent and stochastic gene expression may be due to following reasons:

- It may be due to population transcriptional co operation; means every protein population does not mean to form gene product.
- Check points to assure that cascaded events are adequately synchronized.

- Widespread redundancy in the genes and in regulatory pathways even in normal condition (normal fluctuations) in protein production can be large which is related to regulatory threshold which controls the expression of downstream genes.
- Consequences in wide variation from cell to cell in switching time for controlling protein to activate gene it controls.
- Without a coordinative mechanism, time variation will cause errors synchronization of cellular functions.
- The mechanism to provide coordination can be provided by regulatory check points with conditions provided.
- Regulatory circuits design and molecular details that determine the kinetic parameters must be under selection pressure.
- Redundancy in the circuit efforts the resilience in gene circuit performs with respect to gene mutation and regulation failure caused by erratic protein production.
- Mutation in one or more of gene in regulatory network can increase probability of failure of networks that accounts for robustness of circuit.

. There are various approaches for the inference of gene regulatory networks. Bayesian method, Boolean method and ODE (ordinary differential Equation) method. To perform the network inference, network can be created by using Natural language rules or one can import the constructed model in SBML (Systems Biology Markup Language). Network inference can be made by time series data generated after the simulation of circuit. For Qualitative inference Boolean method is used and Quantitative inference Bayesain method is used.

## Quantative and Qualitative synthetic network

Successful completion of the various genome projects has led to the realization that effective models for predicting cellular behavior must take into account the network interactions that dynamically mediate gene regulation. Since behavior arising from these complex interactions is difficult to predict without quantitative models and qualitative models, there is a need for experimentally validated computational modeling approaches that can be used to understand the complexities of gene regulation

**Bayesain method**

In the formalism of Bayesian Networks, the structure of a genetic circuit system is modeled by a directed acyclic graph $G= \langle V, E \rangle$. The vertices I ∈ V, I = (1….. *n),* represent genes regulation levels and Correspond to random variables *Xi.* In network node are genes and edges are condition dependency regulators in circuit. In Bayesain method relationship between genes in biological system can be studied. Probabilistic framework captures the stochastic nature of biological system. It provides insight into uncertainty in circuit. Posterior probability distribution is quantity derived from Monte Carlo simulation in Bayesian method. Posterior probability distribution over possible design parameter values that can be analyzed for parameter sensitivity and robustness provide credible limits as design parameters. It encodes ability of each design to achieve desired behavior. This method includes network probability matrix, network uncertainty and no of regulators in the circuit. Synthetic gene regulatory network is constructed in *Arabidopsis Thaliana* (Locke, J.C *et al*.,2006) Bayesian network method was used to construct transcription regulatory network by selecting five regulatory genes and their time series data from ODE model. Significant regulators in the network were identified. Regulatory genes identified from Bayesian network on basis of conditional probability distribution were experimentally validated against stress and external perturbation of environment. Network analyses were performed by GRENITS package of R.

GRENITS

GRENTIS is Gene Regulatory Network Inference using Time series .The package offers four network inference statistical models using Dynamic Bayesian Networks and Gibbs Variable Selection: a linear interaction model, two linear interaction models.

**Boolean method**

A Boolean network G (V, F) is defined by a set of nodes (genes) and a list of Boolean functions where V= $\{x_1…x_n\}$ are set of genes and F= $(f_1.......f_n)$. Each represents the state (expression) of gene, where represents the fact that gene is expressed (active) and means it is not expressed (inactive).The list of Boolean functions represents the rules of regulatory interactions between

genes(S Ilya *et al.,2002)*. This method includes conversion of the model of circuit to Boolean network (BN), network analyses and perturbation experiments by generating random networks. It includes Synchronous network where the assumption is that all genes are updated at same time and asynchronous network where genes are updated at different point of time. Methods to analyze the Boolean network includes find the attractors and the transition between the attractor. Also check the robustness of gene regulatory network to noise and mismeasurements. Several genetic networks have been successfully modeled and analyzed using BNs, such as the mammalian cell cycle (Faure *et al.*, 2006), or the yeast cell cycle (Li *et al.*, 2004).

<u>BoolNet</u>

Existing software tools in this field often specialize on certain aspects of BN research, or do not support all three types of networks (e.g. Albert *et al.*, 2008; Klamt *et al.*, 2007; Wuensche, 2009). The R package *BoolNet* provides methods for all major uses of synchronous, asynchronous and probabilistic BNs and includes novel functions for attractor search, robustness analysis and binarisation. Boolean network for the yeast cell cycle and mammalian system was done using BoolNet.

**International and national status**

The first synthetic gene circuits were engineered in bacteria, and more recently have been established in yeast and mammalian systems (Travis S Bayer., 2005; Hasty et al. 2002; Kaern et al. 2003; Sprinzak and Elowitz 2005). The first toggle switch and repressilator are explained below.

<u>Genetic toggle switch:</u>

Construction of a genetic toggle switches—a synthetic, bistable gene-regulatory network—in *Escherichia coli* and provide a simple theory that predicts the conditions necessary for bistability. The toggle is constructed from any two repressible promoters arranged in a mutually inhibitory network. It is flipped between stable states using transient chemical or thermal induction and exhibits a nearly ideal switching threshold. In the experiment the LacI-repressor was used as Repressor 2, repressing the promoter *Ptrc-2* being inducible by IPTG working as Inducer 2. Promoter 2 encodes either a heat inducible cI repressor or anhydortetracycline (aTc) inducible tetR repressor that will repress Promoter 1.If there is expression from Promoter 2 (the *Ptrc-2* promoter) there will be expression of a reporter protein, in this case in the form of the *GFPmut3* gene. The vector design used by Gardner *et al.* is shown in **Figure 2.16.** By using the construct design in *E. coli* strain JM2.300, they created a genetic circuit with two separate stable expression states (bistable) in which could be switched between by adding an inducer (chemical or physical) to the medium.

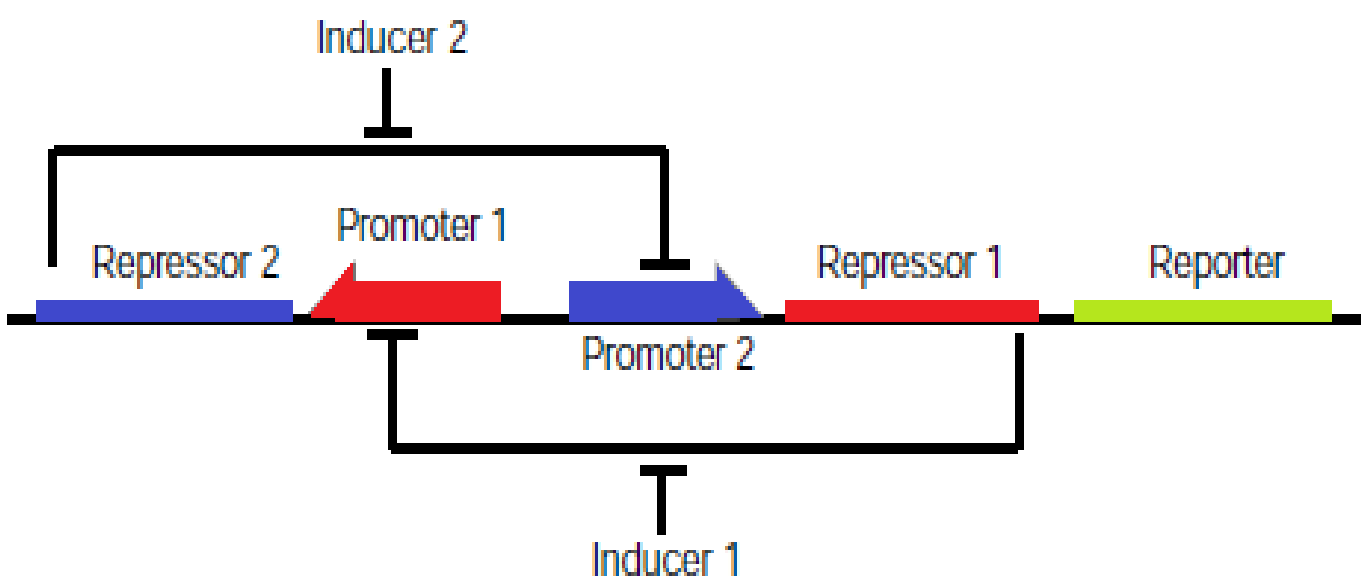

***Fig (5.12)*** a.) Schematic presentation of the genetic toggle switch made by Gardner *et al.* Promoter 1 is repressed by the Inducer 1 inducible Repressor 1. Promoter 2 is repressed by the Inducer 2 inducible Repressor

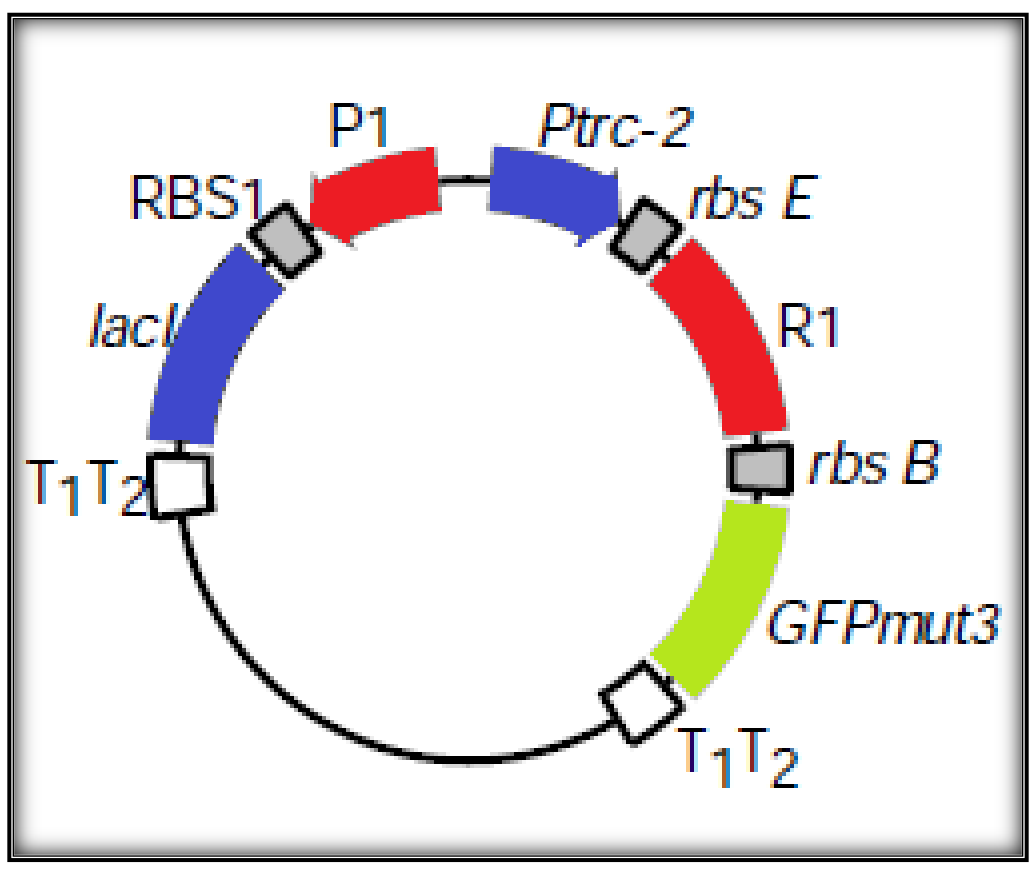

b.)

The vector design applied for demonstration of a genetic toggle switch

Image Courtesy: Timothy S. Gardner., 1999

Repressilator

The repressilator was constructed by Elowitz and Leibler as a synthetic oscillatory network of transcriptional regulators, a biological oscillator. Here it was used three repressors acting in sequence on each other. If it is being transcribed from the *PLtet01* promoter in the beginning there will be produced _ cI repressor and reporter protein. The cI repressor will repress the _ *PR* promoter, and therefore there will be no lacI repressor produced and the RNA polymerase will transcribe from the *PLlac01* promoter, and the tetR repressor will be produced. This repressor will repress both *PLtet01* promoters stopping the production of the reporter protein and the _ cI repressor. This will leave open the _ *PR* promoter and there will be produced lacI repressors. This will stop the production of the tetR repressor, and thereby rend the transcription from the *PLtet01* promoters again producing the reporter protein. . This cyclic fashion of repression gives the circuit an oscillating behavior. The two plasmids illustrated in **Figure 2.17** were contained by a culture of *E. coli* strain MC4100 to produce the oscillating behavior.

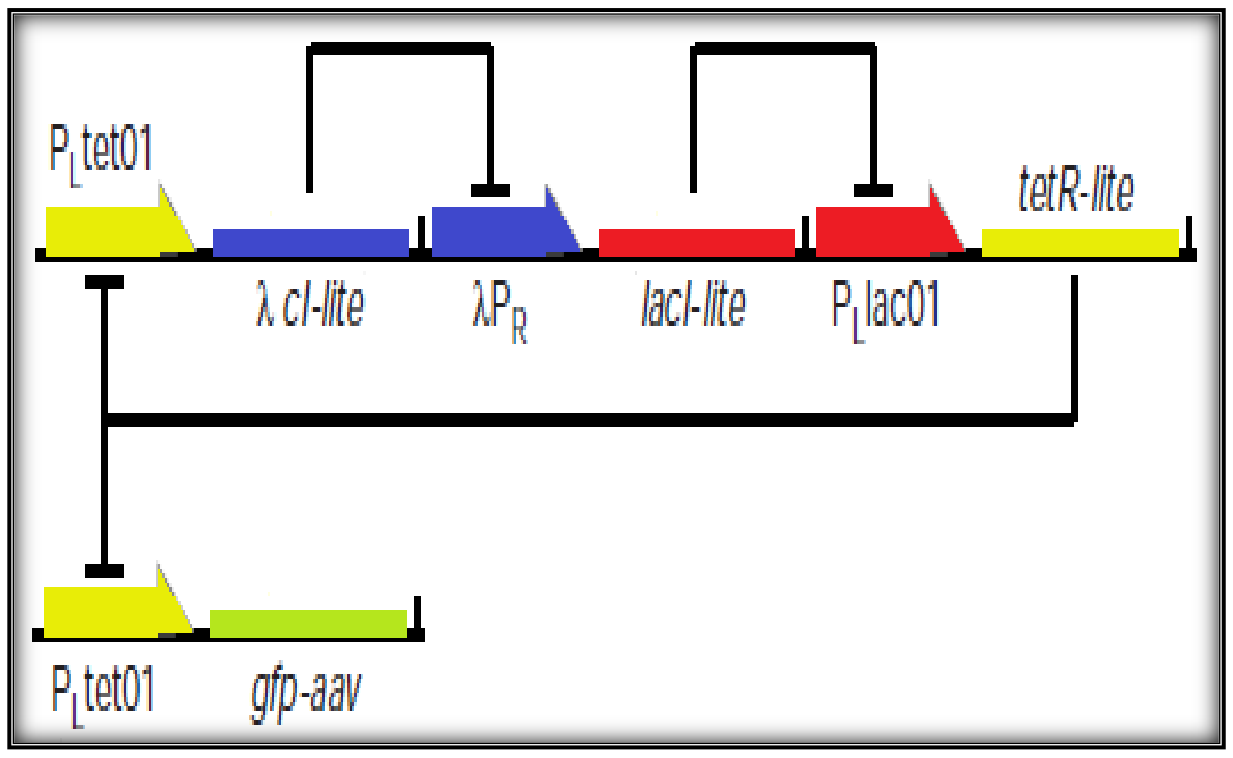

a.) Representation of the repressilator. One of the two *PLtet01* promoters encodes the _ *cI-lite* gene. The _ cI-lite repressor can repress the _ *PR* promoter which encodes the *lacI-lite* gene. The lacI-lite repressor can repress the *PLlac01* promoter which encodes the *tetR-lite* gene. The tetR-lite repressor can repress both the *PLtet01* promoters and thereby the expression of the reporter gene *gfp-aav*.

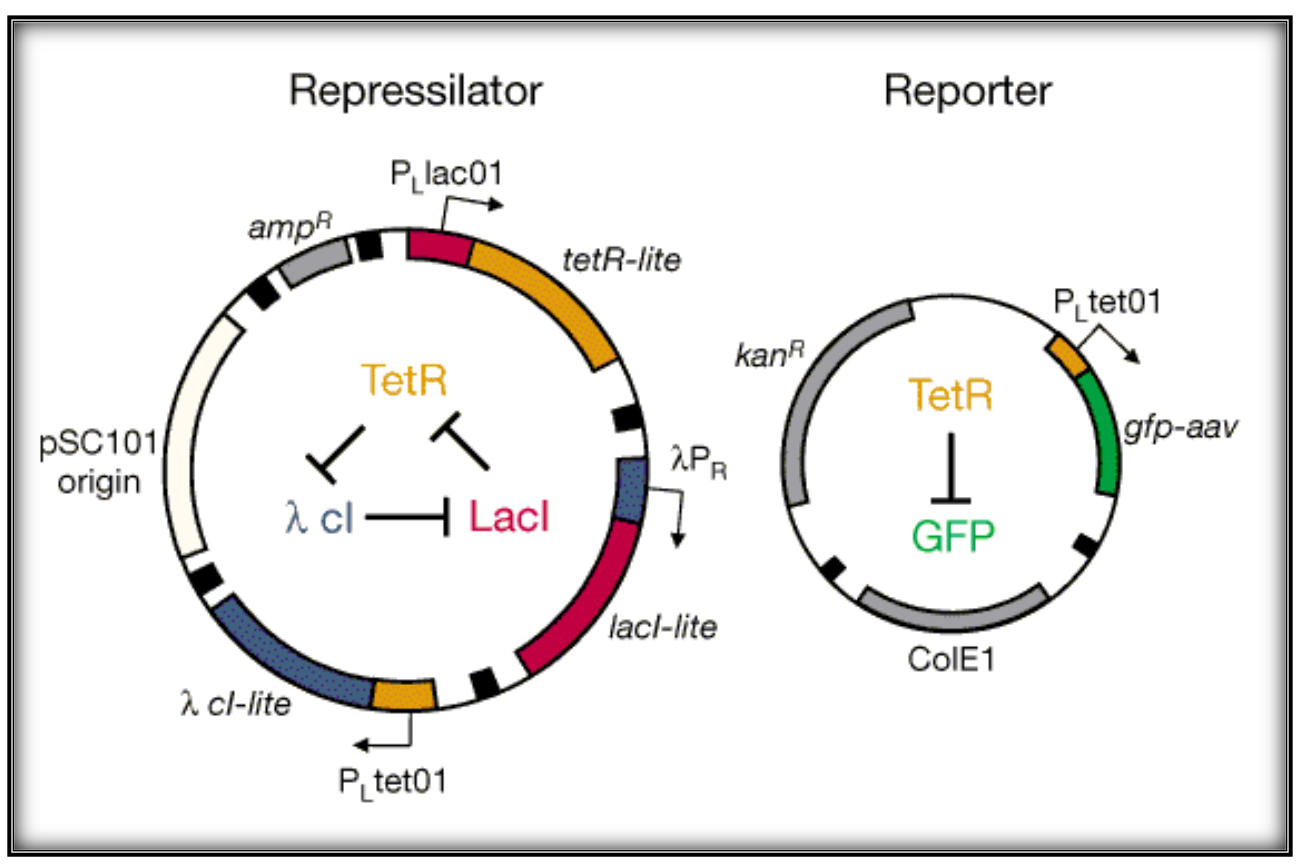

Figure 5.13
***Picture credit***: Elowitz M.B. and Leibler S., 2000.

b.) The vector design applied for the creation of the oscillating repressilator and reporter construct

In Eukaryotes, yeast a bistable positive feedback loop expressed in yeast in which a tetracycline-dependent activator turns on its own expression (Becskei *et al*. 2001) .In this system, cells initially in an "off" state could be flipped onto an "on" state in the presence of tetracycline. In response to a graded increase in tetracycline, a bimodal distribution of cells is created that are either in the on or off state.

Using the same general design concepts as described in the bacterial toggle switch, Kramer and colleagues (Kramer et al. 2004; Kramer and Fussenegger 2005) designed a mammalian (CHO

cell)-based hysteretic toggle switch using streptogramin and macrolide-dependent transcription factors, each fused to a KRAB repression domain. The system switches between two steady states in response to the specific inducers.

To the best of my knowledge and literature survey so far done there is no evidence for the construction of toggle switch or repressilator for the protozoan parasite.

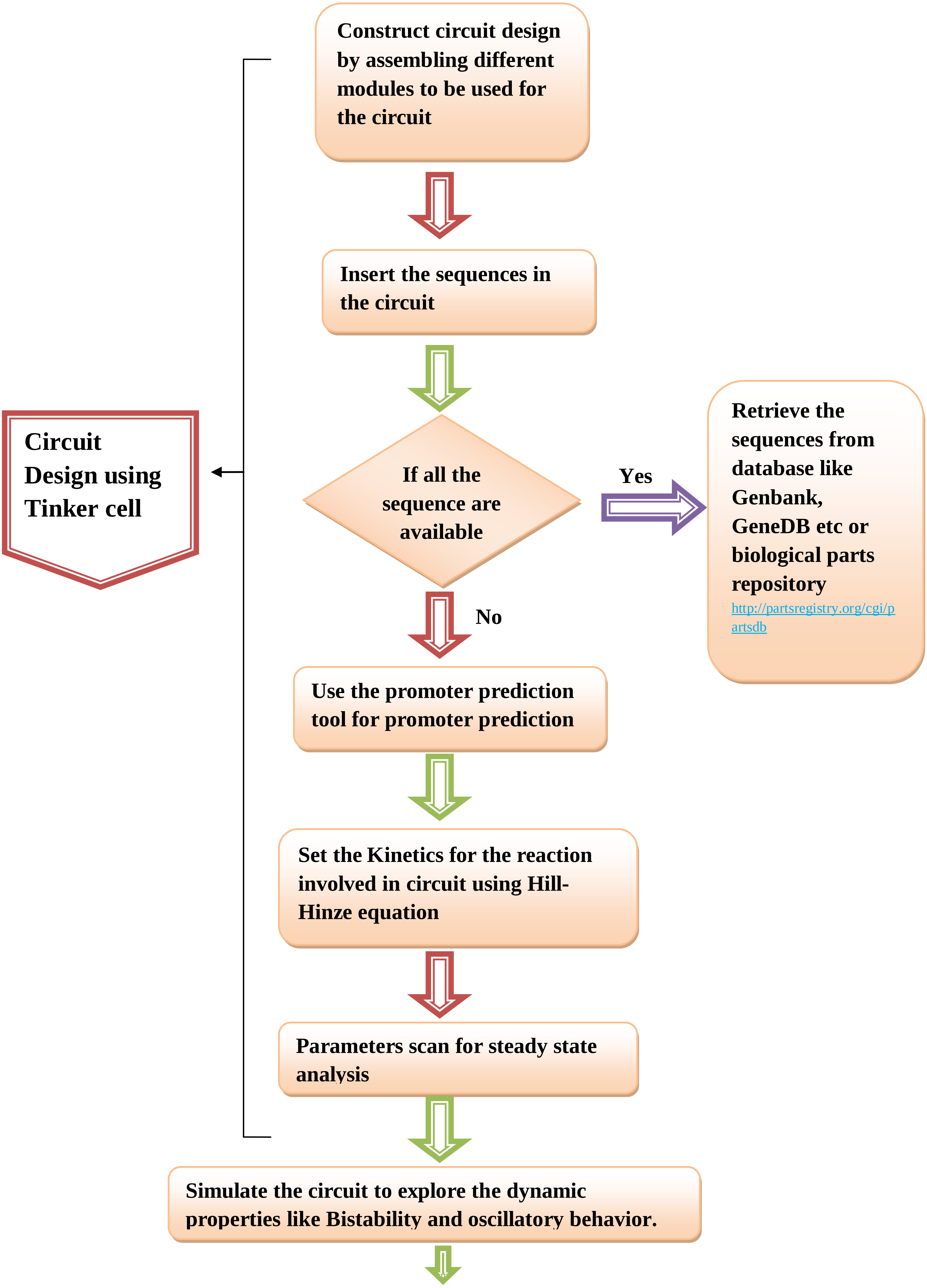

Construct circuit design by assembling different modules to be used for the circuit
Insert the sequences in the circuit
If all the sequence are available
Yes
Retrieve the sequences from database like Genbank, GeneDB etc or biological parts repository
http://partsregistry.org/cgi/partsdb
No
Circuit Design using Tinker cell
Use the promoter prediction tool for promoter prediction
Set the Kinetics for the reaction involved in circuit using Hill-Hinze equation
Parameters scan for steady state analysis
Simulate the circuit to explore the dynamic properties like Bistability and oscillatory behavior.

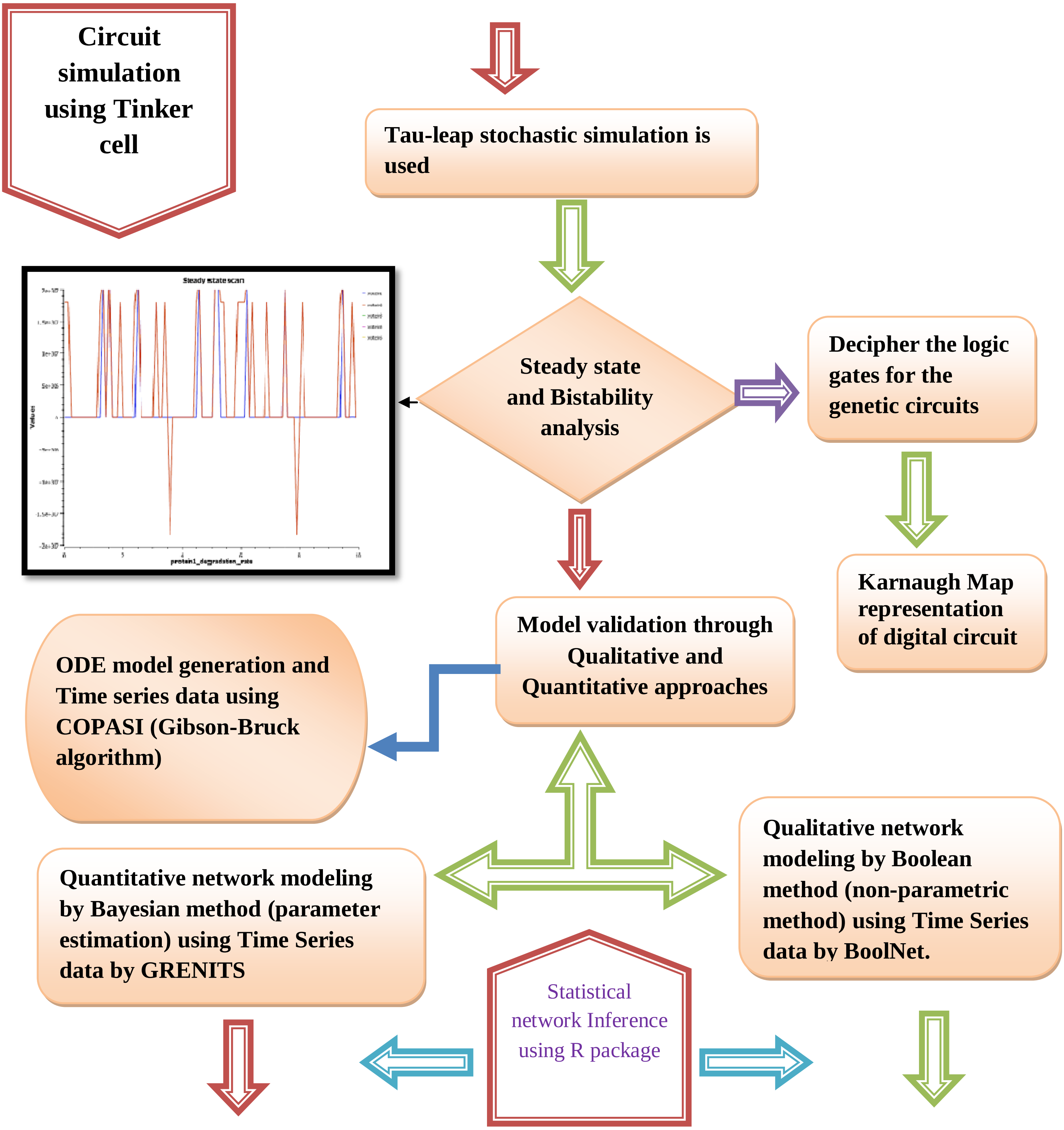
Circuit simulation using Tinker cell
Tau-leap stochastic simulation is used
Steady state scan
protein1_degradation_rate
Steady state and Bistability analysis
Decipher the logic gates for the genetic circuits
Karnaugh Map representation of digital circuit
Model validation through Qualitative and Quantitative approaches
ODE model generation and Time series data using COPASI (Gibson-Bruck algorithm)
Quantitative network modeling by Bayesian method (parameter estimation) using Time Series data by GRENITS
Qualitative network modeling by Boolean method (non-parametric method) using Time Series data by BoolNet.
Statistical network Inference using R package

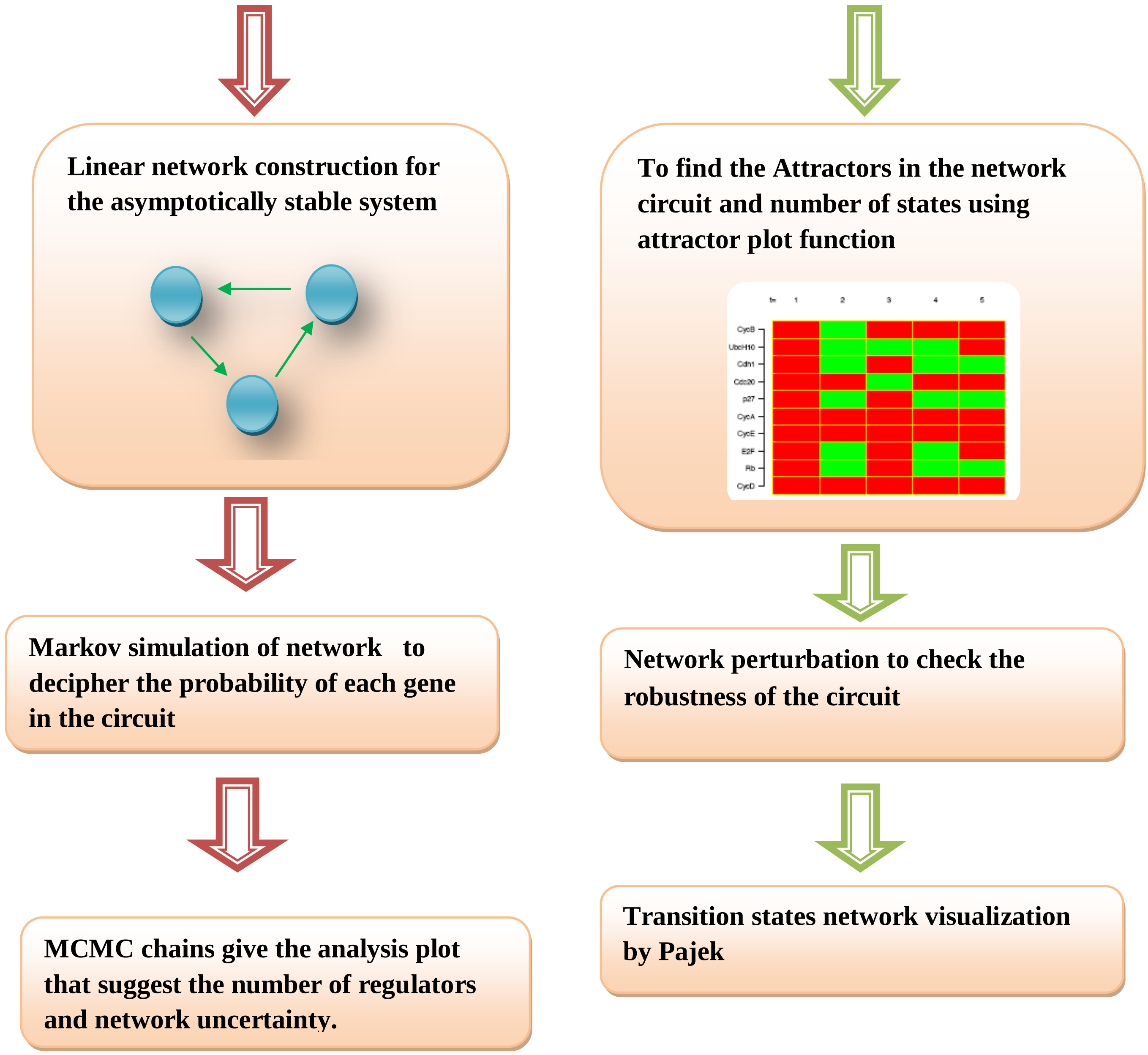

Fig 6.1 **Schematic Representation of the workflow implemented in project**

## Implementation of the circuit design using Tinker cell

Tinker Cell is a visual modeling tool that supports a hierarchy of biological parts. Each part in this hierarchy consists of a set of attributes that define the part, such as sequence or rate constants (Chandran D *et al.*, 2009). Using a computer aided design (CAD) application, it is possible to construct models using available biological "parts" and directly generate the DNA sequence that represents the model, thus increasing the efficiency of design and construction of synthetic networks. Each part in the model can store a large amount of information associated with the part, such as database IDs, annotation, ontology, parameters, equations, sequence, and information required by experimentalists, such as plasmid information or restriction sites found within the part. Parts can be loaded along with all their known information from databases, although this feature depends on the growth of the databases themselves Tinker Cell has extensive support for constructing gene regulatory networks by connecting parts such as promoters, ribosomal binding sites (RBS), and other genetic components listed in the parts catalog. When parts are connected together, Tinker Cell will automatically derive rate equations for transcription and translation by looking at the transcription factors bound to operator sites. Tinker cell consist of three windows: Tools window, main menu bar and parts and connection window. It includes workspace where circuit is designed.

**Circuit design:** From the parts and connection window, select parts that are to be inserted in circuit. Select repressible promoter part, coding region for each genes. On basis of the connection of the modules, reactions are set. For genetic circuit transcription repression reaction is assigned for the mutual repression between the two genes included in the toggle switch. Similarly design the repressilator with three genes connected in cyclic manner and assigns the transcription repression reaction. Coupling between toggle switch and repressilator is done by insertion of the transcription repression reaction between the first gene of the toggle switch and first gene of the repressilator in such a way that gene of the repressilator controls the gene of the toggle switch.

**Insertion the sequence:** To construct the switch for IPCS (Inositol phosphoryl ceramide synthase), sequences homologous to the *L.major* was retrieved by BLAST search.

BLAST: Basic Local Alignment search tool, BLAST finds regions of similarity between biological sequences. BLASTP programs search protein databases using a protein query. BLAST (http://blast.ncbi.nlm.nih.gov/BlastP) requires a query sequence to search for similarity sequences against the target databases or a sequence database containing multiple such sequences. Input sequences are in FASTA format or Genbank format.

Sources of the sequences are:

GeneDB: GeneDB (http://www.genedb.org) is a genome database for prokaryotic and eukaryotic pathogens and closely related organisms. The resource provides a portal to genome sequence and annotation data, which is primarily generated by the Pathogen Genomics group at the Welcome Trust Sanger Institute. It combines data from completed and ongoing genome projects with curated annotation, which is readily accessible from a web based resource. GeneDB now holds the genome sequences of 9 apicomplexans, 3 of them human pathogens; 12 kinetoplastid protozoans, 7 of them human pathogens; 3 parasitic helminths, all human pathogens; and 16 bacterial species, 13 of them either human pathogens or opportunistic human pathogens

UniProt: UniProt is a comprehensive, high quality and freely accessible database of protein sequence and functional information, many of which are derived from the genome sequencing projects. It contains a large amount of information about the biological functions of proteins derived from the research literature (http://www.uniprot.org)

Registry of standard Biological parts: The Registry is a continuously growing collection of genetic parts that can be mixed and matched to build synthetic biology devices and systems developed in 2003 at MIT; the Registry is part of the Synthetic Biology community's efforts to make biology easier to engineer (http://partsregistry.org/cgi/partsdb/Statistics.cgi).

OrthoMCL: OrthoMCL is a genome-scale algorithm for grouping orthologous protein sequences. It provides not only groups shared by two or more species/genomes, but also groups representing species-specific gene expansion families. So it serves as an important utility for automated eukaryotic genome annotation. OrthoMCL starts with reciprocal best hits within each genome as potential in-paralog/recent paralog pairs and reciprocal best hits across any two genomes as potential orthologous pairs. Related proteins are interlinked in a similarity graph.

Selected sequences are inserted in the modules using Text Attributes. DNA sequence viewer used to see the sequences inserted in the modules. Set the target promoter regions for the transcription repression by text attributes.

**Set Parameters:** Set the parameters for all the modules and reaction rates like protein degradation, promoter strength, transcription rate, dissociation constants and hill’s coefficient by using parameter attributes or model summary window.

**Steady state**: For parameter scan, use steady state analysis by steady state program.

**Simulation:** Stochastic simulation for the genetic circuit is done by tau-leap stochastic simulation using stochastic program.

## Logic Circuit design

Decipher the logic gates to be implemented for the circuit. Conversion of the truth table into circuit scheme via Karnaugh map, which is better representation of truth table, was done.

## ODE model

In ODE model the concentration of gene products variables like protein and gene regulation system are represented by Differential Equation**.**

### COPASI

The name COPASI stands for Complex Pathway Simulator. COPASI is an open-source software application for creating and solving mathematical models of biological processes such as metabolic networks, cell-signaling pathways, regulatory networks, infectious diseases, and many others. It includes methods for simulation and analysis of biochemical reaction. It includes methods like ODE solver that converts the reaction into ordinary differential equations. It includes methods like simulation of reaction networks, the computation of steady states and their stability, stoichiometric network analysis, e.g. computing elementary modes (Schuster *et al*., 1999), sensitivity analysis ,metabolic control analysis; Fell (1996); Heinrich and Shuster, 1997), optimization and parameter estimation. Input for the COPASI was the input file exported from the Tinker cell circuit model in SBML format.

SBML format: *Systems Biology Markup Language* is a machine-readable format for describing qualitative and quantitative models of biological systems (The SBML web page: http://sbml.org/index.psp). SBML provides a standard biochemical network model of representation and it promotes inter-operability between tools. SBML is based on XML, Extensible Markup Language, which is a standard language for describing markups languages.

It is a free and open standard with widespread software support and a community of users and developers. SBML can represent many different classes of biological phenomena, including metabolic networks, cell signaling pathways, regulatory networks, infectious diseases, and many others. Time series data for the circuit was developed by using time course method for the stochastic simulation.

## Model validation

### R software

R is a free software environment for statistical computing and graphics. R provides a wide variety of statistical (linear and nonlinear modeling, classical statistical tests, time-series analysis, classification, clustering.) and graphical techniques, and is highly extensible. R is available as Free Software under the terms of the Free Software Foundation's GNU General Public License in source code form. R can be extended (easily) via packages.

### Bioconductor package

Bioconductor is a free, open source and open development software project for the analysis and comprehension of genomic data. Bioconductor is based primarily on the statistical R programming language. It includes packages for the sequence analysis, microarray data analysis, sequence annotation, SNP (single nucleotide polymorphism) analysis, high throughput Assays. GRENITS and BoolNet are packages of Bioconductor.

## GRENITS:

GRENITS is Gene regulatory network inference using time series data. Network inference using ODE time series data was done using GRENITS. It gives the probability of the genes included in the circuit. Probability suggests the regulators of the circuit. It also gives the network uncertainty.

Steps followed in GRENITS are:

| Commands | Description |
|---|---|
| data | • Input Time sereis data |
| LinearNet | • Construt Linear network of the circuit and generates the Markov chains that gives the posterior probability |
| analyse | • Analyse the output of network,produces analyses plot and convergence plot. |
| prob.mat | • Generate the network probability matrix |
| inferred.net | • generate infered network form the regulatory ciruit |

## BoolNet

BoolNet is an R package that provides tools for assembling, analyzing and visualizing synchronous and asynchronous Boolean networks as well as probabilistic Boolean networks. In a biological context, genes can be modeled as Boolean variables (active/expressed or inactive/not expressed), and the transition functions model the dependencies among these genes. In the synchronous model, the assumption is that all genes are updated at the same time. This simplication facilitates the analysis of the networks. For synchronous and asynchronous Boolean networks, the most important tool is the identification of attractors. Attractors are cycles of states and are assumed to be associated with the stable states of cell function. Another possibility of identifying relevant states is the included Markov chain simulation. This method is particularly suited for probabilistic networks and calculates the probability that a state is reached after a certain number of iterations. To test the robustness of structural properties of the networks to noise and mismeasurements, the software also includes extensive support for perturbing networks. Time series data of the circuit is input for the BoolNet. Network constructed in BoolNet can be exported to Latex document and Pajek which is analysis and visualization tool for network.

Steps followed in BoolNet:

| Commands | Description |
|---|---|
| data | •Input Time series data |
| binarizeTimeSeries | •Converts the real values data sets to binary data as required by the network reconstruction algorithm. |
| reconstructNetwork | •Generates probability boolean netwrok and transition functions |
| getAttractors | •Gives the attractors in the ciruit.It includes both type of attractors ;aynchronous and asynchronous. |
| tt | •Gives the transition table |
| plotAttractors | •plots the attractor plot that gives no of states to reach steady state. |
| sim | •generate markov simulation to predict the steady state |
| perturbedNet | •The generation of perturbed copies of a network is a way to test the robustnessof structural properties of the networks to noise and mismeasurements. |
| testNetworkProperties | •plot the robustness plot |
| toPajek | •export network to Pajek |

## Pajek

Pajek is a program for analyzing large networks. Pajek is freeware software and can be downloaded from the URL http://vlado.fmf.uni-lj.si/pub/networks/pajek/.**Pajek** (Slovene word for Spider) is a program, for Windows (32 bit), for analysis of large networks.

## Genetic Circuit

Genetic circuit for IPCS was constructed using the software Tinker cell. Genetic circuit consists of genetic toggle switch coupled with repressilator. Toggle switch has two mutually repressing genes that encode the repressor for the opposite gene. Repressilator includes three genes that code the repressor of next gene in cyclic manner. For the construction of genetic circuit of IPCS, bidirectional reaction has to be considered. For this reason, homologs of IPCS were identified using blastp analysis.

Homologous for IPCS are:

| UniProt ID | GeneDB | Gene symbol | Gene name | Identity (%) | e-value |
|---|---|---|---|---|---|
| **Q38E53** | **Tb09.211.1030** | **SLS1** | **IPC synthase** | **43** | **1e-08** |
| **Q38E56** | **Tb09.211.1000** | **SLS4** | **EPC synthase/ SM synthase** | **44** | **4e-80** |
| **Q38E54** | **Tb09.211.1020** | **SLS2** | **EPC synthase** | **43** | **2e-67** |
| **Q38E55** | **Tb09.211.1010** | **SLS3** | **SM synthase** | **45** | **3e-82** |

**Table 7.1 Homologs of IPCS**

TbSLS1 and TbSLS4 which code for IPC synthase and SM synthase respectively in *T.brucei* were considered for the construction of the genetic toggle switch based on the homology obtained.

Functional relationship between *L.major* IPCS and *T.brucei* IPCS (SLS1) is shown by the Ortholog relationship between them. Orthologs were searched by OrthoMCL database.

**Fig 7.1** Functional relationship between *L.major* and other species was reflected by the graph of ortholog relationship given below.

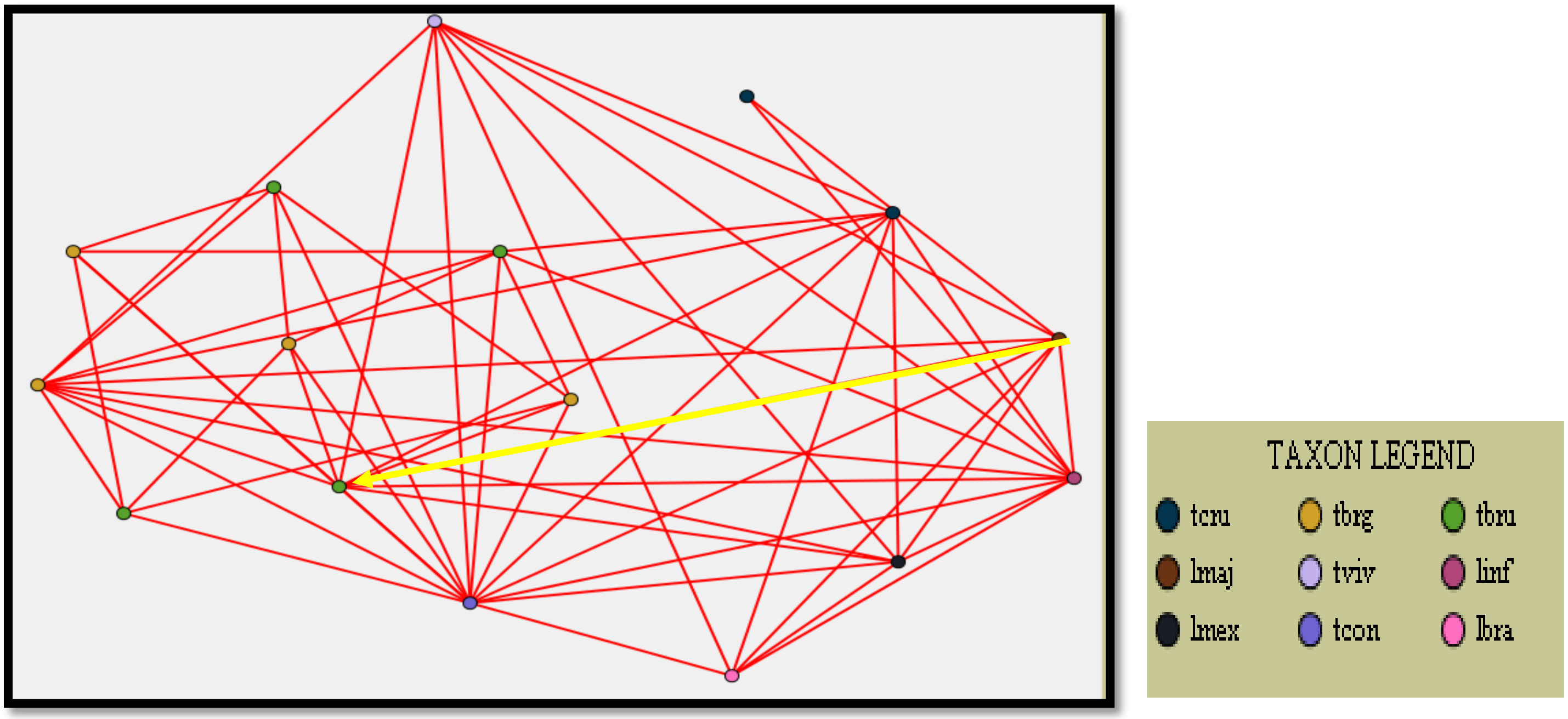

Arrow highlighted shows the 94 % orthology relationship between IPCS of T.*brucei* and L.*major*.

Design of the genetic circuit constructed for IPCS was referred from the genetic toggle switch of E.coli and repressilator of E.coli designed by Gardner and Elowitz respectively. Genetic modules used in construction of genetic circuit are

| NAME | Accesion-ID |
|---|---|
| Protein1:IPCS_1(inositol phosphoryl ceramide synthase 1) | Q38E53 |
| Protein2:IPCS_2(inositol phosphoryl ceramide synthase 2) | Q38E56 |
| Protein3:Lactose-represssor | P03023 |
| Protein4:Tetracycline repressor protein | P04483 |
| Protein5:LAMBD Repressor protein | P03034 |
| rp1:Repressible promoter of IPCS_1 | predicted |
| rp2:Repressible promoter of IPCS_2 | predicted |
| rp3:Repressible promoter of LacR | BBa_R0010 |
| rp4:Repressible promoter of TetR | BBa_R0040 |
| rp5:Repressible promoter of cI | BBa_R0051 |
| tr:Transcription repression | |
| pp:protein product | |
| IPCS_1:coding gene for IPCS_1 | SLS1_TRYB2 |
| IPCS_2:coding gene for IPCS_2 | SLS4_TRYB2 |
| lacI:coding gene for Lactose repressor | BBa_C0012 |
| Tetr;coding gene for Tetracycline repressor | BBa_C0040 |
| lamdar:coding gene for lamda repressor | BBa_C0051 |

TbSLS1 and TbSLS4 were considered as IPCS_1 and IPCS_2 in circuit.

Genetic circuit model for IPCS:

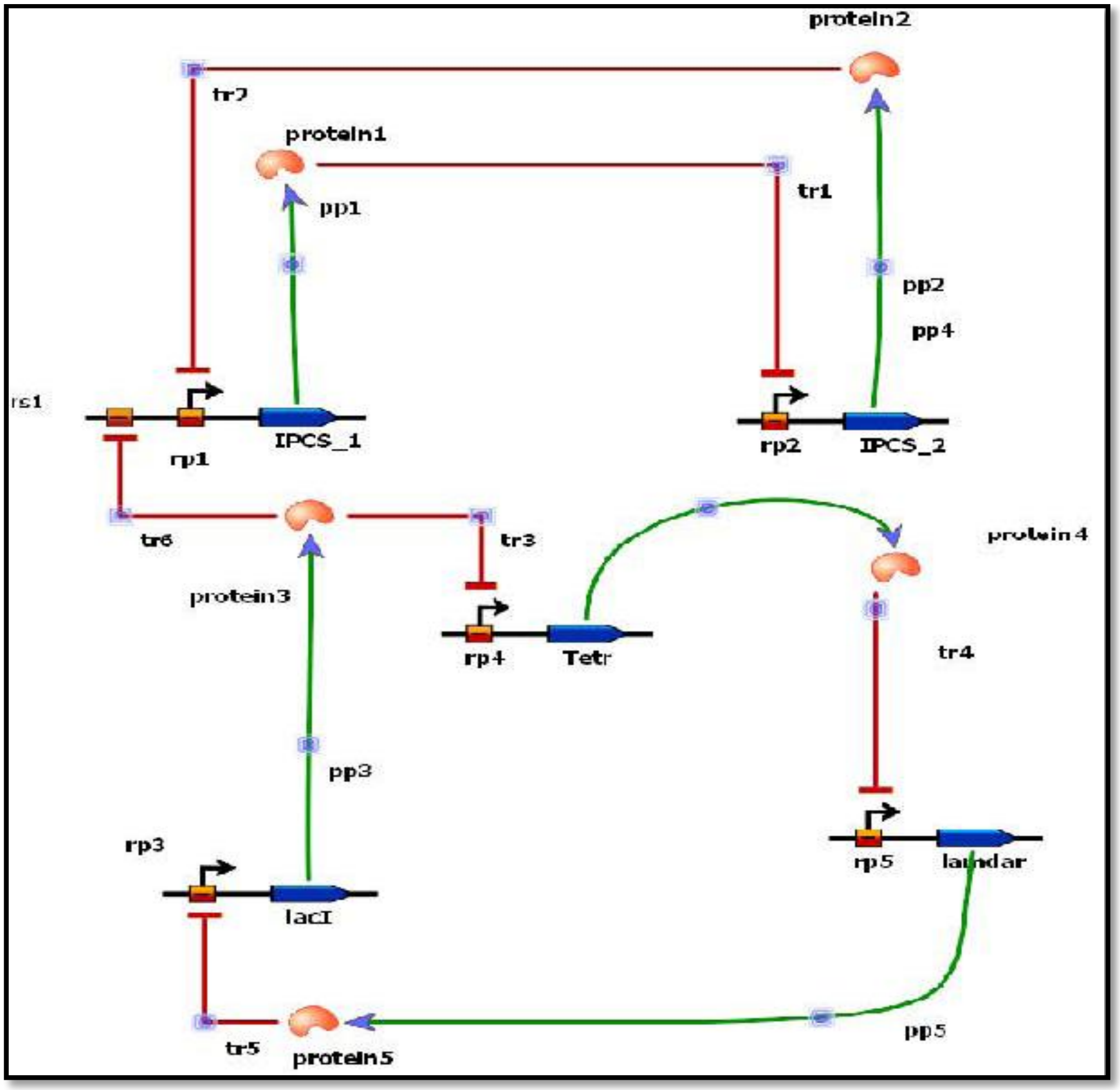

**Fig 7.2** Genetic circuit for IPCS

## Model Summary

Genetic toggle switch consists of IPCS_1 and IPCS_2. Repressilator consist of Lactose repressor (LacI, Tetracycline repressor (Tetr), Lamda repressor (Phage lamda). In circuit, IPCS_1 encodes for the repressor for the opposite gene IPCS_2 and IPCS_2 represses IPCS_1. In the circuit, it is considered that IPCS_1 and IPCS_2 mutually repress each other that reflect the bistable behavior of the genetic switch for the IPCS. Repressilator genes acts in cyclic manner where LacI represses Tetr, Tetr represses LamdaR and LamdaR in turn represses LacI, repressilator tends to produces oscillatory behavior. Coupling between genetic toggle switch and repressilator is developed by LacI and IPCS_1 where LacI regulates the IPCS_1 and represses it.

Parameters considered for the construction of circuit are summarized in table below. Default parameters are initially assigned to the genetic modules on basis of connectivity of the genetic circuit. Initially the Kd values for the genes were set to 1 and hill's coefficient value was set to 2. Parameter scan was done for the steady state analysis and the parameters values considered for the model are summarized in the table below.

| Module no | Promoter strength | Kd | H(Hill/coeffecient) | Degradation rate |
|---|---|---|---|---|
| **IPCS_1** | **5.054** | **2.99** | **4.89** | **0.1835** |
| **IPCS_2** | **5** | **1.9** | **3.116** | **0.087** |
| **LacI** | **11.73** | **1.9** | **1.126** | **0.1867** |
| **Tetr** | **5** | **1.38** | **1.387** | **0.124** |
| **LamdaR** | **5** | **2.029** | **1.54** | **0.125** |

**Table 7.2** Model summary of the genetic circuit

Kd =Dissociation constant

Translation rates and transcription rates are set to 1 for all the genetic modules as the steady state analysis is reflected by degradation rates of the proteins. Kinetics applied to the genetic circuit was by Hill-Hinze's equation for gene product. Kinetics generated for the genetic modules are:

IPCS_1 = rp1.strength * (rs1 * rp1)

IPCS_2 = rp2.strength * rp2

LacI = rp3.strength * rp3

Tetr = rp4.strength * rp4

LamdaR = rp5.strength * rp5

rp1 = 1.0/ ((1+ ((IPCS2protein/tr2.Kd)^tr2.h)))

rp2 = 1.0/ ((1+ ((IPCS1protein/tr1.Kd)^tr1.h)))

rp3 = 1.0/ ((1+ ((LamdaRepressor/tr5.Kd)^tr5.h)))

rp4 = 1.0/ ((1+ ((lactoserepressor/tr3.Kd)^tr3.h)))

rp5 = 1.0/ ((1+ ((TetrRepressor/tr4.Kd)^tr4.h)))

## Steady state Analysis

Steady state scan analysis for the degradation rates of all the proteins was carried out.

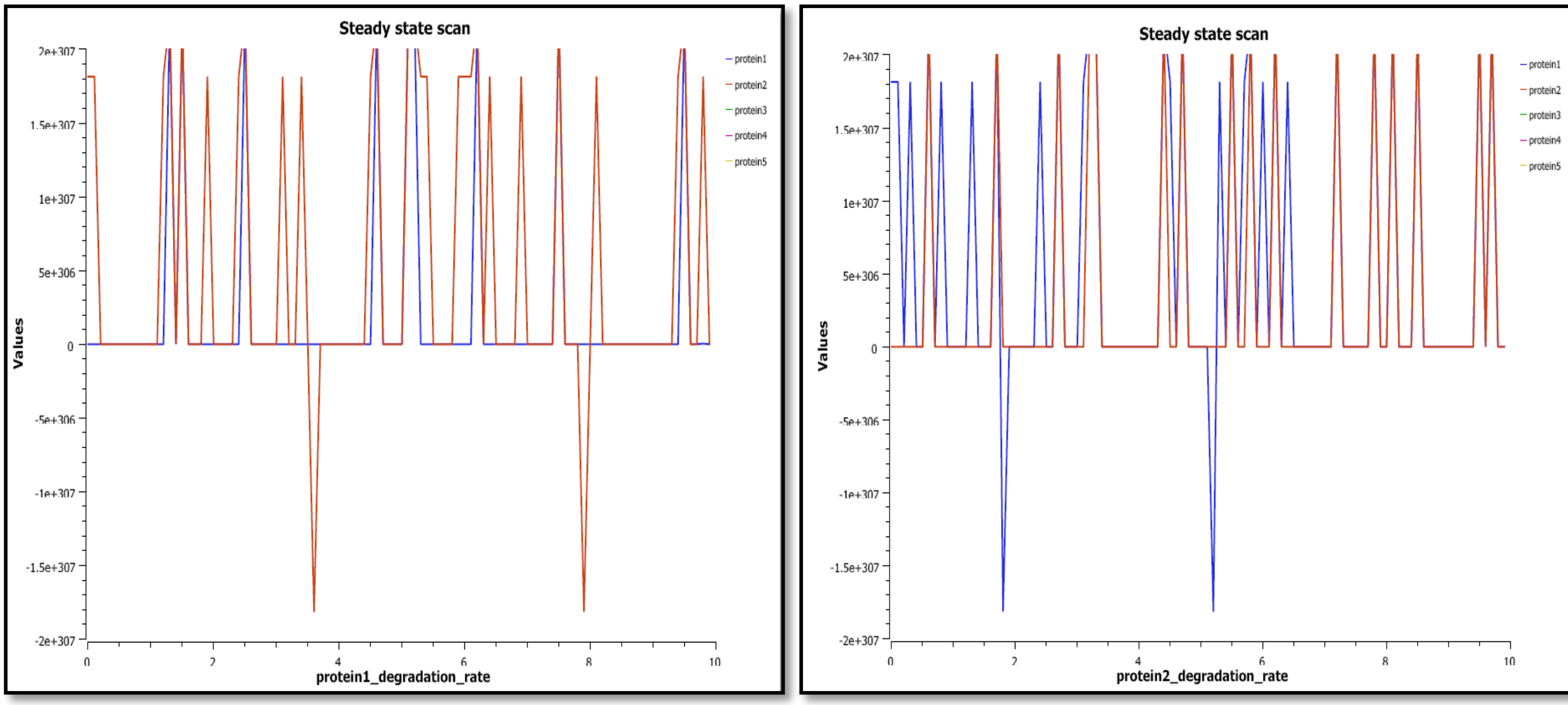

**Fig 7.3 Degradation Rates graphs**

a.)IPCS_1 degradation graph

b.)IPCS_2 degradation graph

Legends of the graphs

IPCS_1:

IPCS_2:

LacI :

Tetr :

LamdaR:

x= time points (in seconds)

y= values of degradation rate

Above are degradation graphs for IPCS_1 and IPCS_2 proteins.In figure 3.3 (a) ,it shows the bistability behavior of the genetic circuit. When the degradation of IPCS_1 protein decreases there is rise in the expression of the opposite toggle gene IPCS_2. There are two points of time where there is peak fall in the IPCS_2 ,where there is switching of the states.This desired graph was obtained at 0.18 degradation rate of IPCS_1. Fig 3.3(b) shows degradation rate of IPCS_2.When there is increase in the level of IPCS_2 at particular interval of time ,there is decrease in the level of IPCS_1.Two points of time in graph shows there is peak fall in the degradation of the IPCS_1 which shows that IPCS_2 represses the IPCS_1 where the flipping of the switch occurs. The desired behavior was obtained at 0.08 degradation rate of IPCS_2.

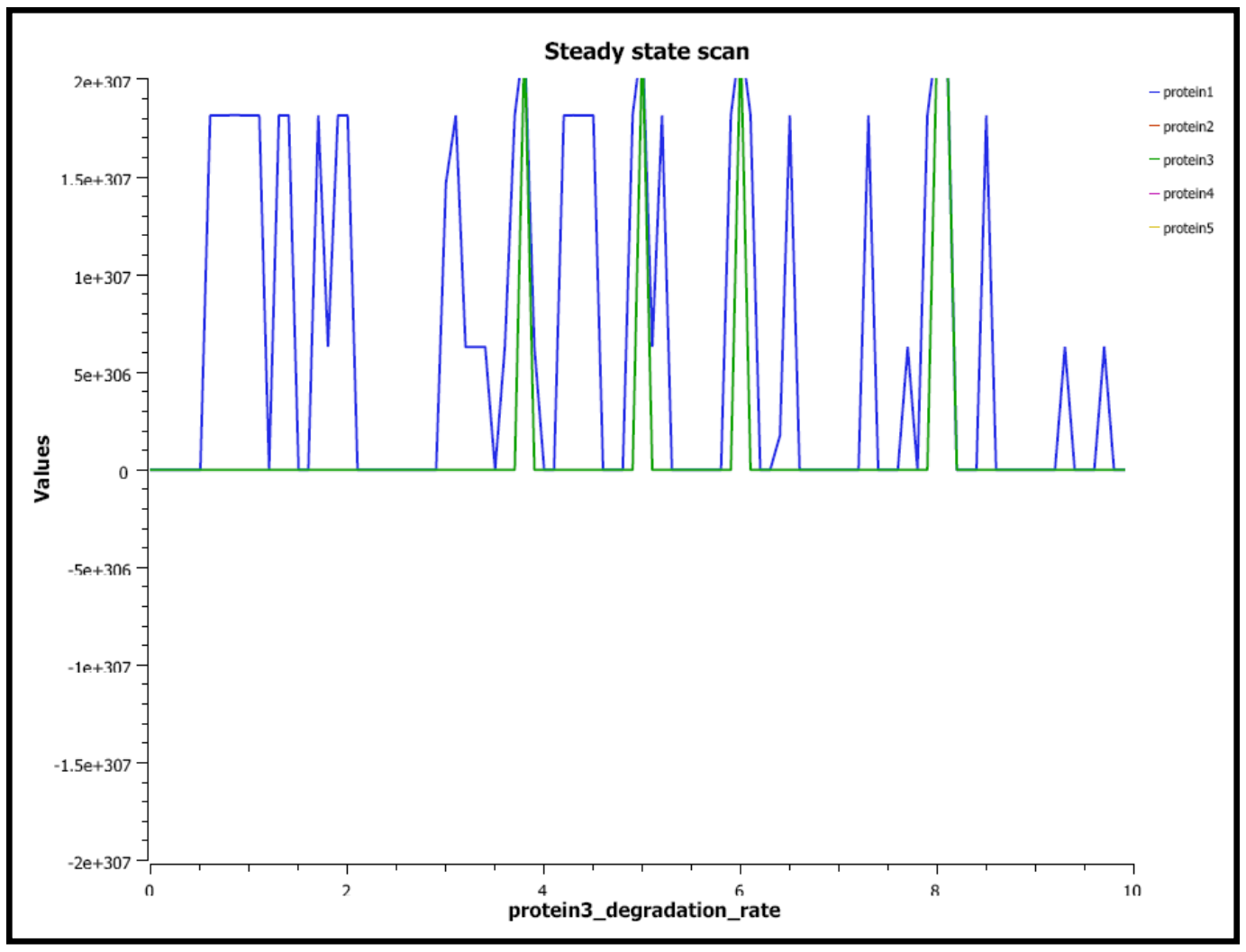

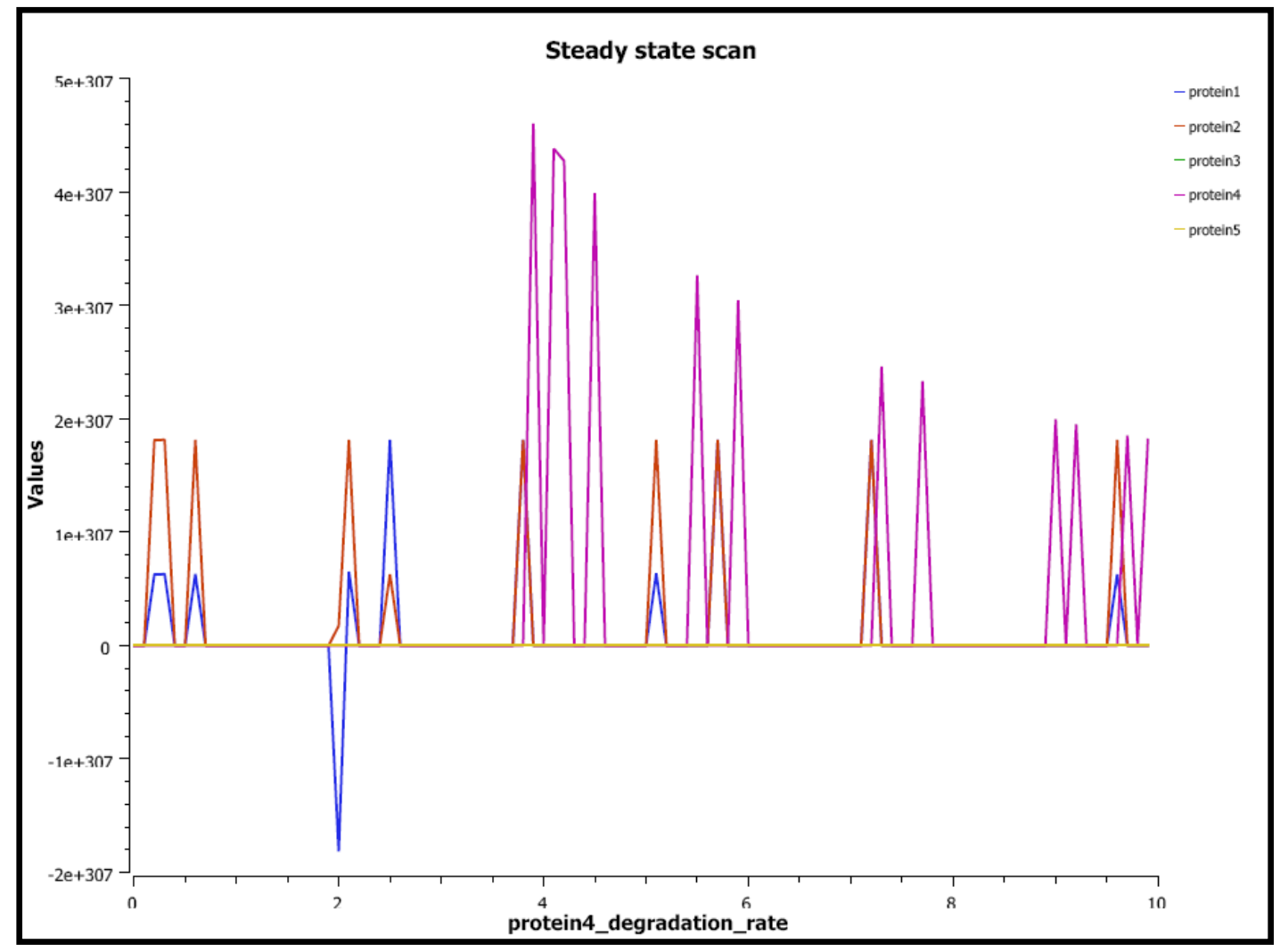

**Fig 7.4 Degradation Rate graphs**

a.)LacI degradation graph b.)Tetr degradation graph

LacI: —— IPCS_1 —— Tetr —— LamdaR ——

In figure 3.4 degradation graphs for the LacI ,which shows the coupling behavior between IPCS_1 and LacI. Intially there is no increase in the level of the LacI because of there is no coupling between LacI and IPCS_1 which results in the increase in the level of IPCS_1.At time interval of 4 ,there in increase in the level of LacI and drop in the level of the IPCS_1.This change in the regulation of IPCS_1 with respective to LacI shows the coupling between genetic toggle switch and repressilator. Fig 3.4 (b) is degradation graph for the Tetr ,there is less oscillation between LamdaR and Tetr but it shows the repression of LamdaR.Remaining results for the change in degradation of proteins are attached in supplementary file.

## Simulation results

Simulation of the genetic circuit was performed using tau-leap stochastic simulation.Simulation was performed for 100 time points.Simulation of genetic circuit was done at differernt concentration.Results of fluctuations in the protein level with respect to change in dissociation constant(Kd) were obtained as follows :

case 1 :change in the Kd values of IPCS_1

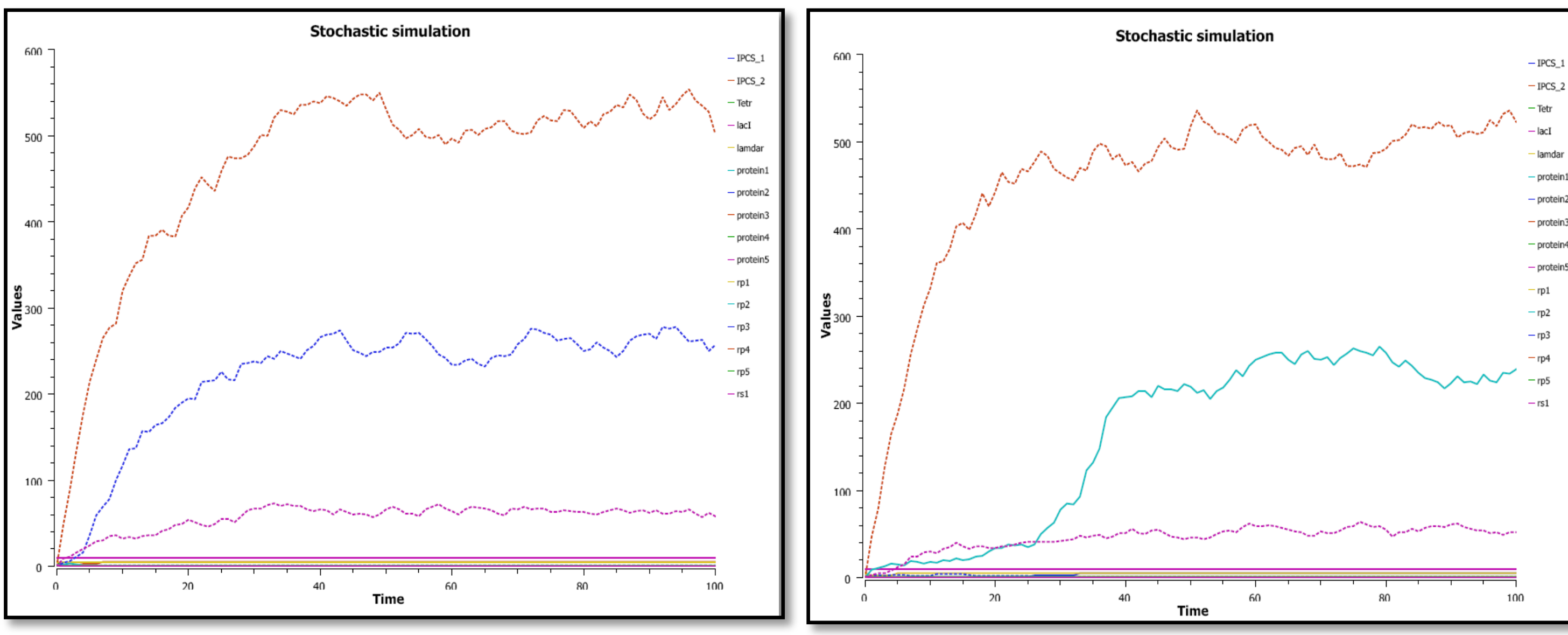

Legends IPCS_1 : — IPCS_2 — LacI — Tetr — LamdaR —

a.) Kd=1 b.) Kd=1.01

**Fig 7.5** Simulation results for IPCS_1 by change in the kc values of IPCS_1

In Fig 3.5 ,there is change in the proteins level with respect to change in the Kd values,when Kd=1 ,level of IPCS_1 is increased but when the value of Kd is rose to 1.01,there is switch in the level of IPCS_2 because of the dissociation between IPCS_1 and regulatory binding site is increased, this leads the circuit to toggle and level of IPCS_2 is raised. Similar fluctuations in protein level is observed when Kd values at regular intervals are changed .The remaining results for the change in Kd values are attached in supplementary file.

Case 2: Simulation for the change in kd values of IPCS_2 were also performed that results in to the fluctuation of levels of IPCS_1

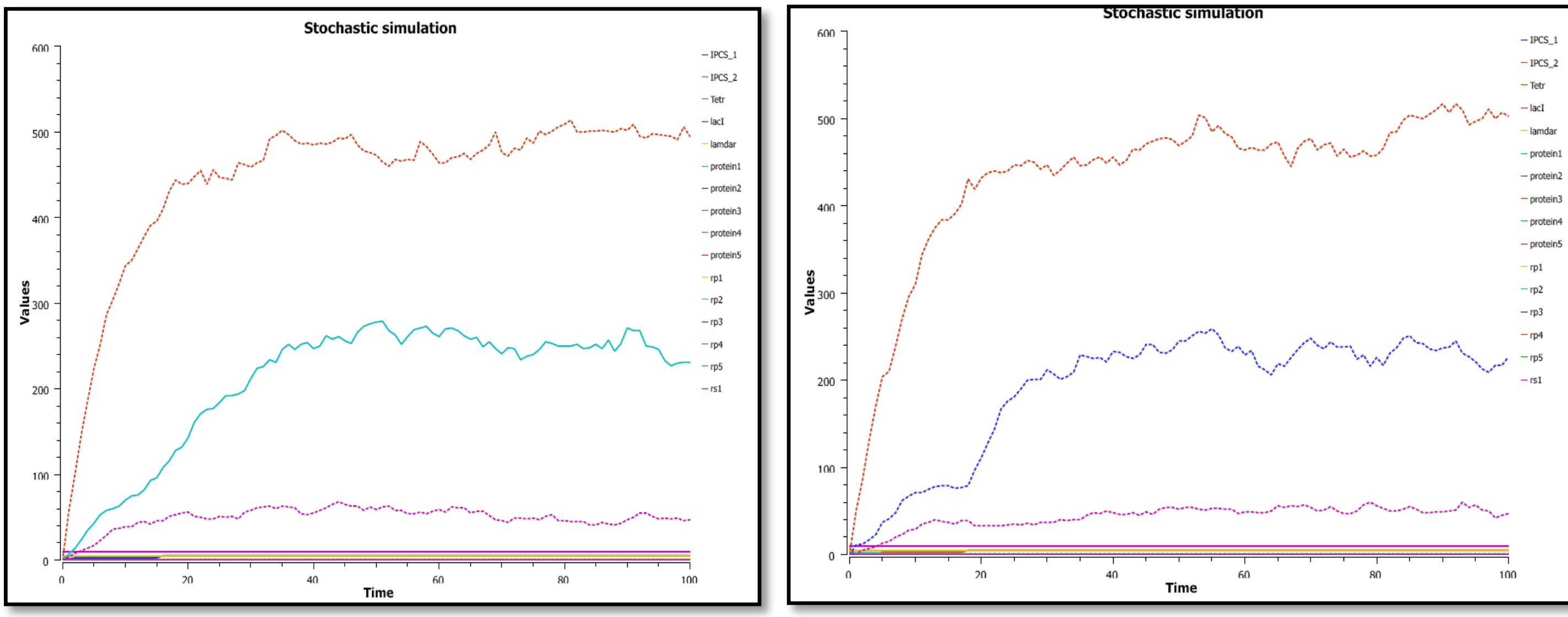

**Fig 7.6** Simulations results for circuit by change in the Kd values of IPCS_2

a.) Kd=1.01 b.) Kd=1.02

In Fig 3.6 , when Kd=1.01 ,level of IPCS_2 is increased but when the value of Kd is rose to 1.01,there switch in the level of IPCS_1 because of the dissociation between IPCS_2 and repressor is increase this leads the circuit to flip and level of IPCS_1 is raised. Similar perturbations in levels of IPCS_1 and IPCS_2 is observed when changes in the Kd at regular intervals are done.The remaining results for the change in Kd values are attached in supporting file. Above results shows the bistable behavior between IPCS_1 and IPCS_2 of the genetic ciruit and coupling between IPCS_1 of toggle switch snd LacI of the repressilator is also shown.

## Digital circuit representation of the Synthetic Genetic circuit

Digital circuits gives insight into the logic gates implemented in the circuit design for the functioning of the circuit. The circuit consists of two genes coding for two repressors. The two repressors mutually repress each other such that a high concentration of one protein inhibits the transcription of the other gene. Any signal that causes the breakdown of the existing repressor protein molecules or increases the transcription of the other repressor protein would cause the flip. A Toggle Switch has two IMPLIES gates connected in such a way that the output of one represses the other. Two IMPLIES gates mutually connected with each other where the ouptut signal of one IMPLIES gate is input signal for the opposite IMPLIES gate.

The repressilator is constructed by integrating an odd number of NOT gates in a circular fashion such that the output of the last gate is the input of the first one. Three NOT gates connected in cyclic manner in which the output signal of first gate is the input signal for the second gate,output signal for the second gate is input signal for the third gate. Output signal for the last gate is input signal for the first gate.It is mandatory to consider the odd numbers of NOT gate in the repressilator because if there are even number of gates then the output generated from the last gate will be high. However, if the number is odd, the output from the last inverter gate will be low and when fed to the first gate generates a high output from the last inverter. Hence the last output signal oscillates between low and high alternately.

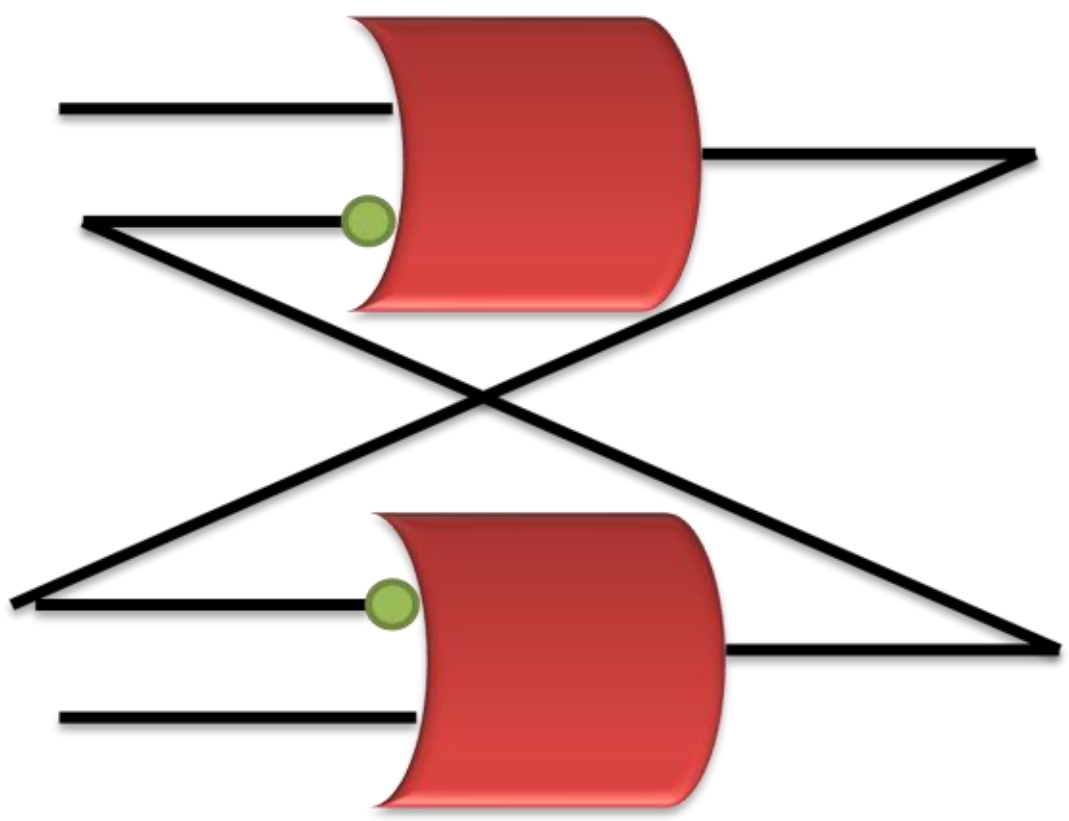

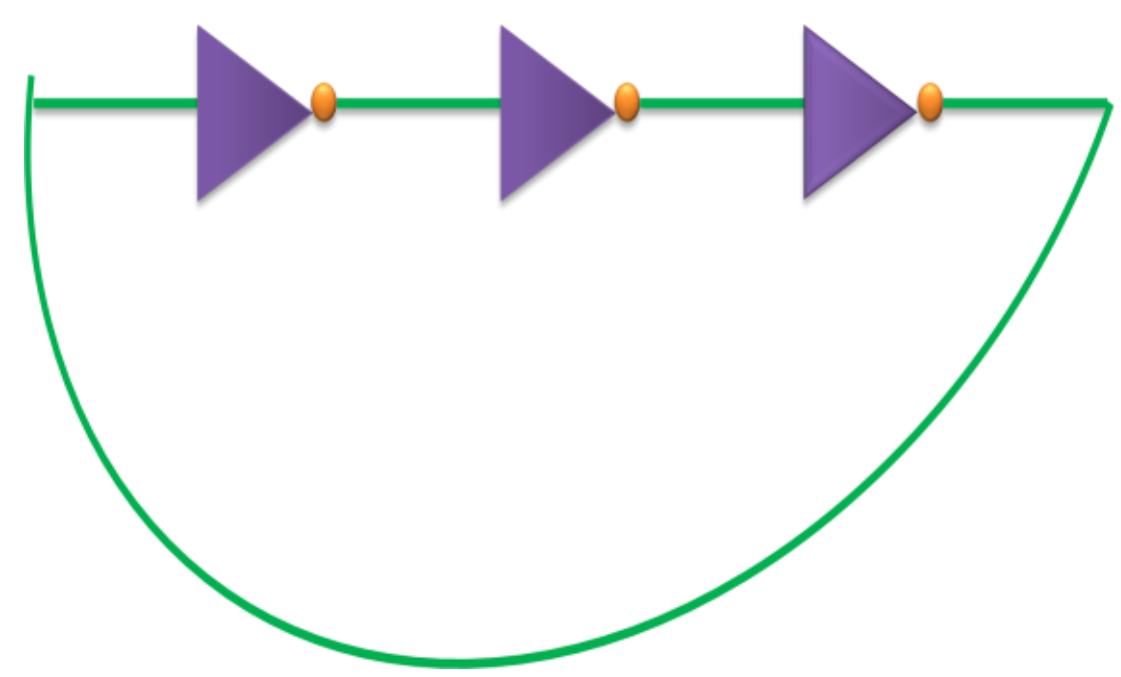

**Fig 7.3 a.)** Logic gates for the genetic toggle switch b.) Logic gates for the repressilator

Logic of the genetic circuit is represented by the truth table. Truth table of logic gates for the genetic circuit of IPCS is represented below.

| **Input(A)** | **Input (B)** | **Input (C)** | **output** |
|---|---|---|---|
| **0** | 0 | 0 | 1 |
| **0** | 0 | 1 | 0 |
| **0** | 1 | 0 | 1 |
| **0** | 1 | 1 | 0 |
| **1** | 0 | 0 | 0 |
| **1** | 0 | 1 | 0 |
| **1** | 1 | 0 | 0 |
| **1** | 1 | 1 | 0 |

**Table 7.3** Truth Table for the genetic circuit. Input A and B are the input signal of genetic toggle switch and C is the input signal of the repressilator.

Karnaugh map is simplified representation of truth table. It gives more insight to study the logic of digital circuit. Coordinates used in the Karnaugh map should map the values in the truth table.

| C | AB | | | | |
|---|---|---|---|---|---|
| | | 00 | 01 | 11 | 10 |
| | 0 | 1 | 1 | 1 | 0 |
| | 1 | 0 | 0 | 0 | 0 |

**Table 7.4** Karnaugh map of the truth table of the genetic circuit constructed

The values of **C** are written on the rows of the Karnaugh map, whereas the values of **A** and **B** lie on its columns. The Karnaugh map method permits to derive both the Sum of product (SOP) and POS (Product of sum) form of the Boolean expression associated with any truth table.

$$SOP= (|A*C) + (|A * |B) \qquad POS=|A* (|B+ C)$$

Karnaugh map help to construct digital circuit. Circuit scheme gives the logic design for how the input signals affect the output of the circuit.

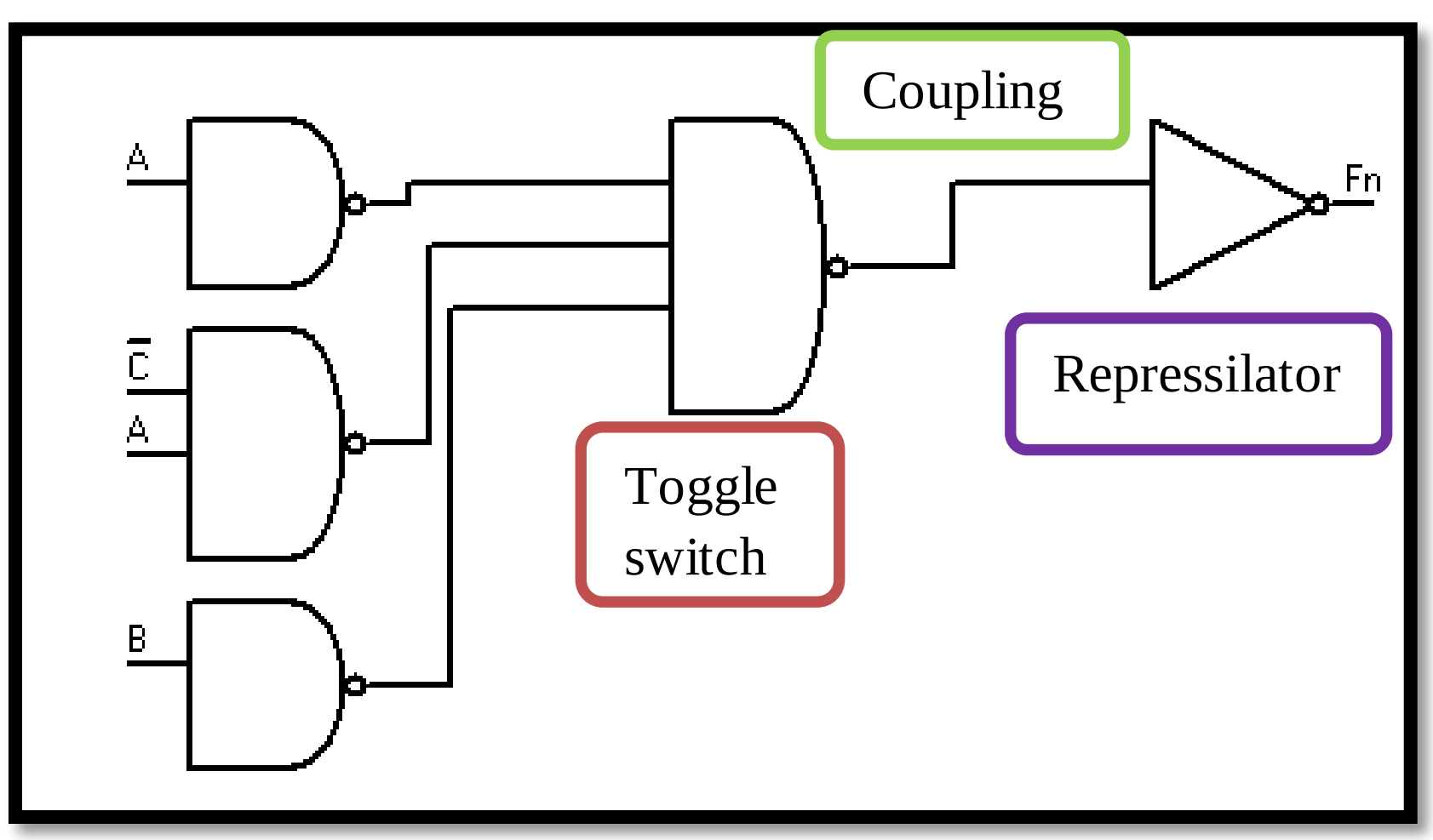

**Fig 7.8** Conversion of a truth table into a digital circuit via the Karnaugh map method.

Digital circuit implements the logic gates deciphered for the genetic circuit. Digital circuit shows the logic gate for the toggle switch and repressilator.

## Genetic circuit model validation

Model validation was done by using Qualitative and Quantitative approaches. For this ODE model was generated for the genetic circuit.ODE model is mathematical representation of the circuit in form of differential equations.ODE model gives more insight into regulatory mechanism for the genetic circuit. ODE model derived for the genetic circuit constructed are as follows:

For IPCS_1

$$\frac{d([IPCS1protein].V_{DefaultCompartment})}{d\,t} = +V_{DefaultCompartment}\cdot\left(\frac{pp1_\,translation_rate.rp1_strength\,.[IPCS\,_1]}{V_{DefaultCompartment}}\right) - V_{DefaultCompartment}\cdot\left(\frac{IPCS1protein_degradation_rate\,.[IPCS1protein]}{V_{DefaultCompartment}}\right)$$

For IPCS_2

$$\frac{d([IPCS2protein].V_{DefaultCompartment})}{d\,t} = +V_{DefaultCompartment}\cdot\left(\frac{pp2_\,translation_rate.rp2_strength\,.[IPCS\,_2]}{V_{DefaultCompartment}}\right) - V_{DefaultCompartment}\cdot\left(\frac{IPCS2protein_degradation_rate\,.[IPCS2protein]}{V_{DefaultCompartment}}\right)$$

For LacI

$$\frac{d([lactoserepressor].V_{DefaultCompartment})}{d\,t} = +V_{DefaultCompartment}\cdot\left(\frac{pp3_translation_rate.rp3_strength.[lacI]}{V_{DefaultCompartment}}\right) - V_{DefaultCompartment}\cdot\left(\frac{lactoserepressor_degradation_rate\,.[lactoserepressor]}{V_{DefaultCompartment}}\right)$$

For Tetr

$$\frac{d([Tetrepressor].V_{DefaultCompartment})}{d\ t} = +V_{DefaultCompartment}\cdot\left(\frac{pp4_translation_rate.1.[Tetr]}{V_{DefaultCompartment}}\right)$$

$$-V_{DefaultCompartment}\cdot\left(\frac{Tetrepressor_degradation_rate\ .[Tetrepressor]}{V_{DefaultCompartment}}\right)$$

For LamdaR

$$\frac{d([lamdarepressor].V_{DefaultCompartment})}{d\ t} = +V_{DefaultCompartment}\cdot\left(\frac{pp5_translation_rate.1.[lamdar]}{V_{DefaultCompartment}}\right)$$

$$-V_{DefaultCompartment}\cdot\left(\frac{lamdarepressor_degradation_rate\ .[lamdarepressor]}{V_{DefaultCompartment}}\right)$$

Equation 3.1 for protein concentration

[IPCS_1]=rp1_strength. [rp1]

[IPCS_2]=rp2_strength. [rp2]

[Tetr] =rp4_strength. [rp4]

[LacI] =rp3_strength. [rp3]

[LamdaR]=rp5_strength. [rp5]

Equation 3.2

[rp1] $= \frac{1}{1+\left(\frac{[IPCS2protein]}{tr2_Kd}\right)^{tr2_h}}$

[rp2] $= \frac{1}{1+\left(\frac{[IPCS1protein]}{tr1_Kd}\right)^{tr1_h}}$

[rp3] $= \frac{1}{1+\left(\frac{[lamdarepressor]}{tr5_Kd}\right)^{tr5_h}}$

[rp4] $= \frac{1}{1+\left(\frac{[lactoserepressor]}{tr3_Kd}\right)^{tr3_h}}$

[rp5] $= \frac{1}{1+\left(\frac{[Tetrepressor]}{tr4_Kd}\right)^{tr4_h}}$

[rs1] $= \frac{1}{1+\left(\frac{[repressor]}{tr6_Kd}\right)^{tr6_h}}$

Equation 3.3 Transcription rate

ODE model has the expression for the concentration of the protein in form of variables. Reactions in which proteins are synthesized is represented by positive sign (+) and reactions in which proteins are degraded are represented by negative sign (-) i.e. Translation rate reaction is indicated by + sign in which the proteins are synthesized and degradation reactions is represented by – sign in which proteins are degraded. Promoter strength is also represented by equation for the coding regions that includes the promoter strength associated with each coding region. The transcription rates reaction which includes Kd values and hills coefficient.

**Network Inference**

Gene regulatory network was constructed for the genetic circuit using Bioconductor packages. Model was simulated using COPASI.Time series data was generated and used for qualitative and quantitative network modeling. Circuit model is asymptotically stable hence Linear network construction was followed. Quantitative network modeling performed by GRENITS gave the probability of each regulator in the regulatory network circuit. Posterior probability was derived by using Monte Carlo Markov Simulation (MCMC). MCMC simulation was run by using default parameters. Simulation generated two markov chains that resulted into link probability of the network generated. Results includes probability matrix for each gene in the regulatory network. Analysis plot and convergence plot was generated for the network circuit. The Convergence plots contain plots associated to the adequate convergence of the Markov chains. This is crucial for further analysis as, if convergence has not been reached, the results are not trustworthy. Analysis plot contains the link probability of each regulator with other in the circuit. Network inference was made for 10 and 100 time series .Inferred network showed the regulatory mechanism in the circuit.

| From | To | Probability |
|---|---|---|
| IPCS_1 | IPCS_1 | 0 |
| IPCS_1 | IPCS_2 | 1 |
| IPCS_1 | Tetr | 0 |
| IPCS_1 | lacI | 0 |
| IPCS_1 | LamdaR | 0 |
| lacI | IPCS_1 | 1 |
| lacI | Tetr | 0 |
| lacI | lacI | 0 |
| lacI | LamdaR | 1 |
| LamdaR | IPCS_1 | 1 |
| LamdaR | IPCS_2 | 0 |

**Table 7.5** Posterior probabilities for each network connection

Above probability matrix has the probability 1 for IPCS_1 and IPCS_2 gives the regulatory mechanism between them. Matrix where probability is 1 shows the regulation between the respective regulators. Probability 0 indicates that there is no regulation between those regulators of genetic circuit.

| | 1 Regulators | 2 Regulators | 3Regulators | 4Regulators | 5 Regulators |
|---|---|---|---|---|---|
| IPCS_1 | 0 | 1 | 1 | 0 | 0 |
| IPCS_2 | 1 | 0 | 0 | 0 | 0 |
| Tetr | 1 | 0 | 0 | 0 | 0 |
| LacI | 1 | 0 | 0 | 0 | 0 |
| LamdaR | 0 | 1 | 0 | 0 | 0 |

**Table 7.6** Posterior probabilities for number of regulators for each gene.1 shows the regulation between the regulators and 0 indicates no regulation

Analysis plot of the genetic circuit shows the link probability which gives insight in to switching behavior between toggle switch and coupling between the LacI repressilator and IPCS_1 of genetic toggle switch as designed in the circuit model.

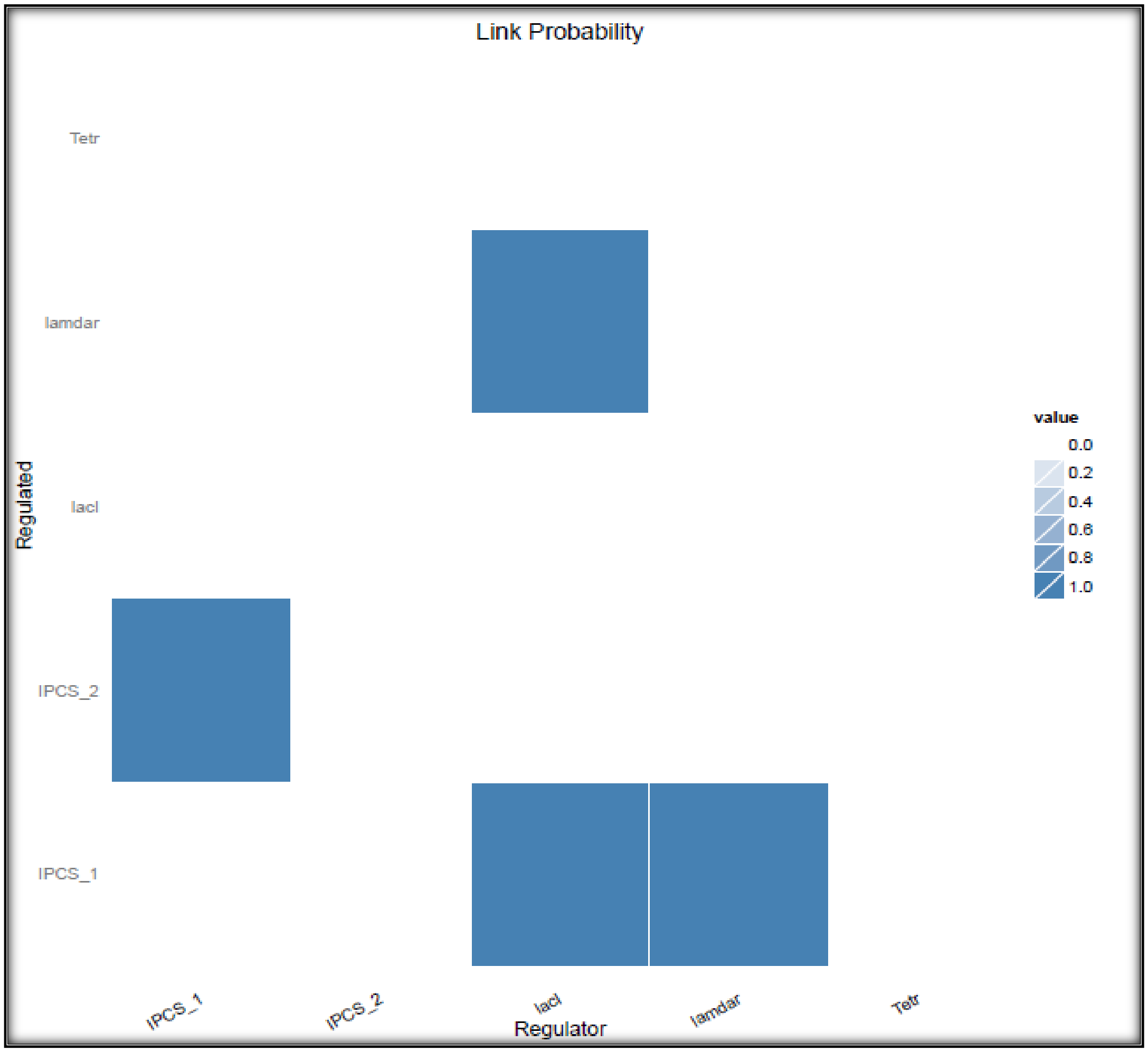

**Fig 7.9** Heat map plot of network link probabilities

Analysis plot showed the probability of 1 (blue color) between IPCS_1 and IPCS_2.Coupling between LacI and IPCS_1 is deciphered by the link probability 1 between them.

Network uncertainty was given by marginal network uncertainty plot. Network uncertainty gave the idea about top regulators in the network. Plot (figure 3.10) reveals that IPCS_1, IPCS_2 and LacI to be top regulators of the genetic circuit constructed for IPCS

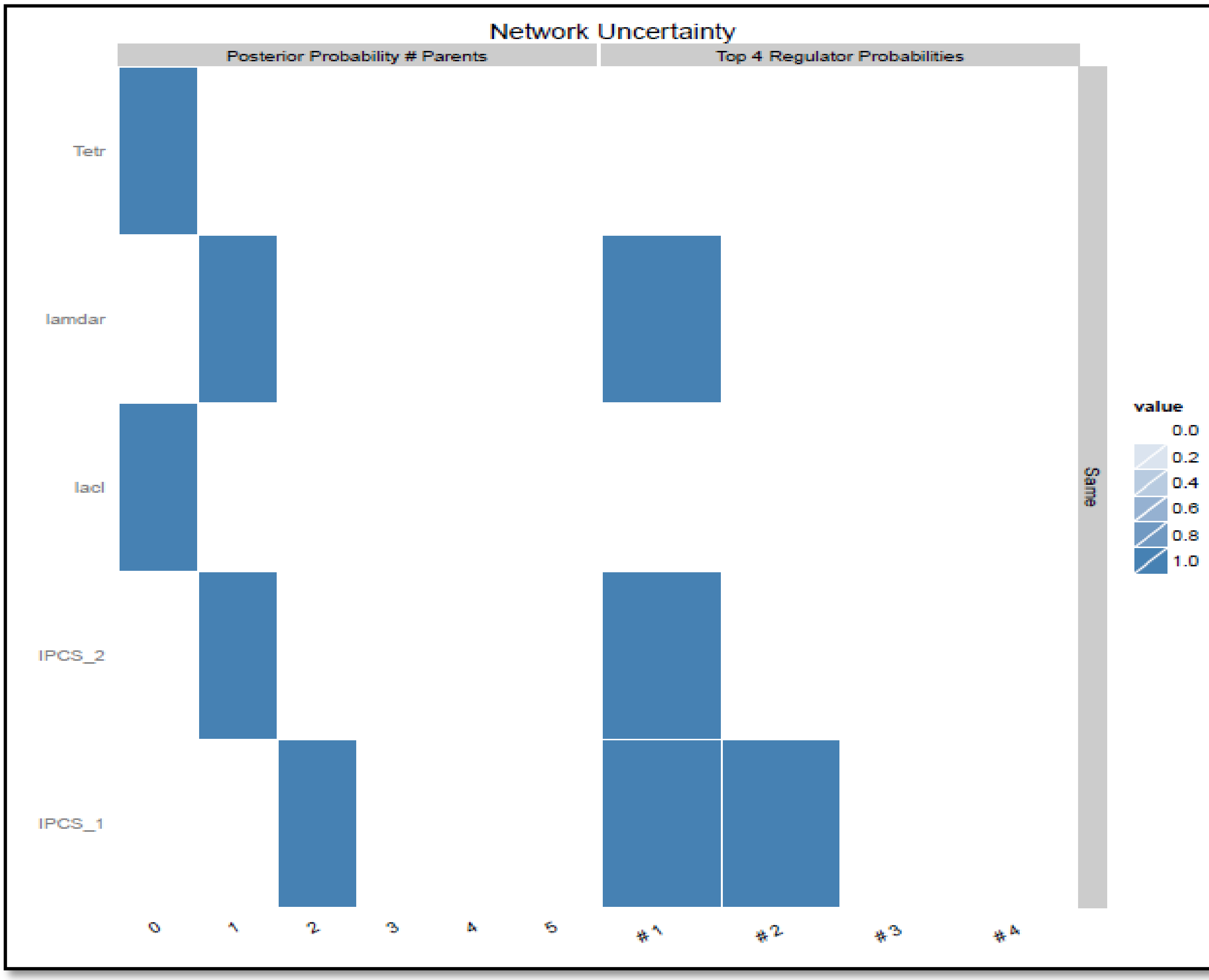

**Fig 7.10** Network uncertainty plot

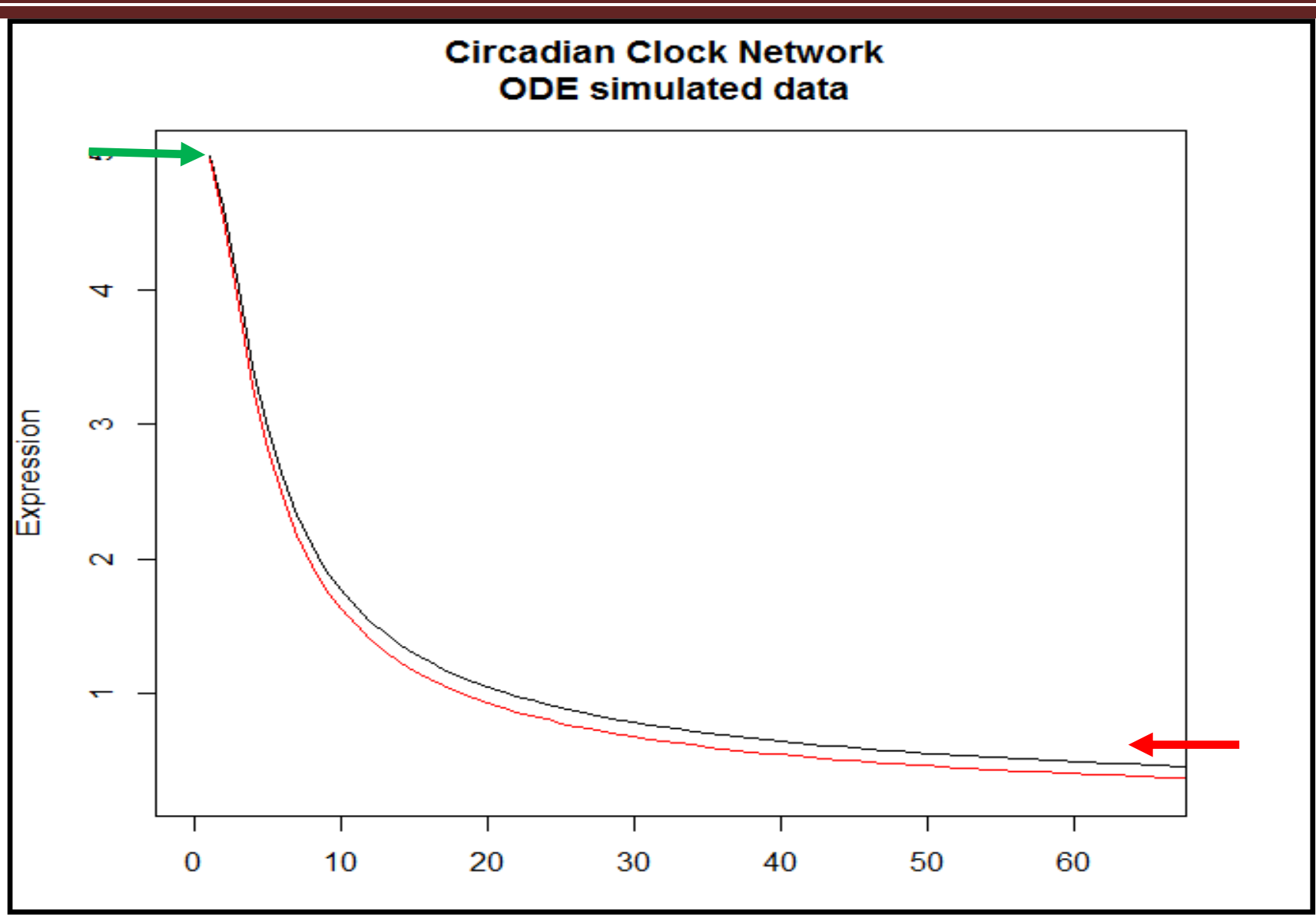

**Fig7.11** Circadian clock network generated for IPCS_1 and IPCS_2 suggest the ON and OFF state of both genes. Top right arrow is ON state and bottom left arrow indicates OFF state.

**Convergence plots**

Basic convergence plots are constructed. The posterior means of each variable are compared.

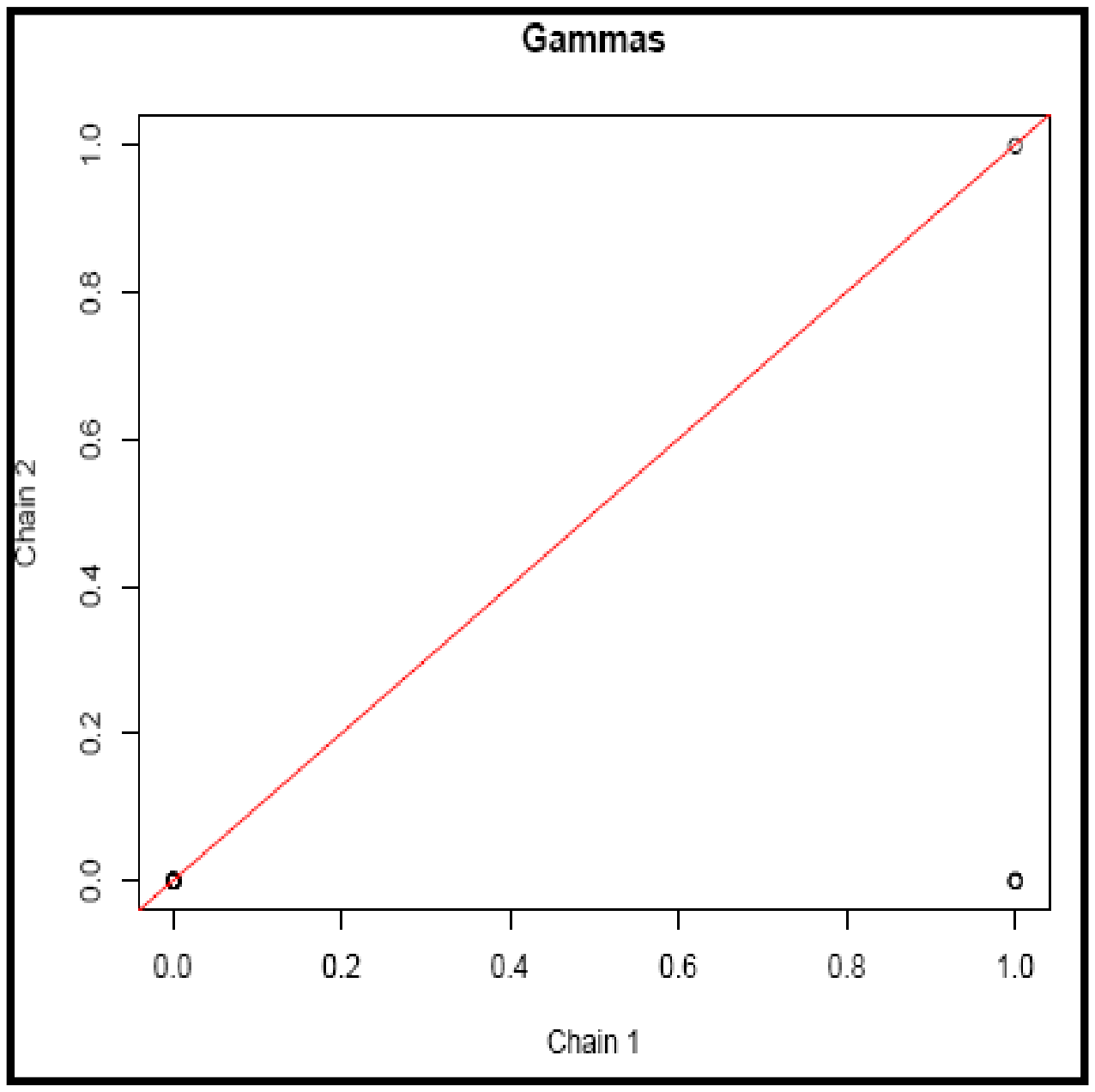

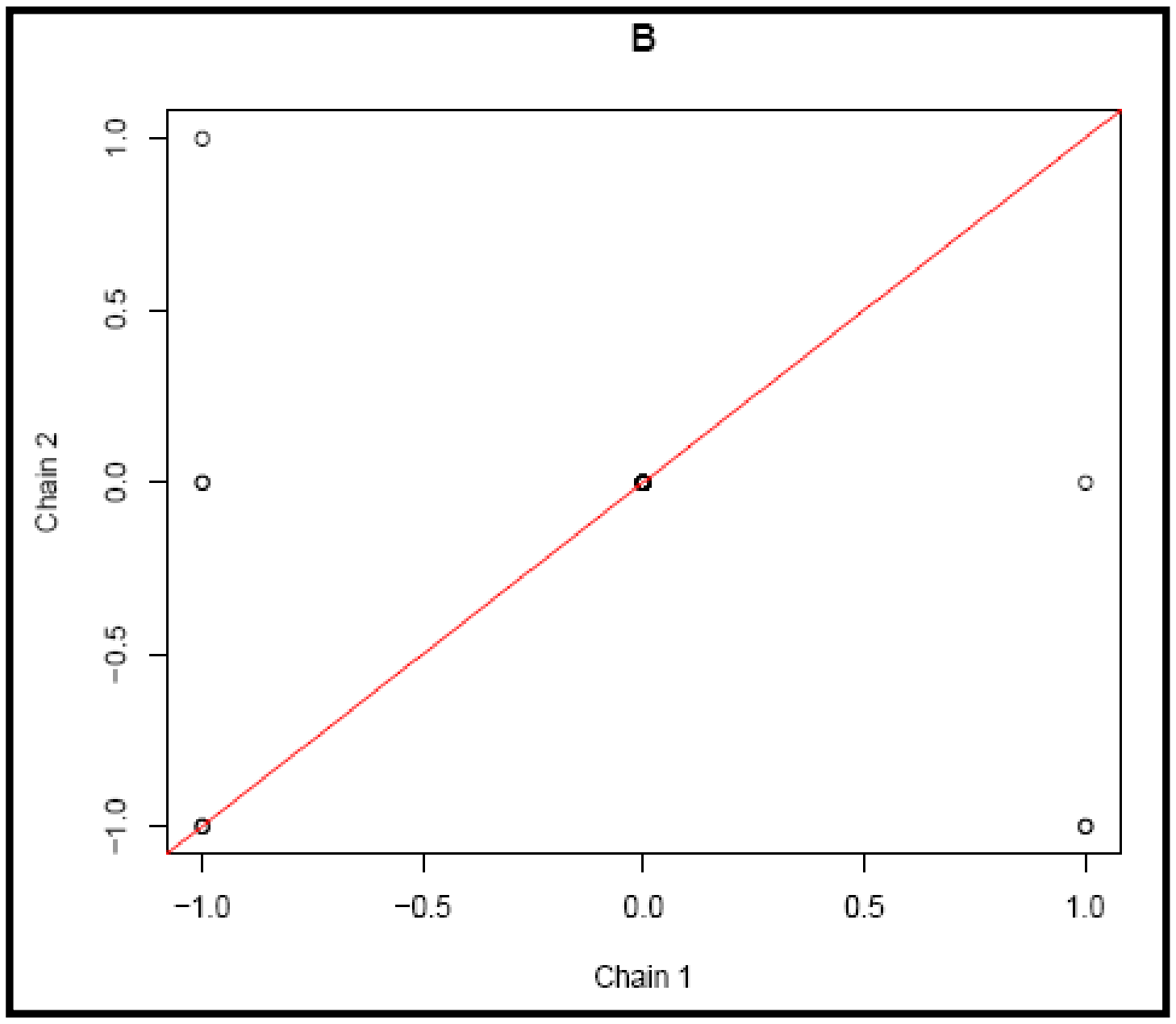

**Fig 7.12**

**a.)** Indicator variable of Gibbs variable selection **b.)** Co-efficient of linear Regression

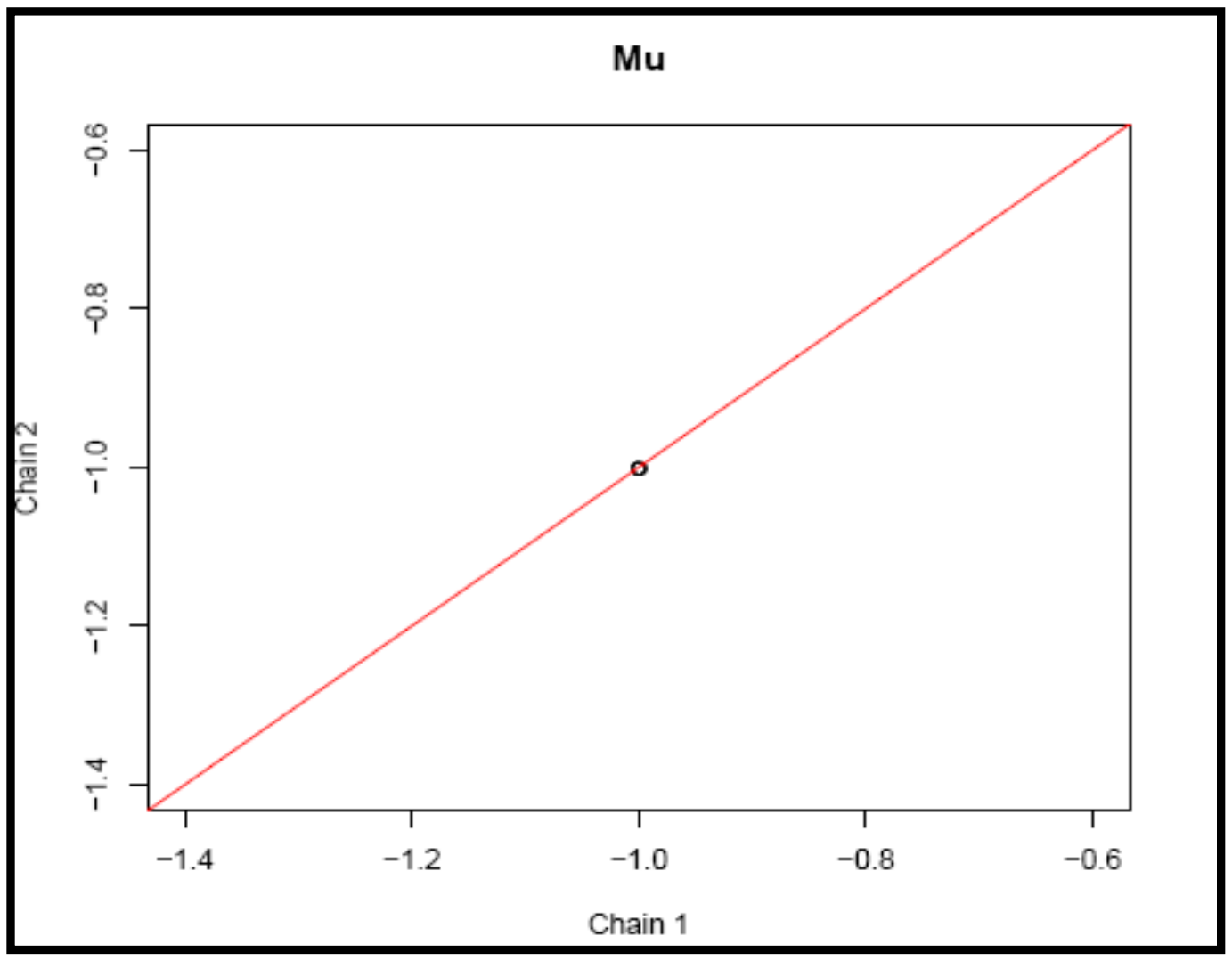

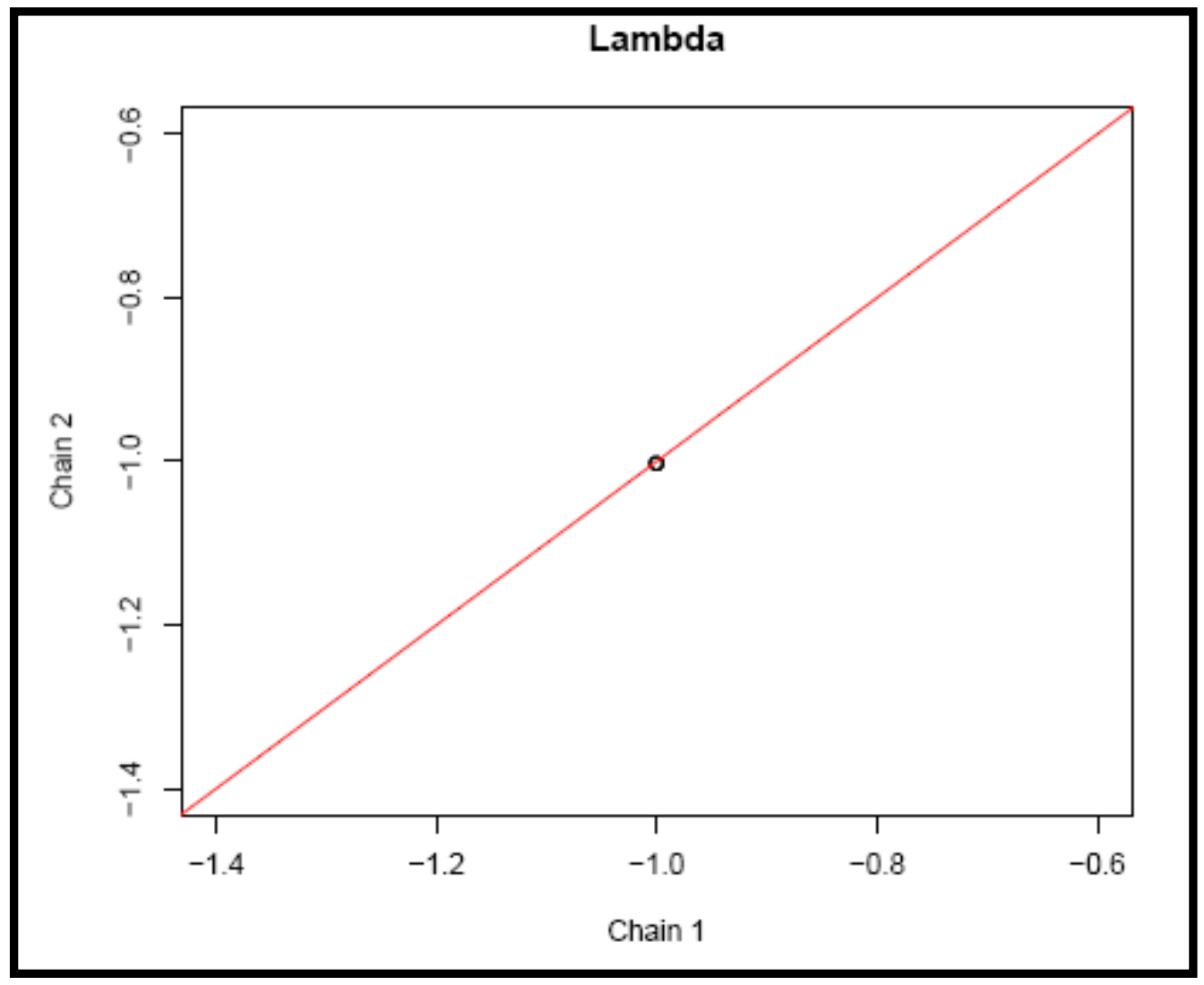

c.) Precision of each regression

d.) Intercept of each regression

Gamma, B, Lamda, Mu are the variables of the Markov chains generated.

**Inferred Network**

Network inferred for the genetic circuit constructed for IPCS displays the regulation between each genetic modules used for the construction of the genetic circuit.

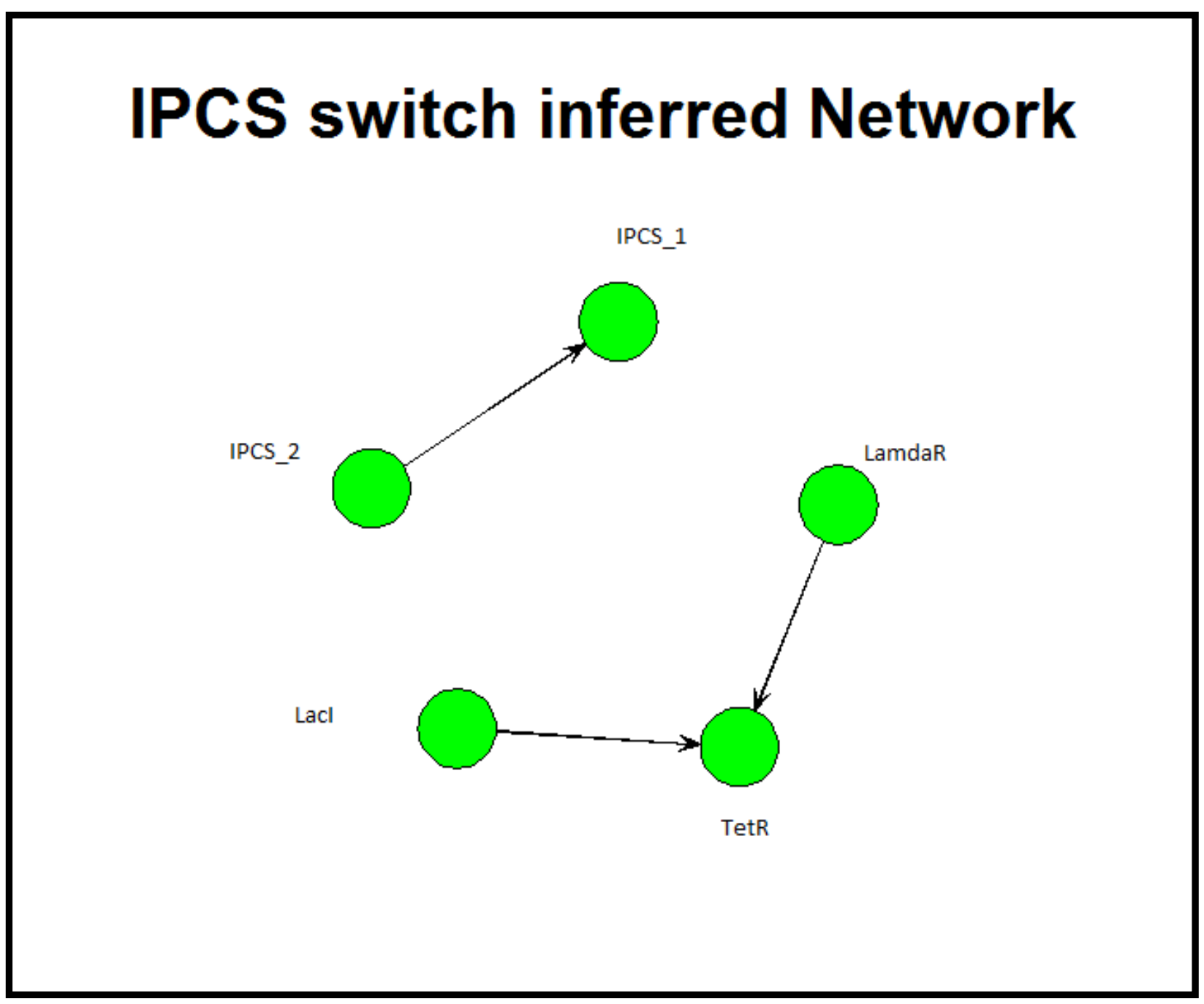

**Fig 7.13** Inferred network of the genetic circuit for IPCS

In Fig 3.12, inferred network display the regulation between IPCS_1 and IPCS_2 which suggest the switching behavior of the toggle switch. Network inference was made at 10 seconds .As there is no regulation between LacI and IPCS_1 it shows no coupling between genetic toggle switch and repressilator which results into repression of IPCS_2 by IPCS_1 (regulation indicated by edge in the network) Lack of coupling, IPCS_1 is ON and IPCS_2 is OFF.

In Fig 3.13 As there is regulation between LacI and IPCS_1 it shows coupling between genetic toggle switch and repressilator which results into repression of IPCS_1 by LacI. Network inference was made at 100 seconds. Coupling between IPCS_1 and LacI leads to IPCS_1 in OFF state and IPCS_2 in ON state.

Probability derived justifies the design of the circuit that has achieved the design objective in addition to associated parameters that has accounted for the designability of the genetic circuit designed.

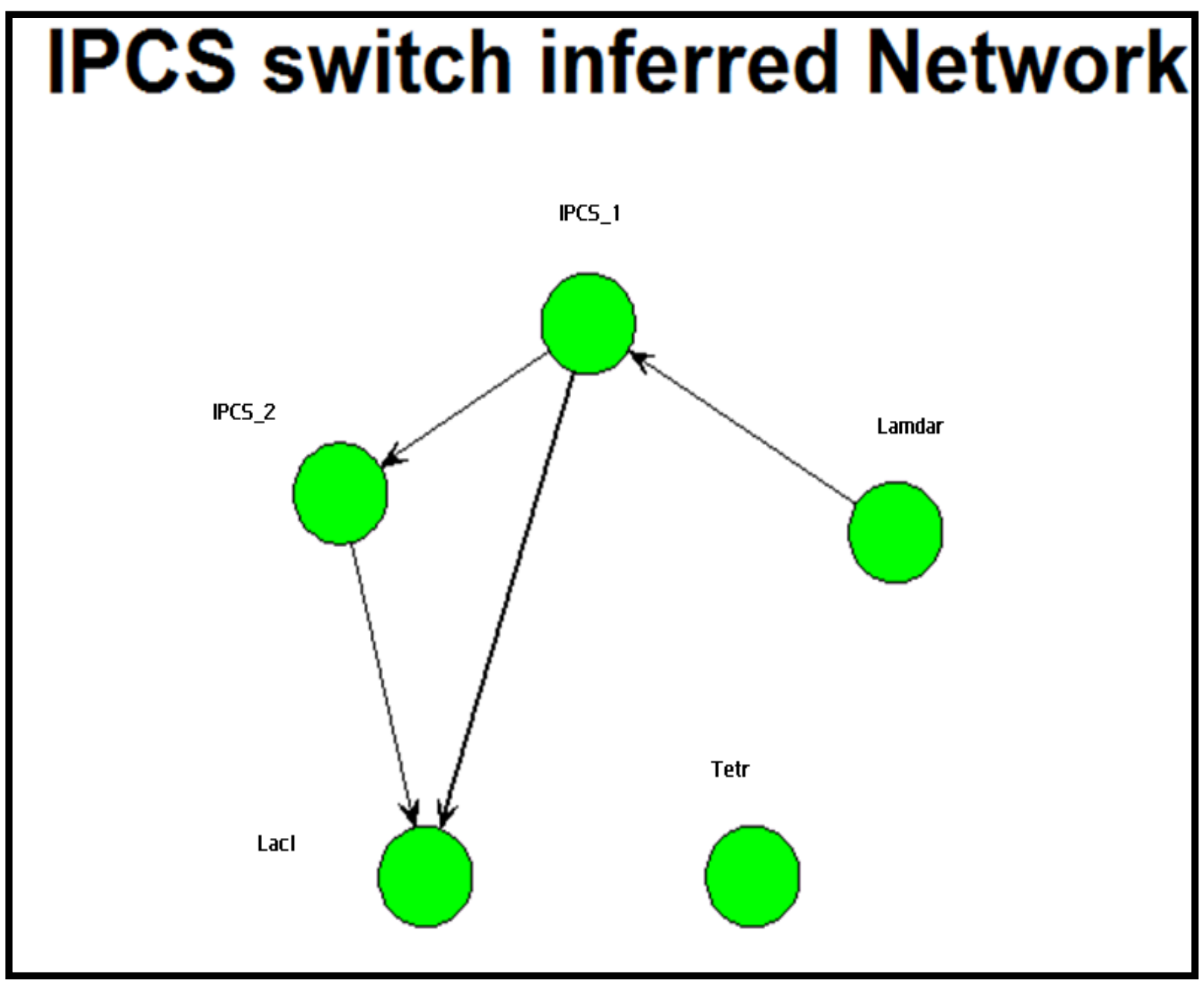

**Fig 7.14** Inferred network of the genetic circuit for IPCS

**Qualitative network modeling**

To test the robustness of the circuit and identification of the number of attractors that circuit exhibit, qualitative Boolean network modeling for genetic circuit is done. After obtaining the steady state of the network, network was perturbed to check the robustness of the circuit. Boolean network for the circuit was constructed using time series. For the construction of Boolean network, binarisation of the time series data was performed.

| | **x1** | **x2** | **x3** | **x4** | **x5** | **x6** | **x7** | **x8** | **x9** | **x10** | **x11** |
|---|---|---|---|---|---|---|---|---|---|---|---|
| Ipcs_1 | 1 | 1 | 1 | 0 | 0 | 0 | 0 | 0 | 0 | 0 | 0 |
| Ipcs_2 | 1 | 1 | 1 | 0 | 0 | 0 | 0 | 0 | 0 | 0 | 0 |
| Tetr | 1 | 1 | 1 | 1 | 0 | 0 | 0 | 0 | 0 | 0 | 0 |
| LacI | 1 | 0 | 0 | 0 | 0 | 0 | 0 | 0 | 0 | 0 | 0 |
| LamdaR | 1 | 0 | 0 | 0 | 0 | 0 | 0 | 0 | 0 | 0 | 0 |

**Table 7.5** Binarized Time series data

Boolean Network

Boolean network was constructed from the time series data by Best fit algorithm for the construction of the network. Boolean network consist of number of genes involved in network circuit and number of states. Number of states present in Boolean network is given by the equation

$$2^n \quad \text{where n=no of genes in the network circuit.}$$

Genetic circuit constructed for IPCS has 5 genes, so number states in the network was 32. Boolean network also includes the transition of one state to other.

Boolean network with 5 genes

Involved genes:

Gene 1 Gene 2 Gene 3 Gene 4 Gene 5

Transition functions:

Gene 1 = <f (Gene 3, Gene 2, Gene 4, Gene 1, Gene 5){11101101101111011111001111110000}>

Gene 2 = <f (Gene 3, Gene 4, Gene 5, Gene 2, Gene 1){00100001000010110100101000101011}>

Gene 3 = <f (Gene 5, Gene 3, Gene 1, Gene 2, Gene 4){10101110001010011000001111000110}>

Gene 4 = <f (Gene 4, Gene 2, Gene 3, Gene 5, Gene 1){11101010011101110011100010000111}>

Gene 5 = <f (Gene 3, Gene 1, Gene 5, Gene 2, Gene 4){00010011000101111110000001111011}>

Above is the Boolean network of the genetic circuit for IPCS. It has transition function which shows the transition of the genes of the circuit. There are 32 states presented by 1 and 0 that specifies the active and inactive respectively that gives the state of each gene during the transition from one state to another. After transitions of states network circuit attains steady states (attractors).Genetic circuit has two attractors that specifies the bistability of the circuit constructed. The two steady state obtained represent the ON and OFF state for each gene in the circuit. 0 specify the OFF state and 1 specify the ON state.

Attractor 2 is a simple attractor consisting of 2 state(s) and has a basin of 30 state(s):

```
|--<----|
V       |
00111   |
|       |
01111   |
|       |
V       |
```

|-->----|

Genes are encoded in the following order: Gene 1 Gene 2 Gene 3 Gene 4 Gene 5

Transition of different states to obtain the steady state is represented by the transition table of the states. It also includes the probability for each transition of the state.

| State | | Next state | Probability |
|---|---|---|---|
| 00000 | => | 01010 | 1 |
| 00001 | => | 10011 | 1 |
| 00010 | => | 00111 | 1 |
| 00011 | => | 11111 | 1 |
| 00100 | => | 11010 | 1 |
| 00101 | => | 00011 | 1 |
| 00110 | => | 10111 | 1 |
| 00111 | => | 01111 | 1 |
| 10101 | => | 00011 | 1 |
| 10110 | => | 00010 | 1 |
| 10111 | => | 01010 | 1 |
| 11000 | => | 10111 | 1 |
| 11001 | => | 11011 | 1 |
| 11010 | => | 11000 | 1 |
| 11011 | => | 10000 | 1 |
| 11100 | => | 00111 | 1 |
| 11101 | => | 01011 | 1 |
| 11110 | => | 01010 | 1 |

Transition table for the genetic circuit .In the last state the 0 represent the OFF state for IPCS_1 and ON state for IPCS_2 and ON and OFF state for the other genes respectively. This reveals the switching of genetic toggle switch of circuit where IPCS_1 was repressed by IPCS_2 that exhibits OFF and ON state respectively. Attractors of the circuit are represented by the attractor plot which includes ON and OFF state for each gene in the network.

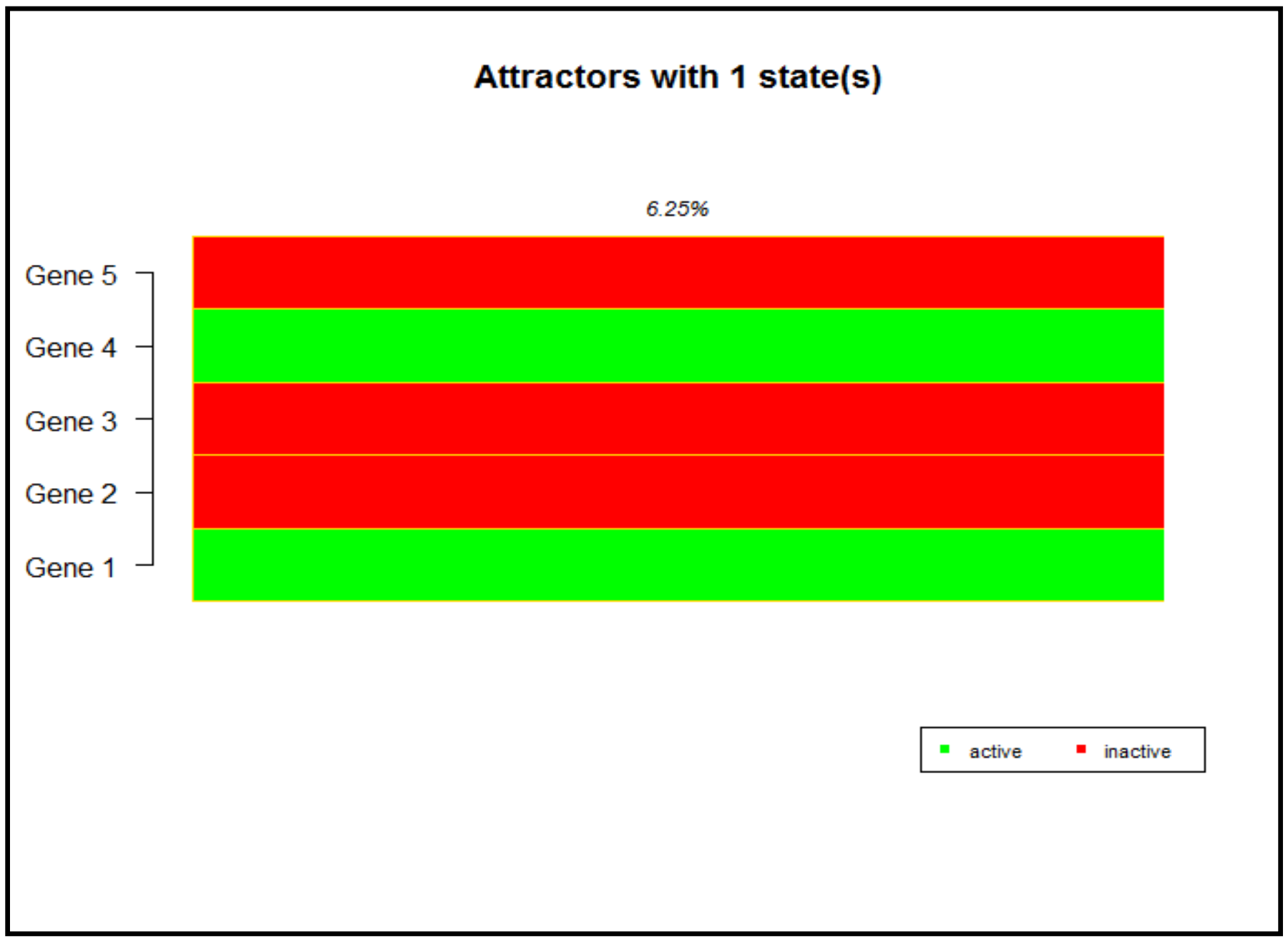

**Fig 7.15** Attractor plot for the **i**nitial attractor of the network.

Legends

Gene1 - IPCS_1

Gene2 - IPCS_2

Gene3 – LacI

Gene4 – Tetr

Gene5 - LamdaR

In initial state, IPCS_1 is ON i.e. active state and IPCS_2 is OFF i.e. inactive state. LacI is OFF so there is no coupling and that results in the repression of IPCs_2 by IPCS_1. After the transition of the state's two steady states were obtained represented by the attractor plot that justifies the dynamic behavior of the circuit to act as bistable genetic switch.

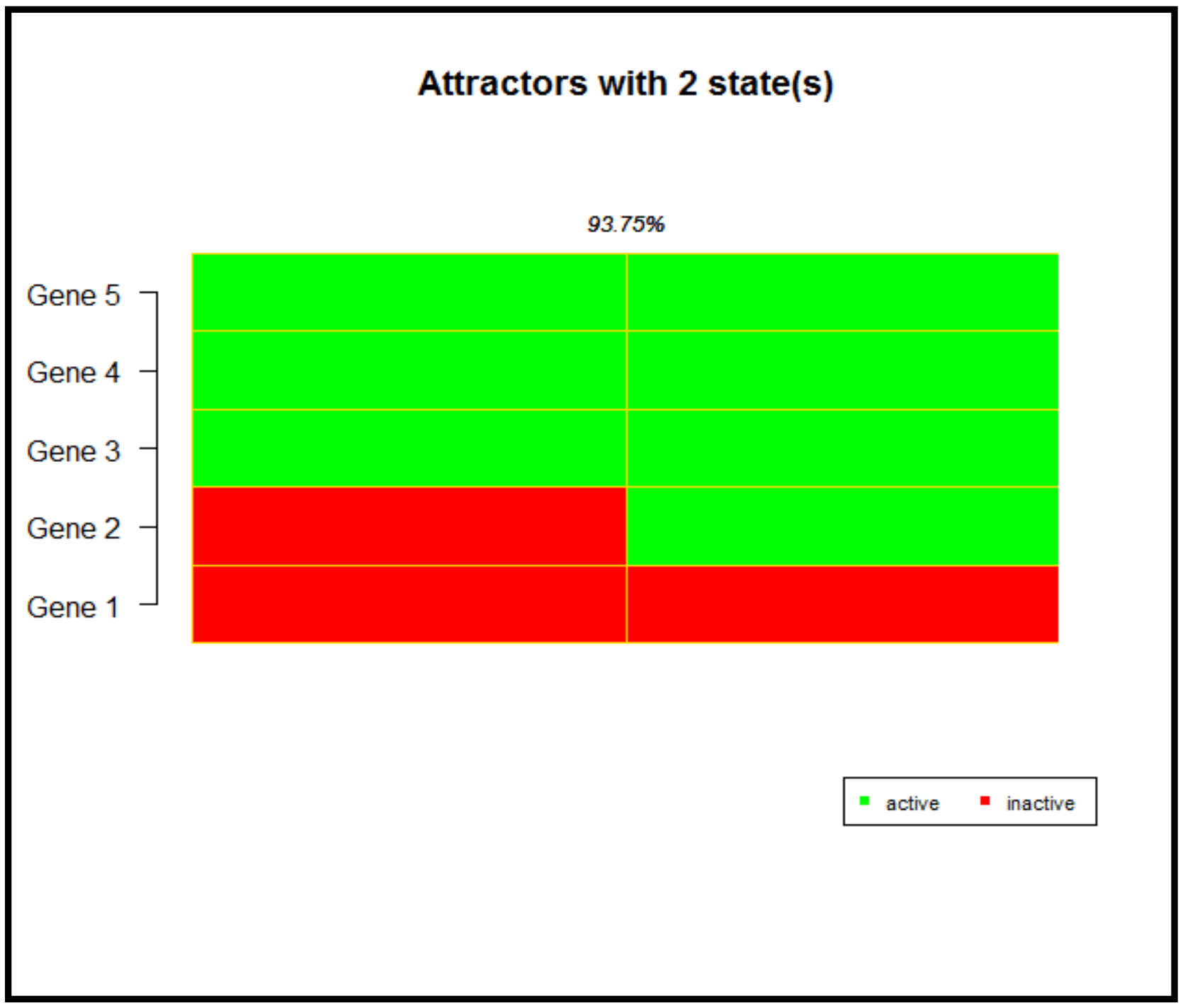

**Fig 7.16** Attractor plot of the two steady states

Attractor plot displays the two steady states of the genetic circuit. The percentage to attain the steady states of the circuit is 93.75 % which reflects the robustness of the attractors. Attractor plot represent the OFF state of IPCS_1 and ON state of IPCS_2 which depicts the switching behavior of the circuit as IPCS_2 represses IPCS_1 which was not in the initial state as mentioned in above figure 3.14.Both the attractor plot represents how the circuit toggles between two states where the one represses the other mutually. This is in line with the hypothesis laid with the constructed genetic circuit for IPCS. Visualization of a sequence of states can be represented by the path of attractors plot. The columns of the table represent consecutive states of the time series. The last state is the steady-state attractor of the network. This gives the inactive and active state of each gene at different time series points. Figure 3.16 represent the active and inactive state of each gene that leads to attractor.

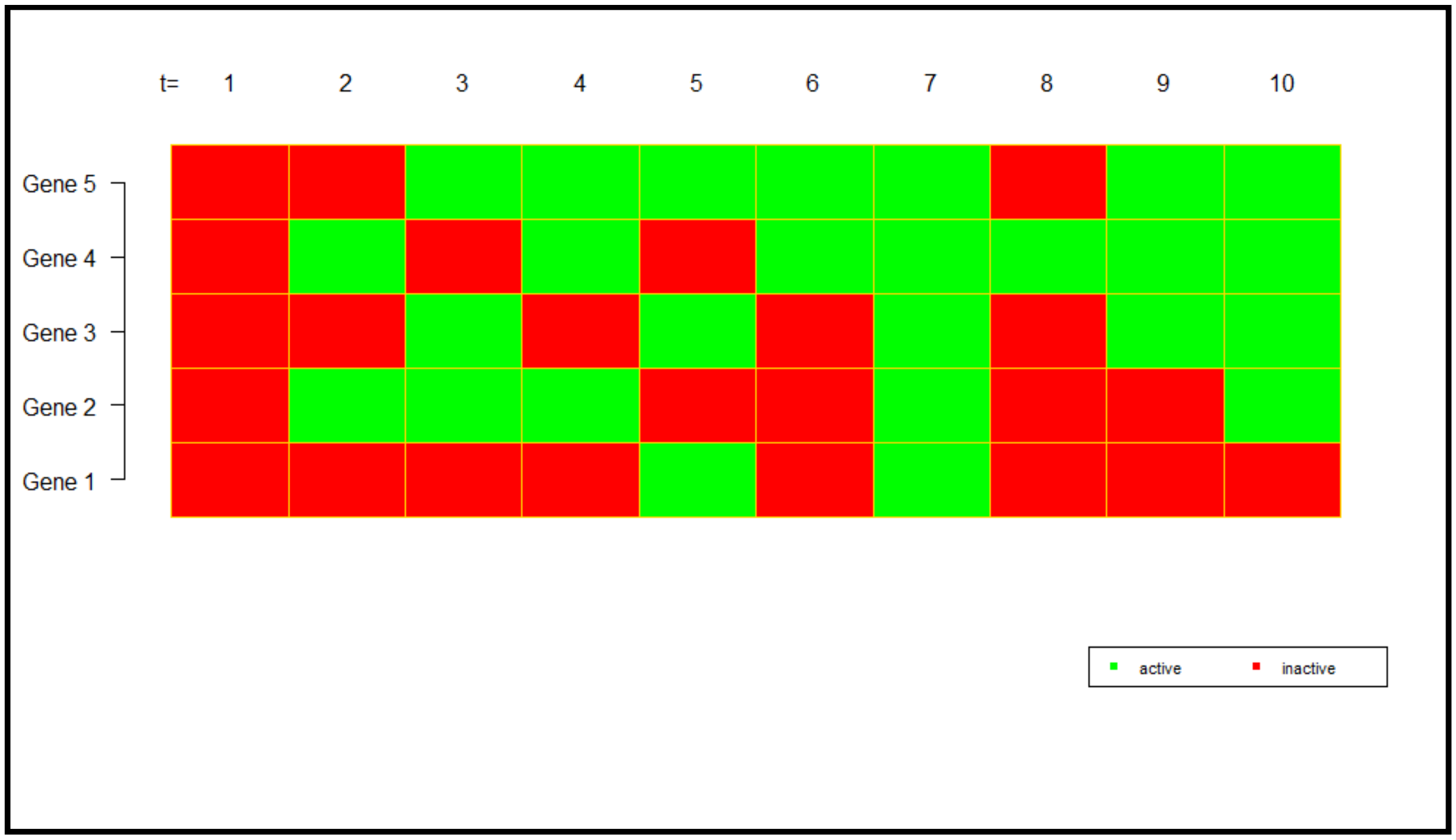

**Fig 7.17** sequences of path to attractor

**Robustness**

Perturbation experiments of the network circuit were performed to test the robustness of structural properties of the networks to noise and mismeasurements The percentage of robustness the biological network was higher than most of the percentages of the random network, this suggests that the biological network circuit constructed exhibits a higher robustness. Figure 3.17 represent the robustness plot for the genetic circuit. The percentage obtained were greater than 50 % which confirms the robustness of the genetic circuit constructed. Network for the genetic circuit was exported to Pajek for the visualization of the transition states in the network. Figure 3.18 represents the transition of different states. Different layout for the network visualization is available at Pajek. Network state graph was visualized using by Kamada-Kawai layout.

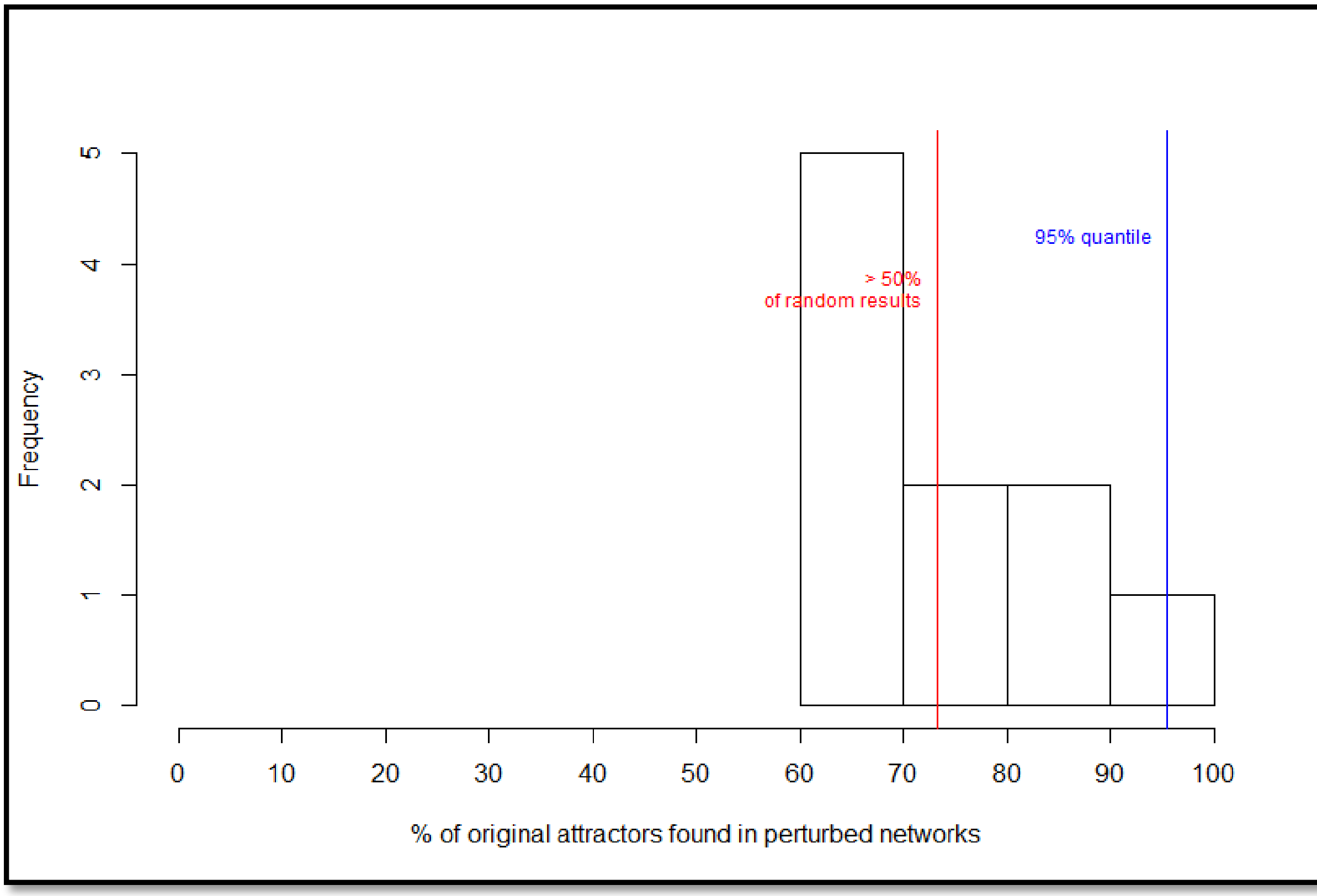

**Fig 7.18: Robustness plot of the genetic circuit.**

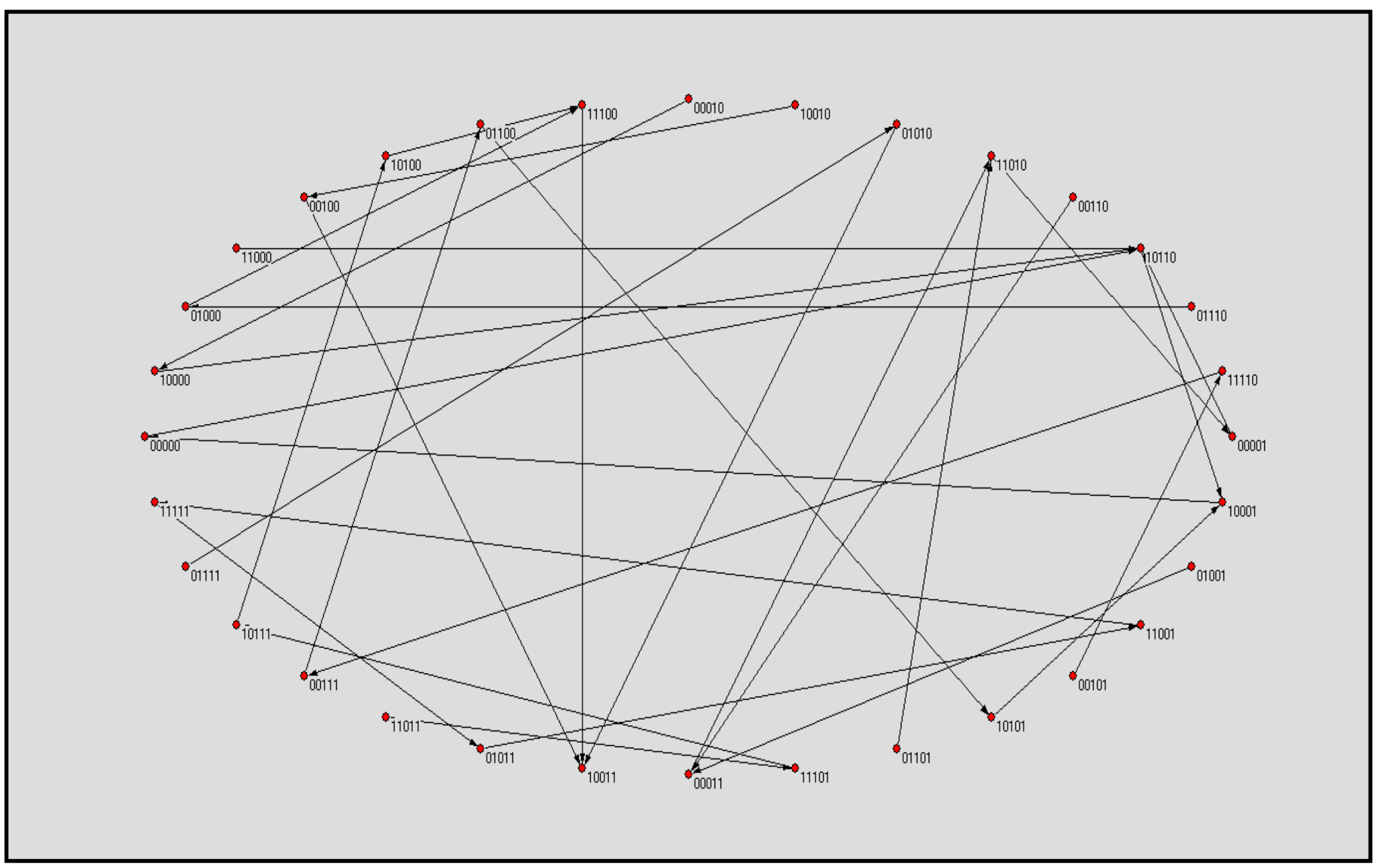

**Fig 7.19 S**tate graph of the transition of states

Robustness of the circuit and the attractors obtained for the genetic circuit justify the design of the circuit for the IPCS *of L.major*. Genetic circuit designed for the IPCS was subjected to simulation and validated by both qualitative and quantative approaches. Results revealed the bistability of the genetic circuit and also the hypothesis laid down while designing the genetic circuit was validated.

The recent dramatic increase in the number of studies bridging synthetic and systems biology is driven by the desire to quantitatively engineer and understand biological systems. The precision resulting from this synergy eliminates much of the uncertainty and failure associated with biological design. This allows for more meaningful conclusions to be drawn from experimental studies. The development of models that capture ever-increasing biological complexity will continue to inspire and guide predictive design of living systems at all scales. Moreover, the ability to precisely perturb these systems will further enable high-resolution insight into biological networks. Such advances hold great promise for deciphering the mechanisms underlying development and disease which facilitates the development of versatile, robust, and useful organisms for industrial biotechnology. Systems approaches may aid in the creation of synthetic circuits which was the underlying aim of the current thesis and synthetic constructs may help uncover novel underlying biological control strategies. This may help the design and implementation of new parts such as synthetic probes and parts that can allow for controlled gene expression. The approaches also facilitate how these synthetic devices can create precise perturbations that can be effectively used to probe the cellular function. Collectively, these prospective advances help mitigate the great complexity and uncertainty currently impeding the study and design of synthetic circuits. Collectively, these prospective advances help mitigate the great complexity and uncertainty currently impeding the study and design of synthetic circuits embedded in cellular systems.

In order to understand the functioning of organisms on the molecular level, we need to know which genes are expressed, when and where in the organism, and to which extent. The regulation of gene expression is achieved through genetic regulatory systems structured by networks of interactions between DNA, RNA, proteins, and small molecules. As most genetic regulatory networks of interest involve many components connected through interlocking positive and negative feedback loops, an intuitive understanding of their dynamics is hard to obtain. Building circuits and studying their behavior in cells is a major goal of synthetic biology in order to evolve a deeper understanding of biological design principles from the bottom up. Collectively, these developments enable the precise control of cellular state for systems studies and the discovery of novel parts, control strategies, and interactions for the design of robust synthetic function .As a consequence, formal methods and computer tools for the modeling and simulation of genetic regulatory networks (genetic circuits) will be indispensable. There are reports and evidences for the synthetic circuit in prokaryotes and eukaryotes like *E.coli* and yeast systems but to the best of my knowledge there are no reports in the literature suggesting the synthetic circuit construction for protozoan parasites. The project serves to the first attempt made to use synthetic biology approach for the construction of genetic circuit for the protozoan parasite *Leishmania*. Leishmania is protozoan parasite responsible for the diseases Leishmaniasis. Inspite of many threuptic approaches exist for this diseases, there is no cure for the diseases. Severity of these lesions are widely distributed all over the world especially non tropical regions of the world. This project reviews formalisms that have been employed in mathematical biology and bioinformatics, synthetic biology to describe genetic regulatory systems, in particular a directed graphs, Bayesian networks, In addition, it discusses how these formalisms have been used in the simulation of the behavior of actual regulatory systems**.** This project aims to construct the genetic circuit for IPCS in L.major, simulation and validation. Genetic circuit was constructed using Tinker cell. Genetic circuit was simulated by using Tauleap stochastic simulation simulation was followed by circuit validation using qualitative and quantitative analysis methods. Qualitative was done by using Boolean method and quantitative was done by using Bayesian method with the aid of ODE of the constructed genetic circuit for IPCS. Qualitative analysis gives hints as to which parameters offer the best success in achieving a desired behavior or whether a certain design can exhibit the wanted function at all. Genetic circuit constructed for IPCS was converted into digital circuit by using Karnaugh map representation of the truth table for the logic circuit deciphered for the constructed genetic circuit. Genetic circuit constructed and validated may be used ad research tool for the protozoan parasite. Genetic circuit for IPCS can be used as threuptic tool for the Leishmaniasis.

# Supplementary file

Genetic circuit constructed for the IPCS of *L.major*. SBML format of the circuit constructed using Tinker cell.

```
<?xml version="1.0" encoding="UTF-8"?>

<sbml xmlns="http://www.sbml.org/sbml/level2" level="2" version="1">

  <model id="repressiblepromotermodel_TIC" name="repressiblepromotermodel.TIC">

    <listOfCompartments>

      <compartment id="DefaultCompartment" name="DefaultCompartment" size="1" constant="true"/>

    </listOfCompartments>

    <listOfSpecies>

      <species id="IPCS1protein" name="IPCS1protein" compartment="DefaultCompartment" initialConcentration="0.05" hasOnlySubstanceUnits="false" boundaryCondition="false" constant="false"/>

      <species id="IPCS2protein" name="IPCS2protein" compartment="DefaultCompartment" initialConcentration="0.05" hasOnlySubstanceUnits="false" boundaryCondition="false" constant="false"/>

      <species id="IPCS_1" name="IPCS_1" compartment="DefaultCompartment" initialConcentration="10" hasOnlySubstanceUnits="false" boundaryCondition="true" constant="false"/>

      <species id="IPCS_2" name="IPCS_2" compartment="DefaultCompartment" initialConcentration="10" hasOnlySubstanceUnits="false" boundaryCondition="true" constant="false"/>

      <species id="TetrRepressor" name="TetrRepressor" compartment="DefaultCompartment" initialConcentration="0.1" hasOnlySubstanceUnits="false" boundaryCondition="false" constant="false"/>

      <species id="Tetrgene" name="Tetrgene" compartment="DefaultCompartment" initialConcentration="10" hasOnlySubstanceUnits="false" boundaryCondition="true" constant="false"/>

      <species id="lacI" name="lacI" compartment="DefaultCompartment" initialConcentration="10" hasOnlySubstanceUnits="false" boundaryCondition="false" constant="false"/>
```

```xml
<species id="lactoserepressor" name="lactoserepressor" compartment="DefaultCompartment" initialConcentration="0.05" hasOnlySubstanceUnits="false" boundaryCondition="false" constant="false"/>

<species id="lamdar" name="lamdar" compartment="DefaultCompartment" initialConcentration="10" hasOnlySubstanceUnits="false" boundaryCondition="true" constant="false"/>

<species id="lamdarepressor" name="lamdarepressor" compartment="DefaultCompartment" initialConcentration="0.05" hasOnlySubstanceUnits="false" boundaryCondition="false" constant="false"/>

<species id="rp1" name="rp1" compartment="DefaultCompartment" initialConcentration="10" hasOnlySubstanceUnits="false" boundaryCondition="true" constant="false"/>

<species id="rp2" name="rp2" compartment="DefaultCompartment" initialConcentration="10" hasOnlySubstanceUnits="false" boundaryCondition="true" constant="false"/>

<species id="rp3" name="rp3" compartment="DefaultCompartment" initialConcentration="10" hasOnlySubstanceUnits="false" boundaryCondition="true" constant="false"/>

<species id="rp4" name="rp4" compartment="DefaultCompartment" initialConcentration="10" hasOnlySubstanceUnits="false" boundaryCondition="true" constant="false"/>

<species id="rp5" name="rp5" compartment="DefaultCompartment" initialConcentration="10" hasOnlySubstanceUnits="false" boundaryCondition="true" constant="false"/>

<species id="rs1" name="rs1" compartment="DefaultCompartment" initialConcentration="10" hasOnlySubstanceUnits="false" boundaryCondition="true" constant="false"/>

</listOfSpecies>

<listOfParameters>

<parameter id="rp1_strength" name="rp1_strength" value="5" constant="true"/>

<parameter id="rp2_strength" name="rp2_strength" value="5" constant="true"/>

<parameter id="rp3_strength" name="rp3_strength" value="5" constant="true"/>

<parameter id="rp4_strength" name="rp4_strength" value="5" constant="true"/>

<parameter id="rp5_strength" name="rp5_strength" value="5" constant="true"/>
```

```xml
<parameter id="TetrRepressor_degradation_rate" name="TetrRepressor_degradation_rate" value="0.5" constant="true"/>
<parameter id="IPCS1protein_degradation_rate" name="IPCS1protein_degradation_rate" value="0.1" constant="true"/>
<parameter id="tr1_Kd" name="tr1_Kd" value="1" constant="true"/>
<parameter id="tr1_h" name="tr1_h" value="2" constant="true"/>
<parameter id="tr2_Kd" name="tr2_Kd" value="1" constant="true"/>
<parameter id="tr2_h" name="tr2_h" value="2" constant="true"/>
<parameter id="tr3_Kd" name="tr3_Kd" value="1" constant="true"/>
<parameter id="tr3_h" name="tr3_h" value="2" constant="true"/>
<parameter id="tr4_Kd" name="tr4_Kd" value="1" constant="true"/>
<parameter id="tr4_h" name="tr4_h" value="2" constant="true"/>
<parameter id="tr5_Kd" name="tr5_Kd" value="1" constant="true"/>
<parameter id="tr5_h" name="tr5_h" value="2" constant="true"/>
<parameter id="tr6_Kd" name="tr6_Kd" value="1" constant="true"/>
<parameter id="tr6_h" name="tr6_h" value="2" constant="true"/>
<parameter id="pp1_translation_rate" name="pp1_translation_rate" value="1" constant="true"/>
<parameter id="pp2_translation_rate" name="pp2_translation_rate" value="1" constant="true"/>
<parameter id="pp3_translation_rate" name="pp3_translation_rate" value="1" constant="true"/>
<parameter id="pp4_translation_rate" name="pp4_translation_rate" value="1" constant="true"/>
<parameter id="pp5_translation_rate" name="pp5_translation_rate" value="1" constant="true"/>
<parameter id="IPCS2protein_degradation_rate" name="IPCS2protein_degradation_rate" value="0.1" constant="true"/>
<parameter id="lamdarepressor_degradation_rate" name="lamdarepressor_degradation_rate" value="0.1" constant="true"/>
```

```
      <parameter id="lactoserepressor_degradation_rate"
name="lactoserepressor_degradation_rate" value="0.1" constant="true"/>
    </listOfParameters>
    <listOfRules>
      <assignmentRule variable="rp1">
        <math xmlns="http://www.w3.org/1998/Math/MathML">
          <apply>
            <divide/>
            <cn> 1 </cn>
            <apply>
              <plus/>
              <cn type="integer"> 1 </cn>
              <apply>
                <power/>
                <apply>
                  <divide/>
                  <ci> IPCS2protein </ci>
                  <ci> tr2_Kd </ci>
                </apply>
                <ci> tr2_h </ci>
              </apply>
            </apply>
          </apply>
        </math>
      </assignmentRule>
      <assignmentRule variable="rp2">
        <math xmlns="http://www.w3.org/1998/Math/MathML">
```

```
    <apply>
     <divide/>
     <cn> 1 </cn>
     <apply>
      <plus/>
      <cn type="integer"> 1 </cn>
      <apply>
       <power/>
       <apply>
        <divide/>
        <ci> IPCS1protein </ci>
        <ci> tr1_Kd </ci>
       </apply>
       <ci> tr1_h </ci>
      </apply>
     </apply>
    </apply>
   </math>
  </assignmentRule>
  <assignmentRule variable="rp3">
   <math xmlns="http://www.w3.org/1998/Math/MathML">
    <apply>
     <divide/>
     <cn> 1 </cn>
     <apply>
      <plus/>
      <cn type="integer"> 1 </cn>
```

```
        <apply>
          <power/>
          <apply>
            <divide/>
            <ci> lamdarepressor </ci>
            <ci> tr5_Kd </ci>
          </apply>
          <ci> tr5_h </ci>
        </apply>
      </apply>
    </apply>
  </math>
</assignmentRule>
<assignmentRule variable="rp4">
  <math xmlns="http://www.w3.org/1998/Math/MathML">
    <apply>
      <divide/>
      <cn> 1 </cn>
      <apply>
        <plus/>
        <cn type="integer"> 1 </cn>
        <apply>
          <power/>
          <apply>
            <divide/>
            <ci> lactoserepressor </ci>
            <ci> tr3_Kd </ci>
```

```
        </apply>
        <ci> tr3_h </ci>
      </apply>
    </apply>
  </apply>
</math>
</assignmentRule>
<assignmentRule variable="rp5">
<math xmlns="http://www.w3.org/1998/Math/MathML">
  <apply>
    <divide/>
    <cn> 1 </cn>
    <apply>
      <plus/>
      <cn type="integer"> 1 </cn>
      <apply>
        <power/>
        <apply>
          <divide/>
          <ci> TetrRepressor </ci>
          <ci> tr4_Kd </ci>
        </apply>
        <ci> tr4_h </ci>
      </apply>
    </apply>
  </apply>
</math>
```

```
</assignmentRule>
<assignmentRule variable="Tetrgene">
  <math xmlns="http://www.w3.org/1998/Math/MathML">
    <apply>
      <times/>
      <ci> rp4_strength </ci>
      <ci> rp4 </ci>
    </apply>
  </math>
</assignmentRule>
<assignmentRule variable="rs1">
  <math xmlns="http://www.w3.org/1998/Math/MathML">
    <apply>
      <divide/>
      <cn> 1 </cn>
      <apply>
        <plus/>
        <cn type="integer"> 1 </cn>
        <apply>
          <power/>
          <apply>
            <divide/>
            <ci> lactoserepressor </ci>
            <ci> tr6_Kd </ci>
          </apply>
          <ci> tr6_h </ci>
        </apply>
```

```
      </apply>
    </apply>
  </math>
</assignmentRule>
<assignmentRule variable="IPCS_1">
  <math xmlns="http://www.w3.org/1998/Math/MathML">
    <apply>
      <times/>
      <ci> rp1_strength </ci>
      <ci> rp1 </ci>
    </apply>
  </math>
</assignmentRule>
<assignmentRule variable="IPCS_2">
  <math xmlns="http://www.w3.org/1998/Math/MathML">
    <apply>
      <times/>
      <ci> rp2_strength </ci>
      <ci> rp2 </ci>
    </apply>
  </math>
</assignmentRule>
<assignmentRule variable="lamdar">
  <math xmlns="http://www.w3.org/1998/Math/MathML">
    <apply>
      <times/>
      <ci> rp5_strength </ci>
```

```
      <ci> rp5 </ci>
     </apply>
    </math>
   </assignmentRule>
  </listOfRules>
  <listOfReactions>
   <reaction id="pp1_v1" name="pp1_v1" reversible="false" fast="false">
    <listOfProducts>
     <speciesReference species="IPCS1protein" stoichiometry="1"/>
    </listOfProducts>
    <listOfModifiers>
     <modifierSpeciesReference species="IPCS_1"/>
    </listOfModifiers>
    <kineticLaw>
     <math xmlns="http://www.w3.org/1998/Math/MathML">
      <apply>
       <times/>
       <ci> pp1_translation_rate </ci>
       <ci> rp1_strength </ci>
       <ci> IPCS_1 </ci>
      </apply>
     </math>
    </kineticLaw>
   </reaction>
   <reaction id="pp1_v2" name="pp1_v2" reversible="false" fast="false">
    <listOfReactants>
     <speciesReference species="IPCS1protein" stoichiometry="1"/>
```

```
    </listOfReactants>
    <kineticLaw>
      <math xmlns="http://www.w3.org/1998/Math/MathML">
        <apply>
          <times/>
          <ci> IPCS1protein_degradation_rate </ci>
          <ci> IPCS1protein </ci>
        </apply>
      </math>
    </kineticLaw>
  </reaction>
  <reaction id="pp2_v1" name="pp2_v1" reversible="false" fast="false">
    <listOfProducts>
      <speciesReference species="IPCS2protein" stoichiometry="1"/>
    </listOfProducts>
    <listOfModifiers>
      <modifierSpeciesReference species="IPCS_2"/>
    </listOfModifiers>
    <kineticLaw>
      <math xmlns="http://www.w3.org/1998/Math/MathML">
        <apply>
          <times/>
          <ci> pp2_translation_rate </ci>
          <ci> rp2_strength </ci>
          <ci> IPCS_2 </ci>
        </apply>
      </math>
```

```
  </kineticLaw>
</reaction>
<reaction id="pp2_v2" name="pp2_v2" reversible="false" fast="false">
  <listOfReactants>
    <speciesReference species="IPCS2protein" stoichiometry="1"/>
  </listOfReactants>
  <kineticLaw>
    <math xmlns="http://www.w3.org/1998/Math/MathML">
      <apply>
        <times/>
        <ci> IPCS2protein_degradation_rate </ci>
        <ci> IPCS2protein </ci>
      </apply>
    </math>
  </kineticLaw>
</reaction>
<reaction id="pp3_v1" name="pp3_v1" reversible="false" fast="false">
  <listOfProducts>
    <speciesReference species="lactoserepressor" stoichiometry="1"/>
  </listOfProducts>
  <listOfModifiers>
    <modifierSpeciesReference species="lacI"/>
  </listOfModifiers>
  <kineticLaw>
    <math xmlns="http://www.w3.org/1998/Math/MathML">
      <apply>
        <times/>
```

```
        <ci> pp3_translation_rate </ci>
        <ci> rp3_strength </ci>
        <ci> lacI </ci>
      </apply>
    </math>
  </kineticLaw>
</reaction>
<reaction id="pp3_v2" name="pp3_v2" reversible="false" fast="false">
  <listOfReactants>
    <speciesReference species="lactoserepressor" stoichiometry="1"/>
  </listOfReactants>
  <kineticLaw>
    <math xmlns="http://www.w3.org/1998/Math/MathML">
      <apply>
        <times/>
        <ci> lactoserepressor_degradation_rate </ci>
        <ci> lactoserepressor </ci>
      </apply>
    </math>
  </kineticLaw>
</reaction>
<reaction id="pp4_v1" name="pp4_v1" reversible="false" fast="false">
  <listOfProducts>
    <speciesReference species="TetrRepressor" stoichiometry="1"/>
  </listOfProducts>
  <listOfModifiers>
    <modifierSpeciesReference species="Tetrgene"/>
```

```
  </listOfModifiers>
  <kineticLaw>
    <math xmlns="http://www.w3.org/1998/Math/MathML">
      <apply>
        <times/>
        <ci> pp4_translation_rate </ci>
        <cn> 1 </cn>
        <ci> Tetrgene </ci>
      </apply>
    </math>
  </kineticLaw>
</reaction>
<reaction id="pp4_v2" name="pp4_v2" reversible="false" fast="false">
  <listOfReactants>
    <speciesReference species="TetrRepressor" stoichiometry="1"/>
  </listOfReactants>
  <kineticLaw>
    <math xmlns="http://www.w3.org/1998/Math/MathML">
      <apply>
        <times/>
        <ci> TetrRepressor_degradation_rate </ci>
        <ci> TetrRepressor </ci>
      </apply>
    </math>
  </kineticLaw>
</reaction>
<reaction id="pp5_v1" name="pp5_v1" reversible="false" fast="false">
```

```
    <listOfProducts>
      <speciesReference species="lamdarepressor" stoichiometry="1"/>
    </listOfProducts>
    <listOfModifiers>
      <modifierSpeciesReference species="lamdar"/>
    </listOfModifiers>
    <kineticLaw>
      <math xmlns="http://www.w3.org/1998/Math/MathML">
        <apply>
          <times/>
          <ci> pp5_translation_rate </ci>
          <cn> 1 </cn>
          <ci> lamdar </ci>
        </apply>
      </math>
    </kineticLaw>
  </reaction>
  <reaction id="pp5_v2" name="pp5_v2" reversible="false" fast="false">
    <listOfReactants>
      <speciesReference species="lamdarepressor" stoichiometry="1"/>
    </listOfReactants>
    <kineticLaw>
      <math xmlns="http://www.w3.org/1998/Math/MathML">
        <apply>
          <times/>
          <ci> lamdarepressor_degradation_rate </ci>
          <ci> lamdarepressor </ci>
```

</apply>

</math>

</kineticLaw>

</reaction>

</listOfReactions>

</model>

</sbml>

**Steady state graphs**

Steady state graphs for the degradation rates at different concentration

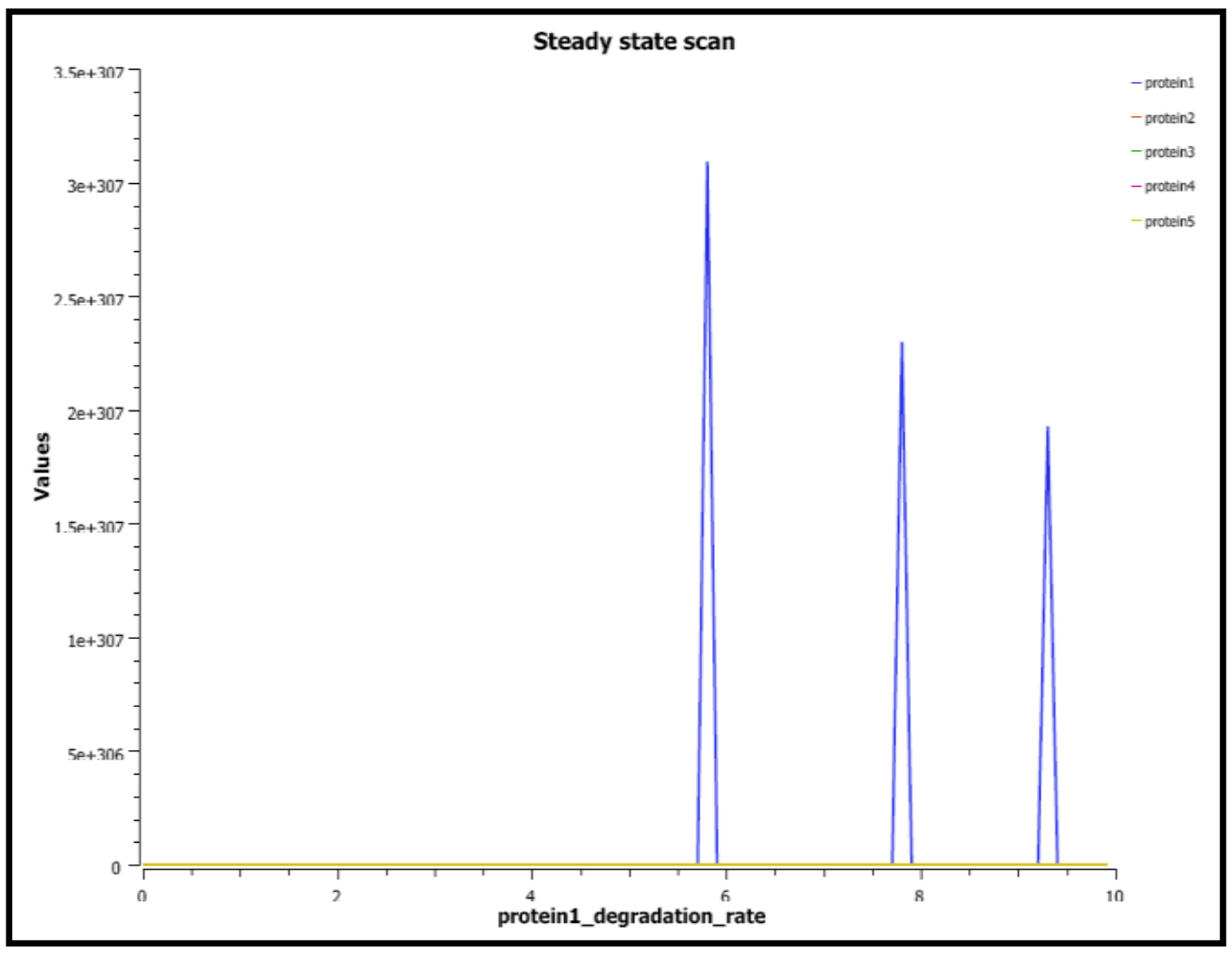

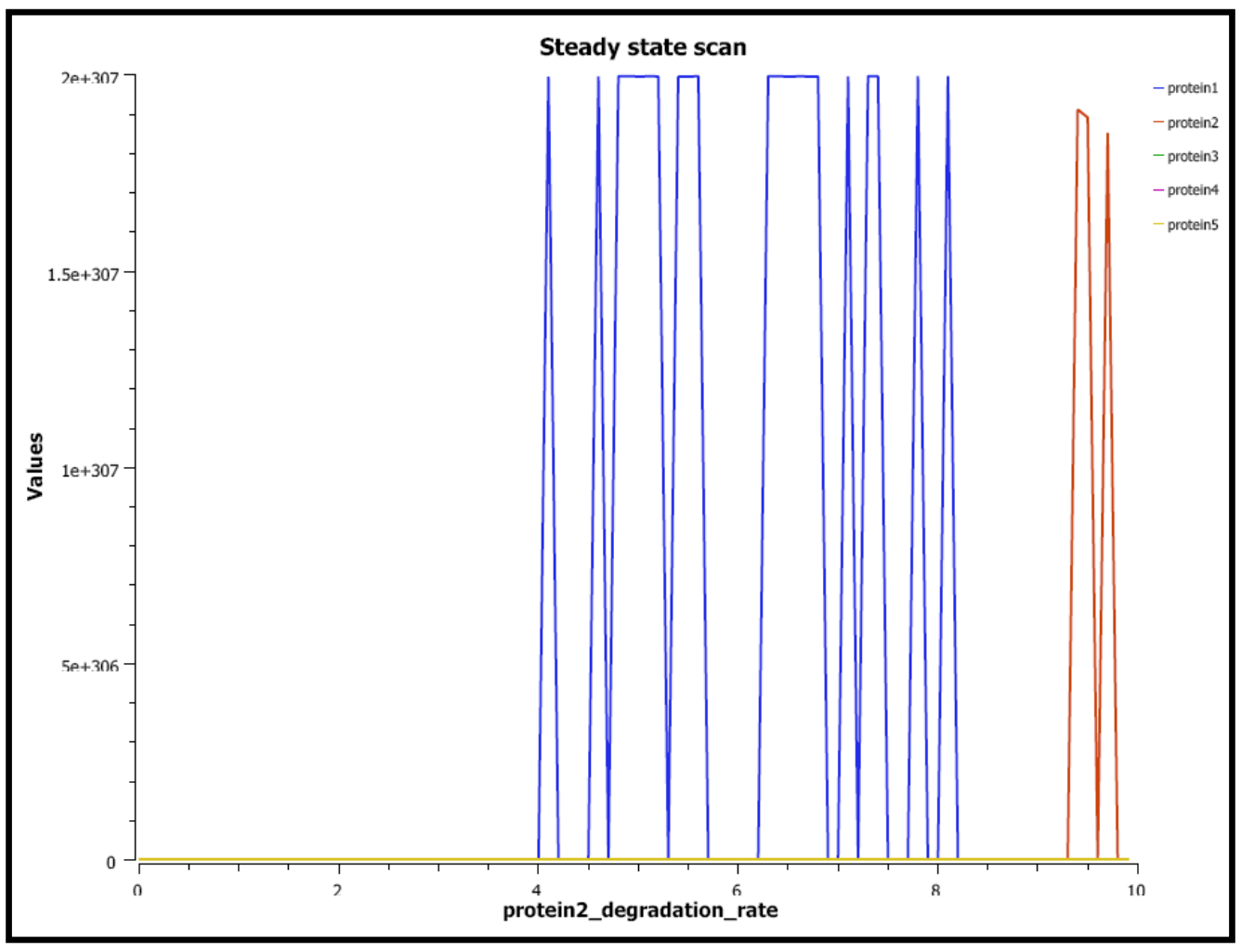

Fig 11.1 a.). Degradation of IPCS_1

b.) Degradation of IPCS_2

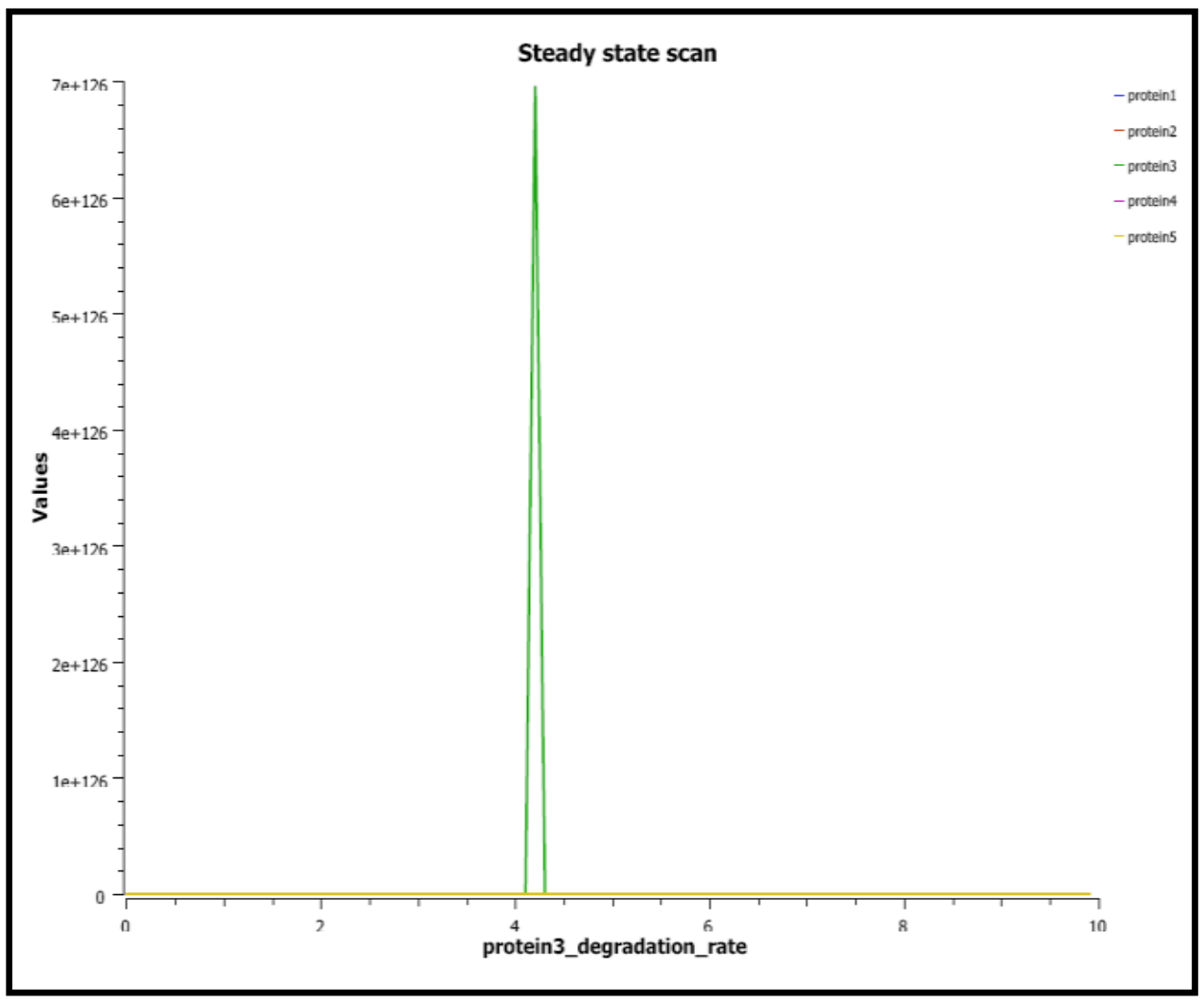

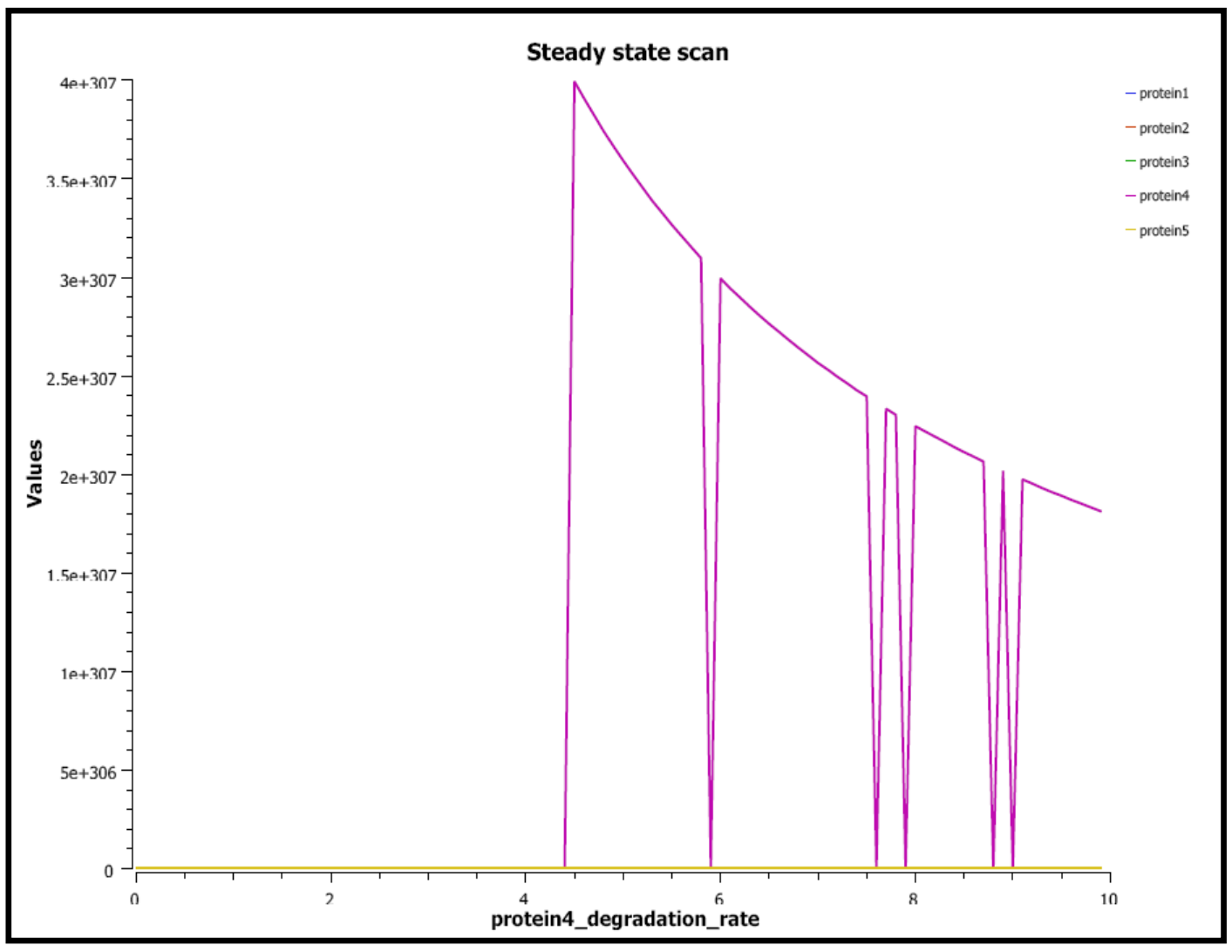

C.) degradation graph of LacI

d.) degradation graph of Tetr

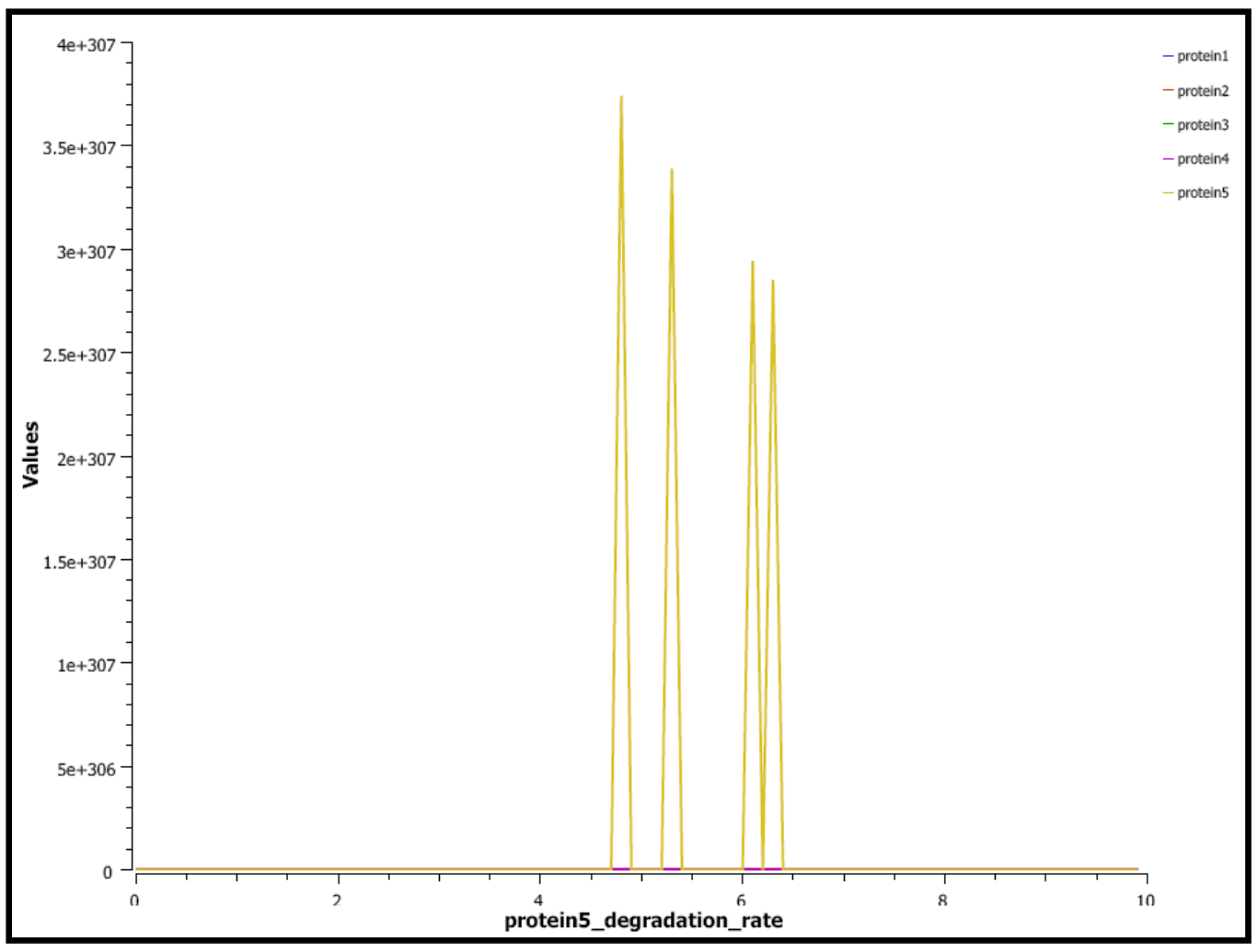

e.) degradation graph of Lamdar

**Simulation results of circuit by change in Kd values of IPCS_1 (Case 1 results)**

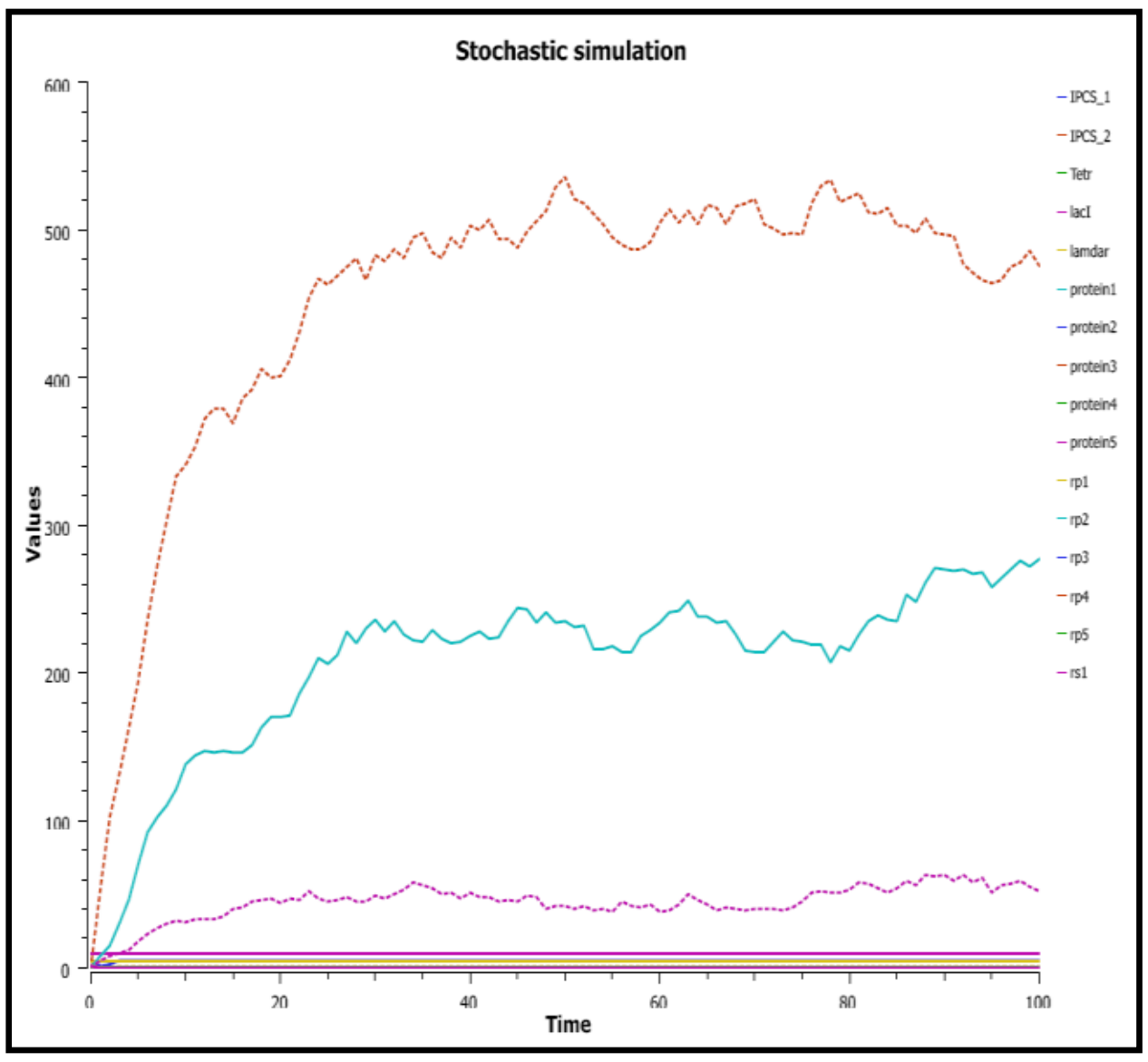

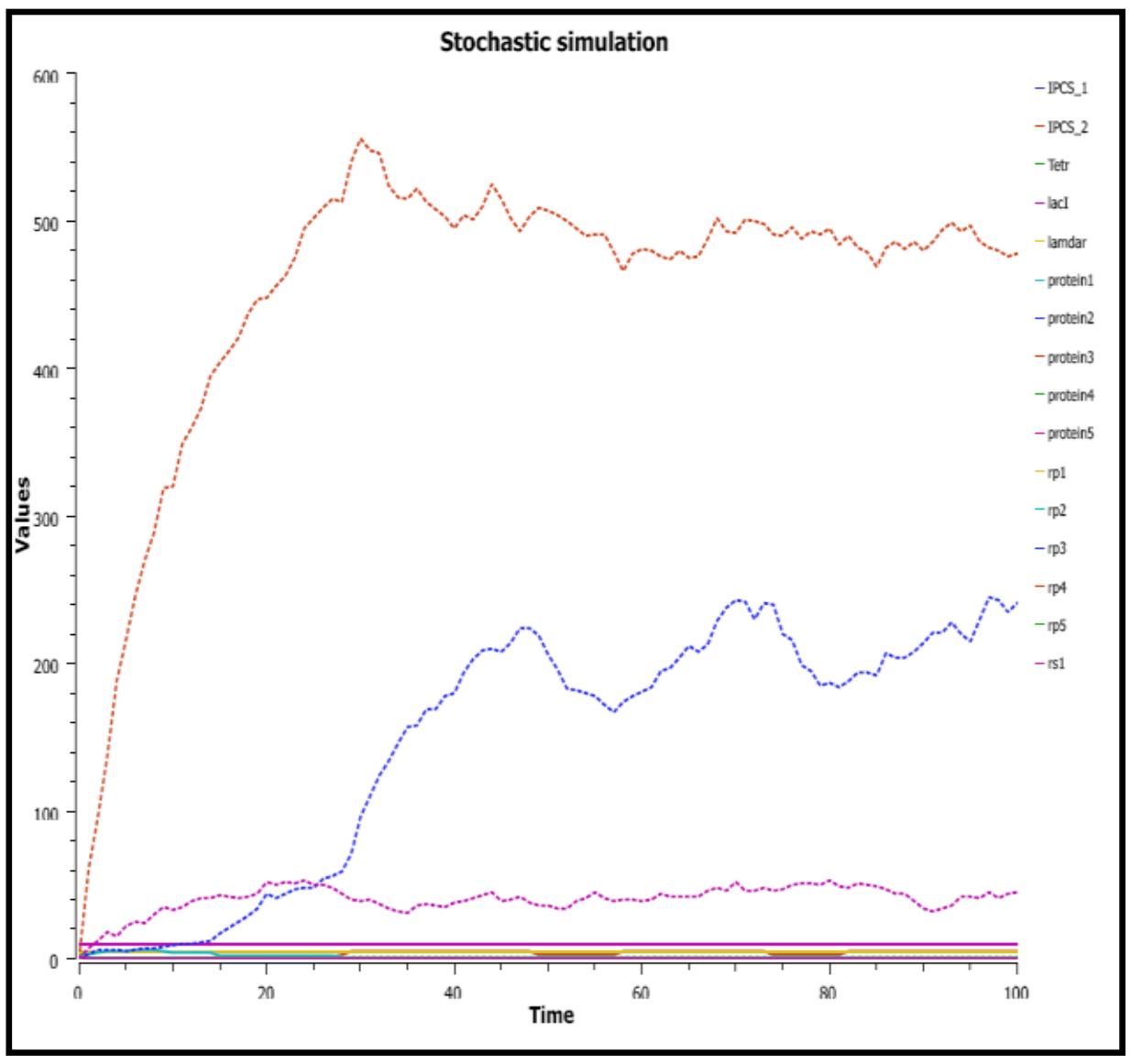

Fig 11.2 : a.)Kd =1.02 b.) Kd =1.03

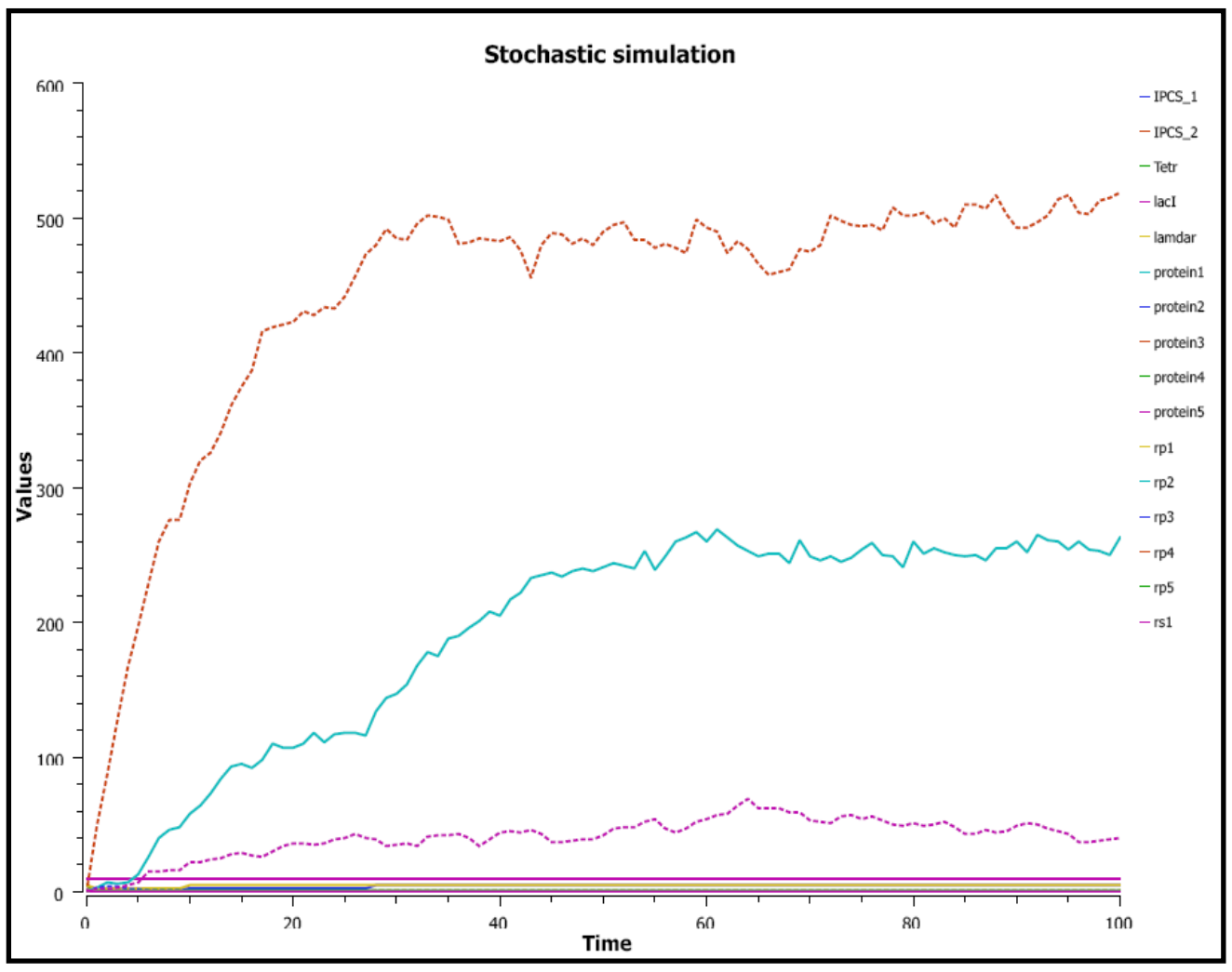

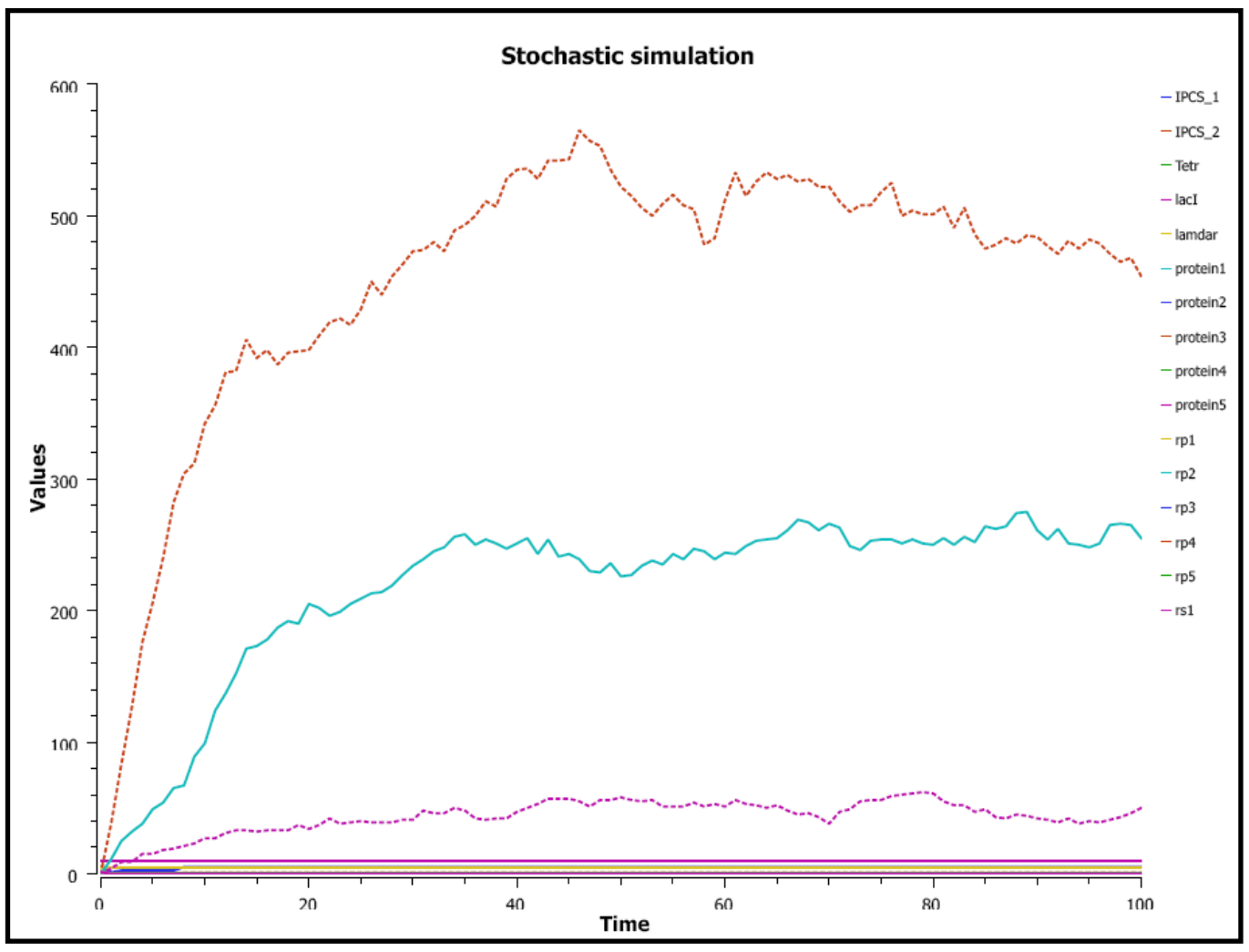

c.) Kd =1.04 d.) Kd=1.05

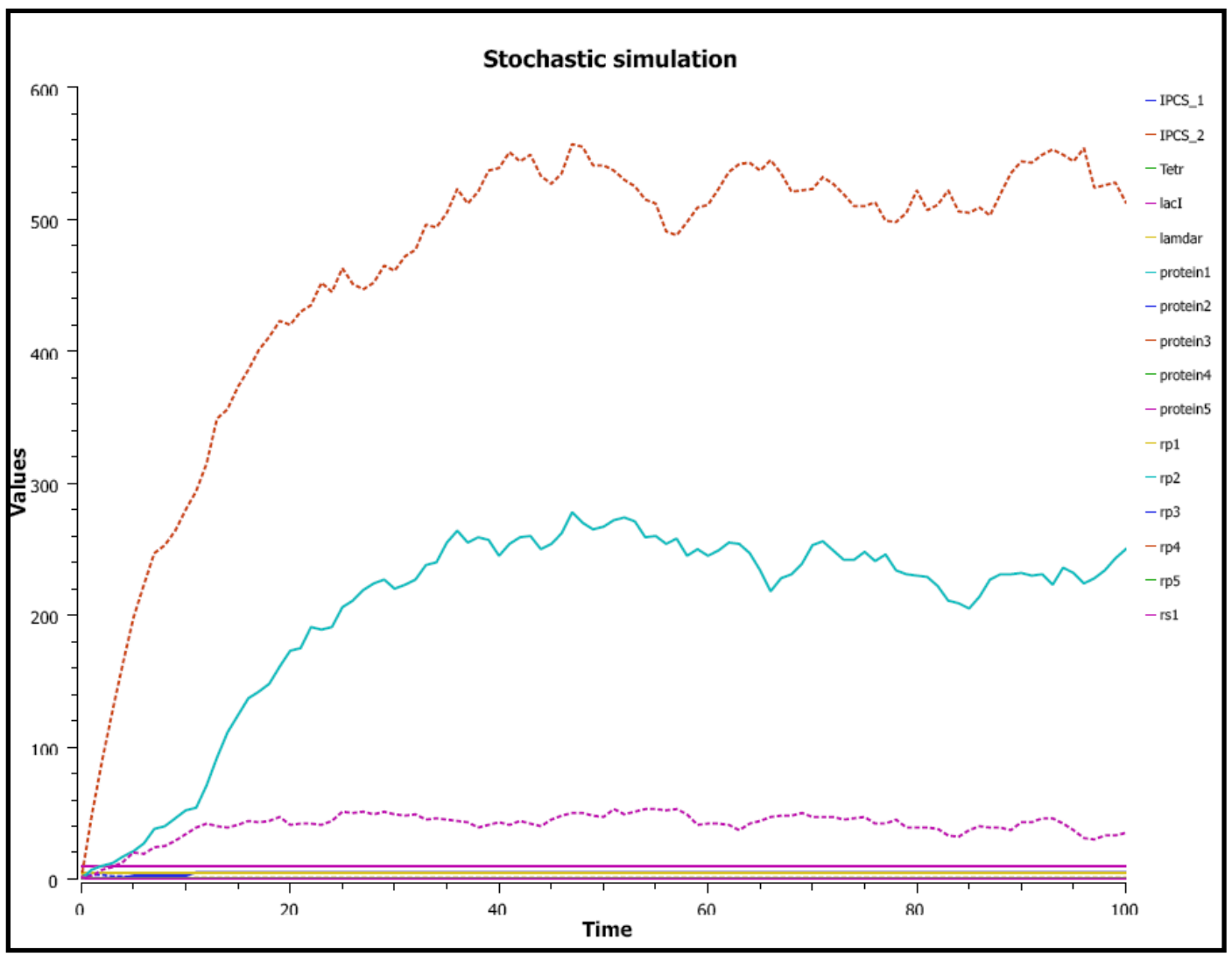

e.) Kd = 1.06

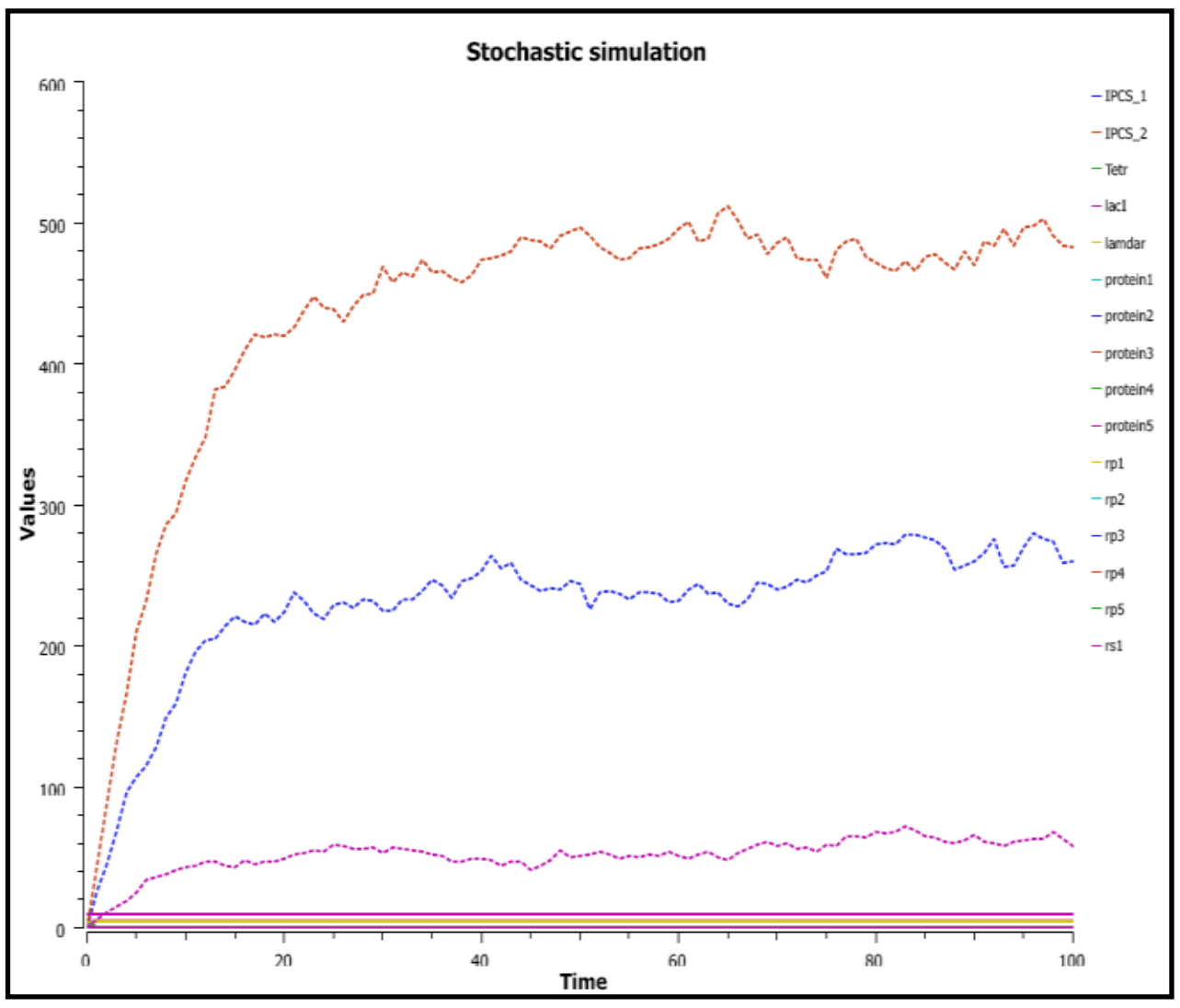

f.) Kd =1.07

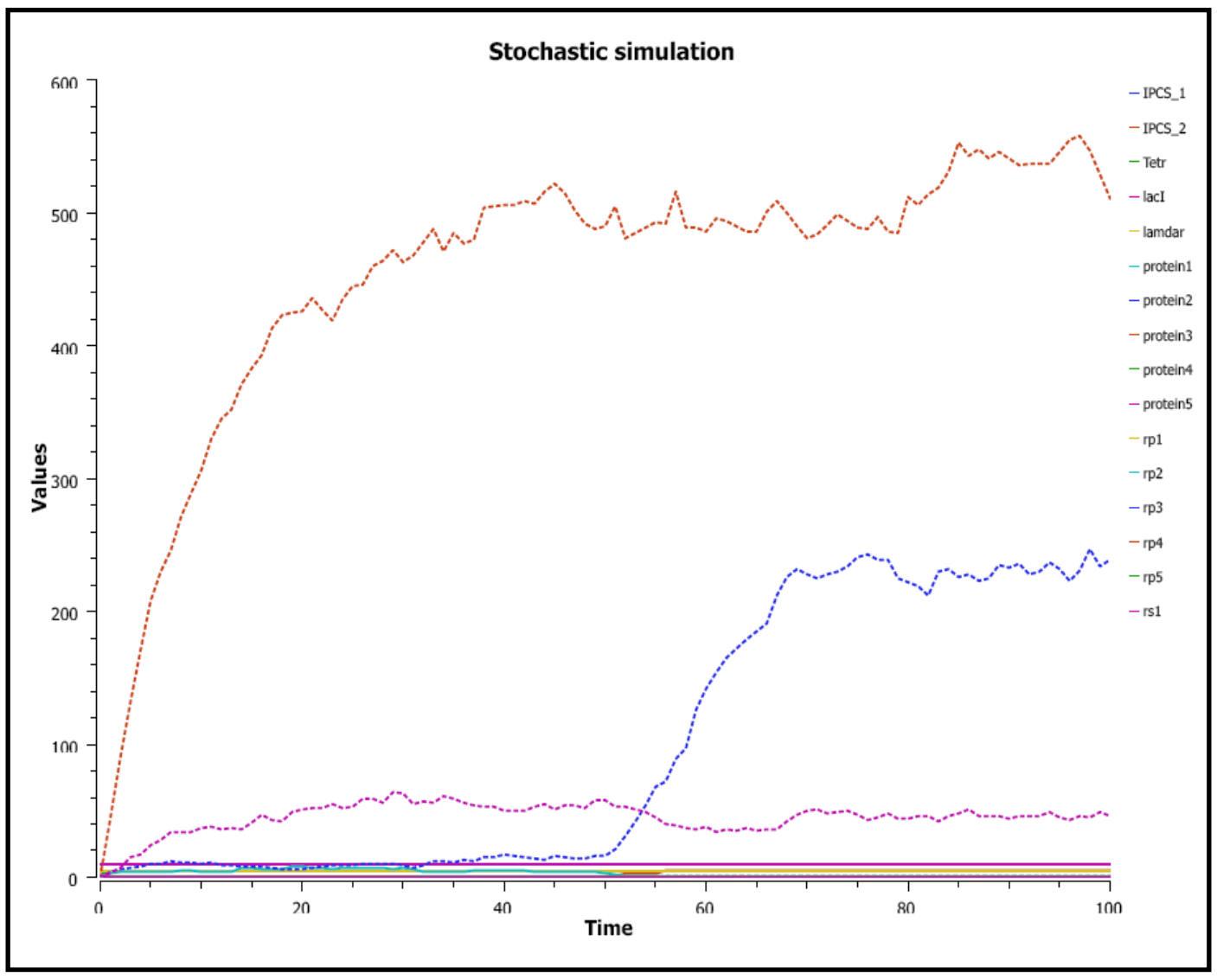

g.) Kd=1.08

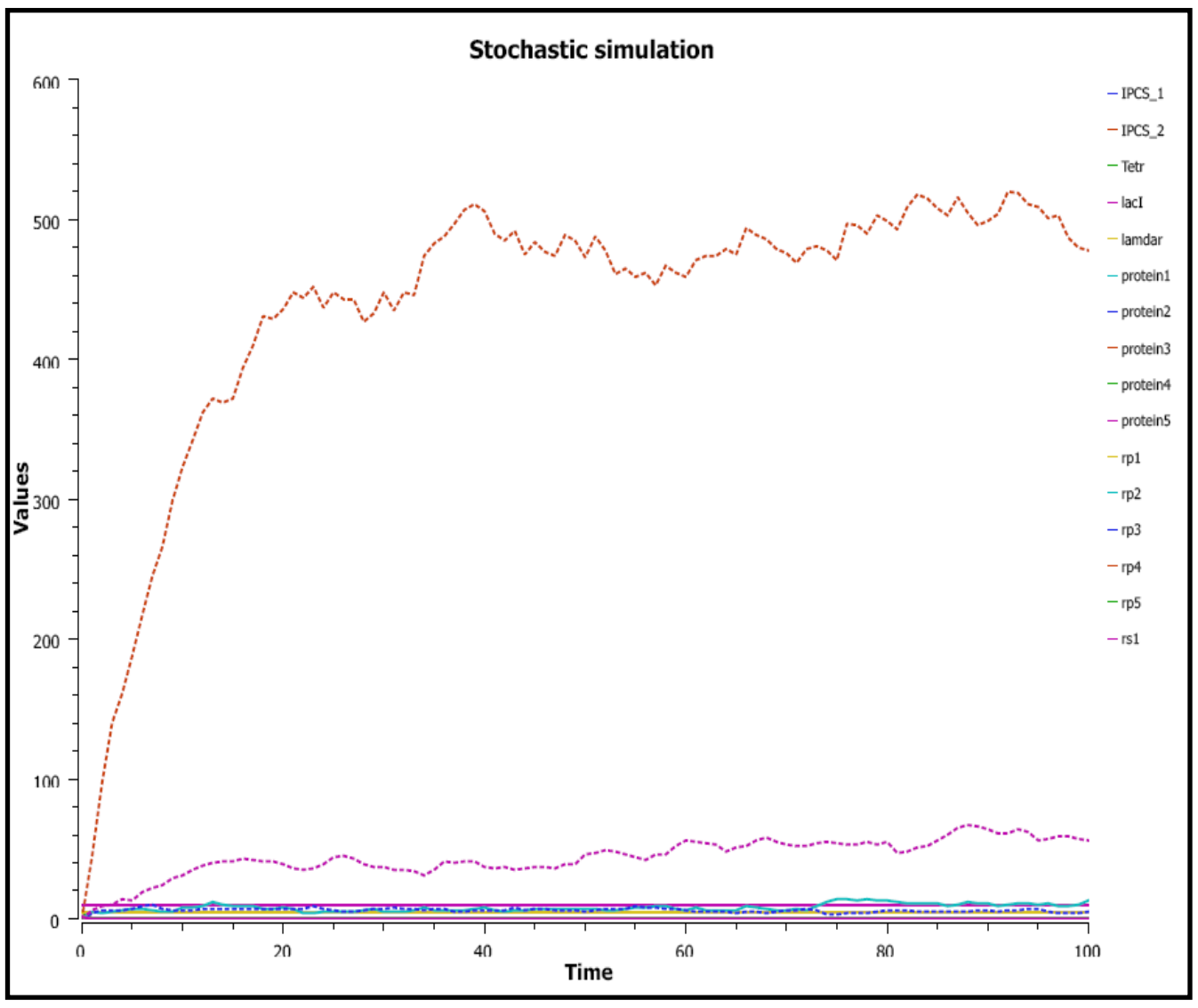

h.) Kd=1.09

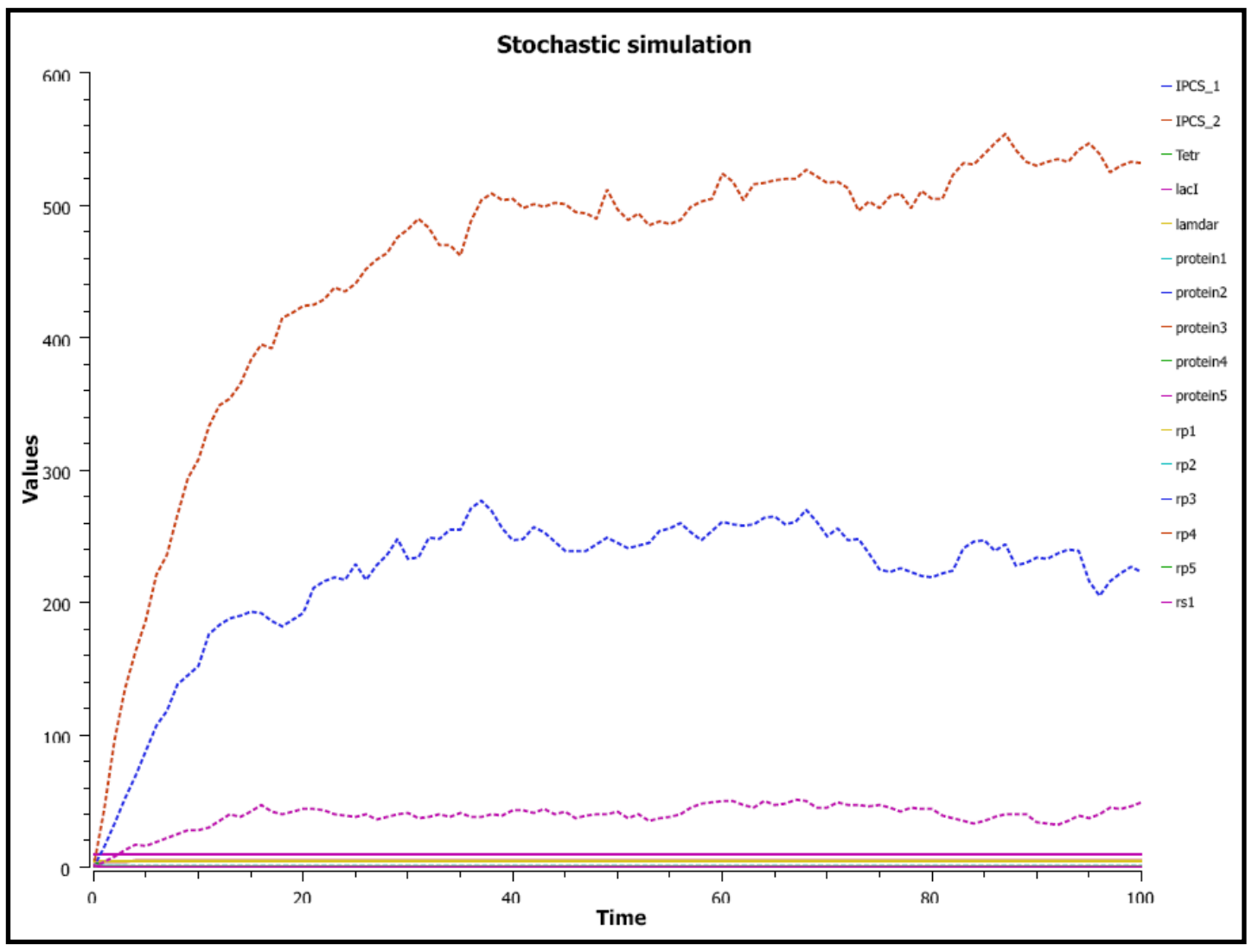

i.) Kd=2

**Simulation results of circuit by change in Kd values of IPCS_2 (Case 2 results)**

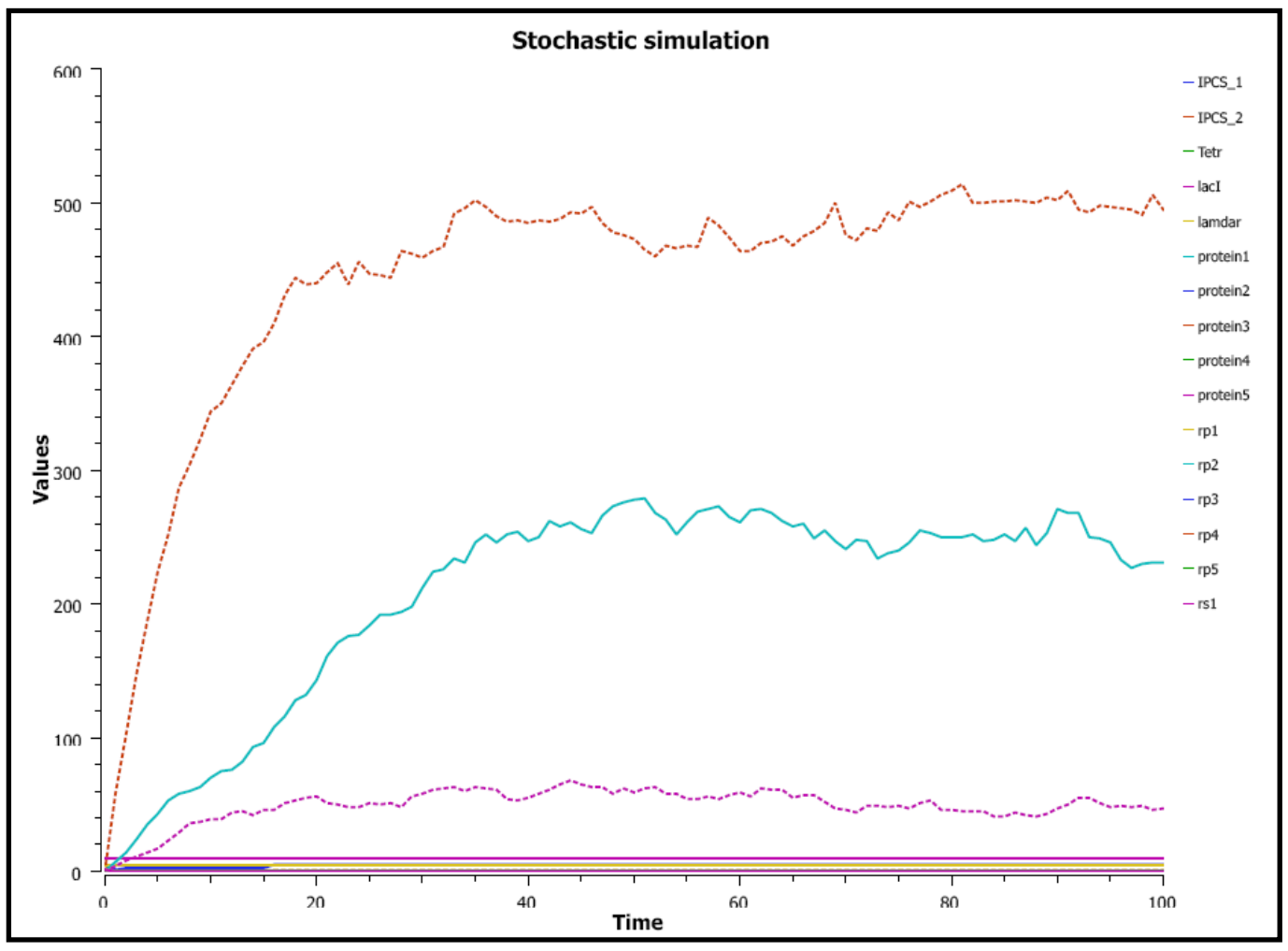

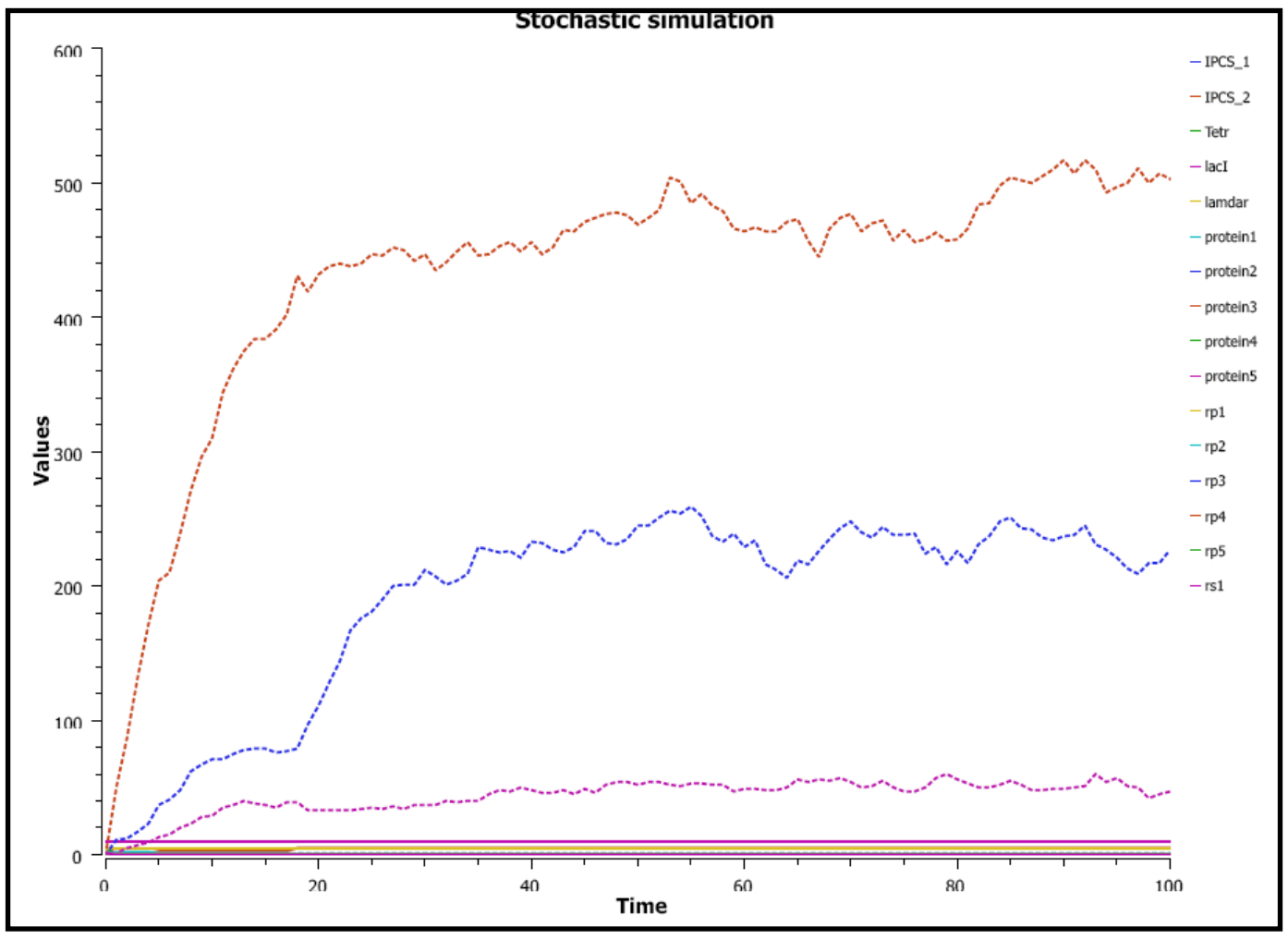

Fig 11.3 a.)Kd=1.01

b.) Kd=1.02

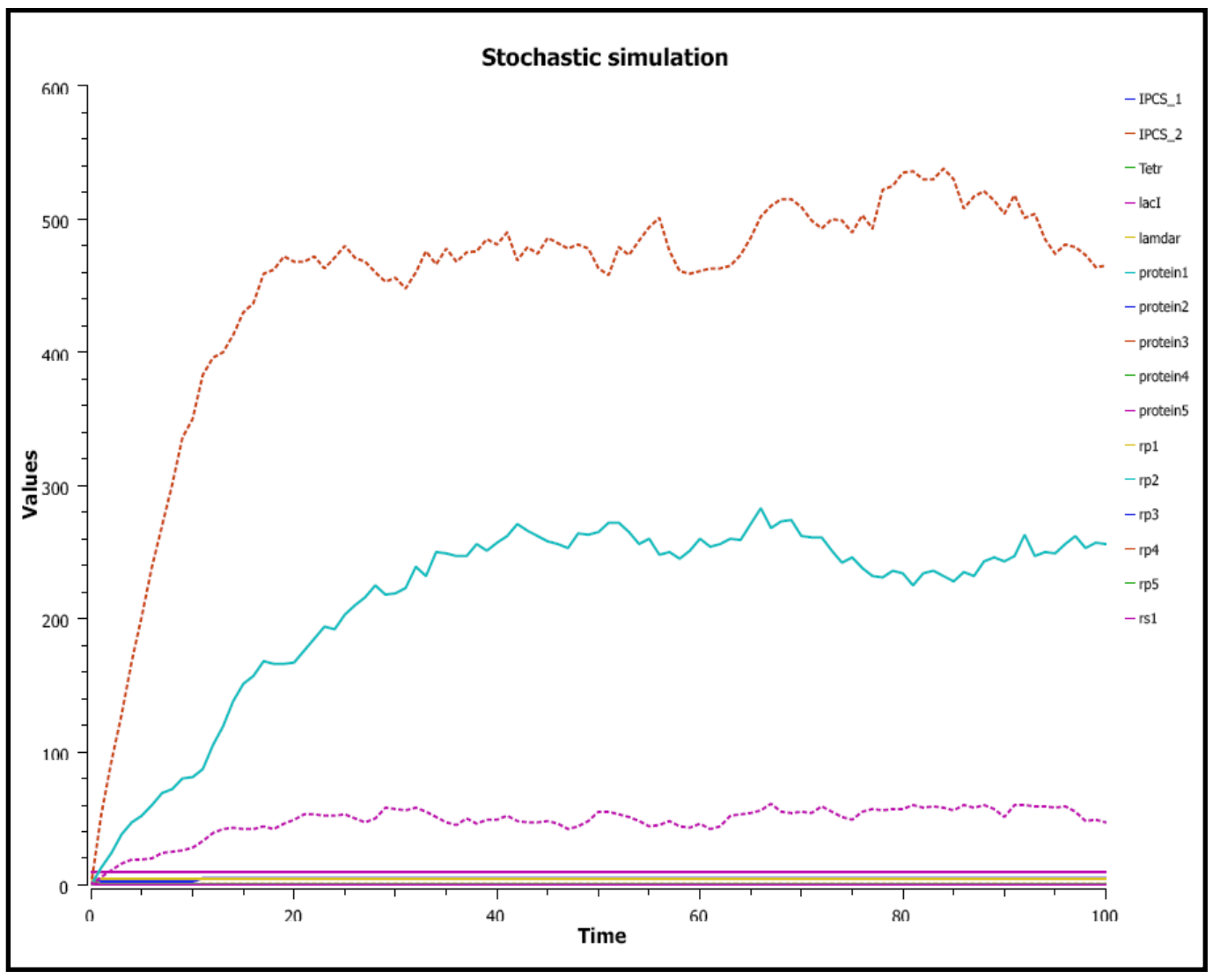

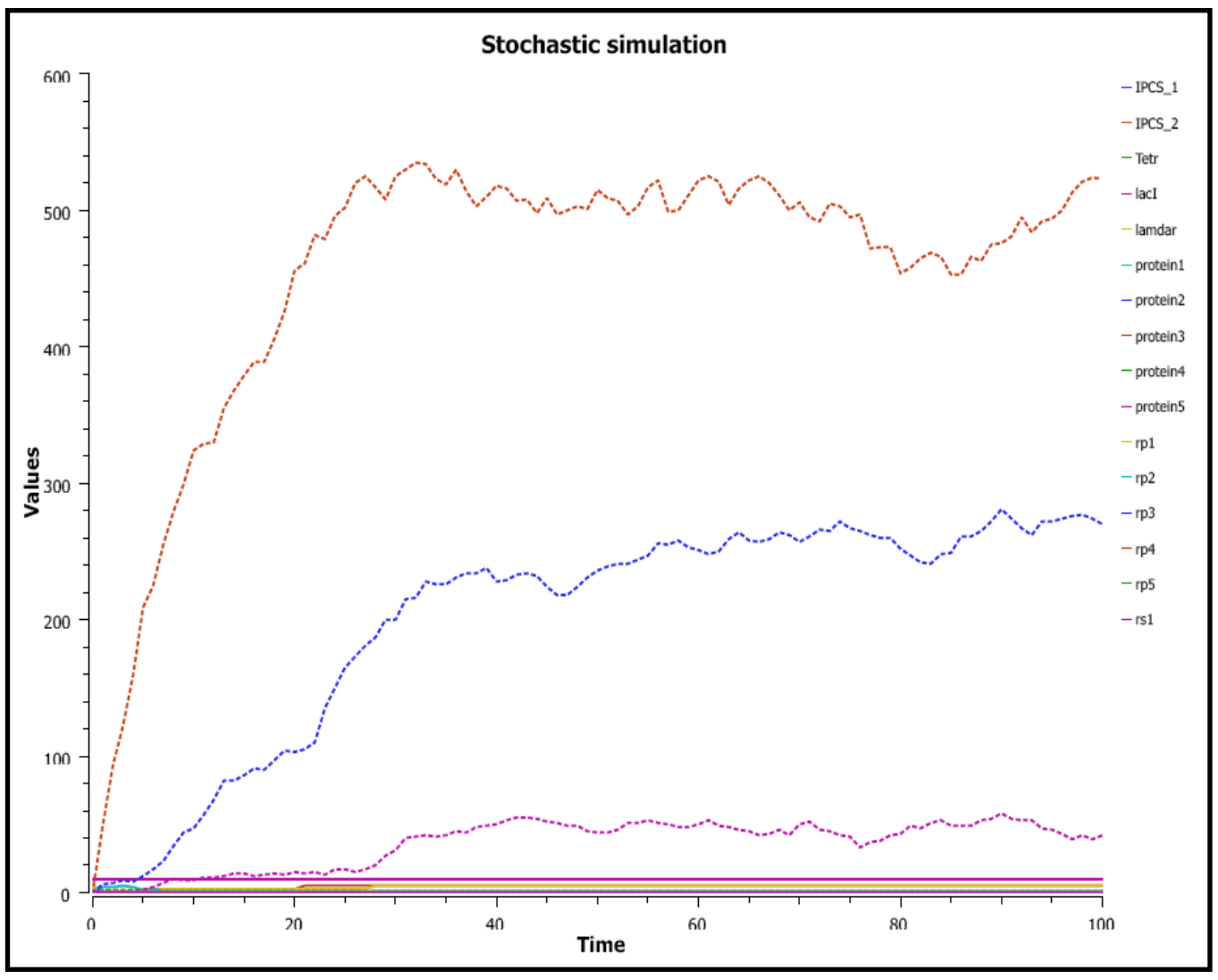

c.) Kd=1.03

d.) Kd=1.04

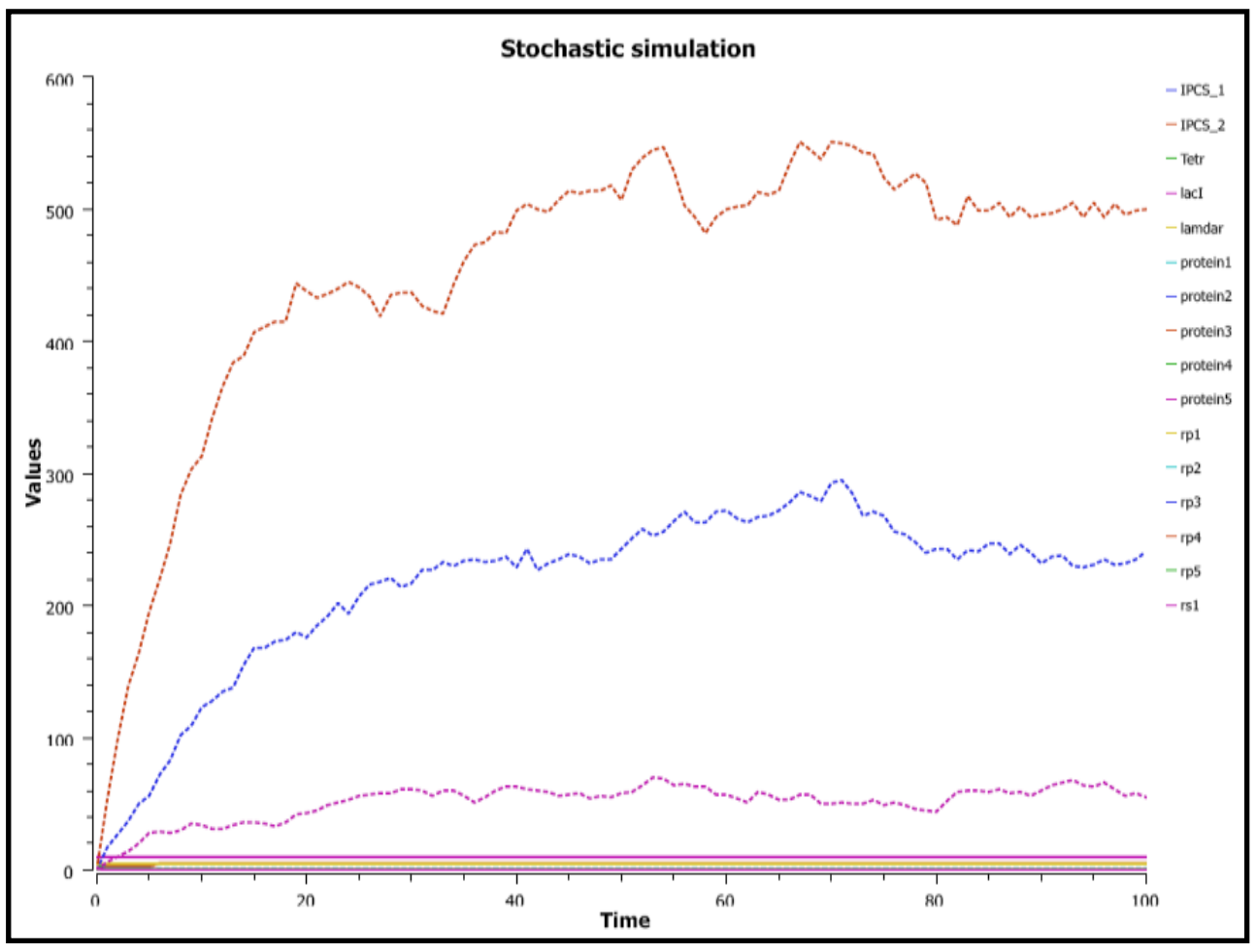

e.) Kd = 1.05

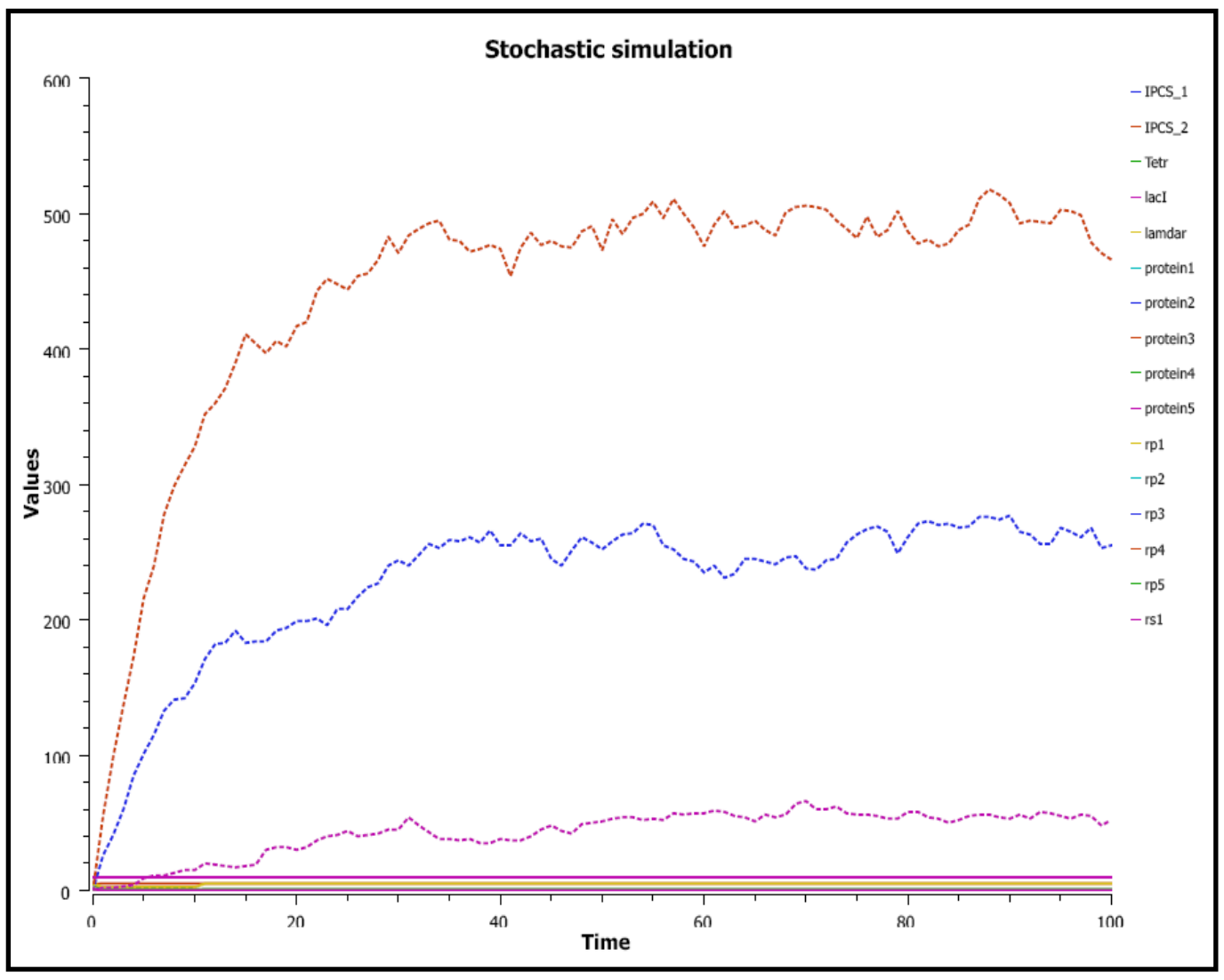

f.) Kd =1.06

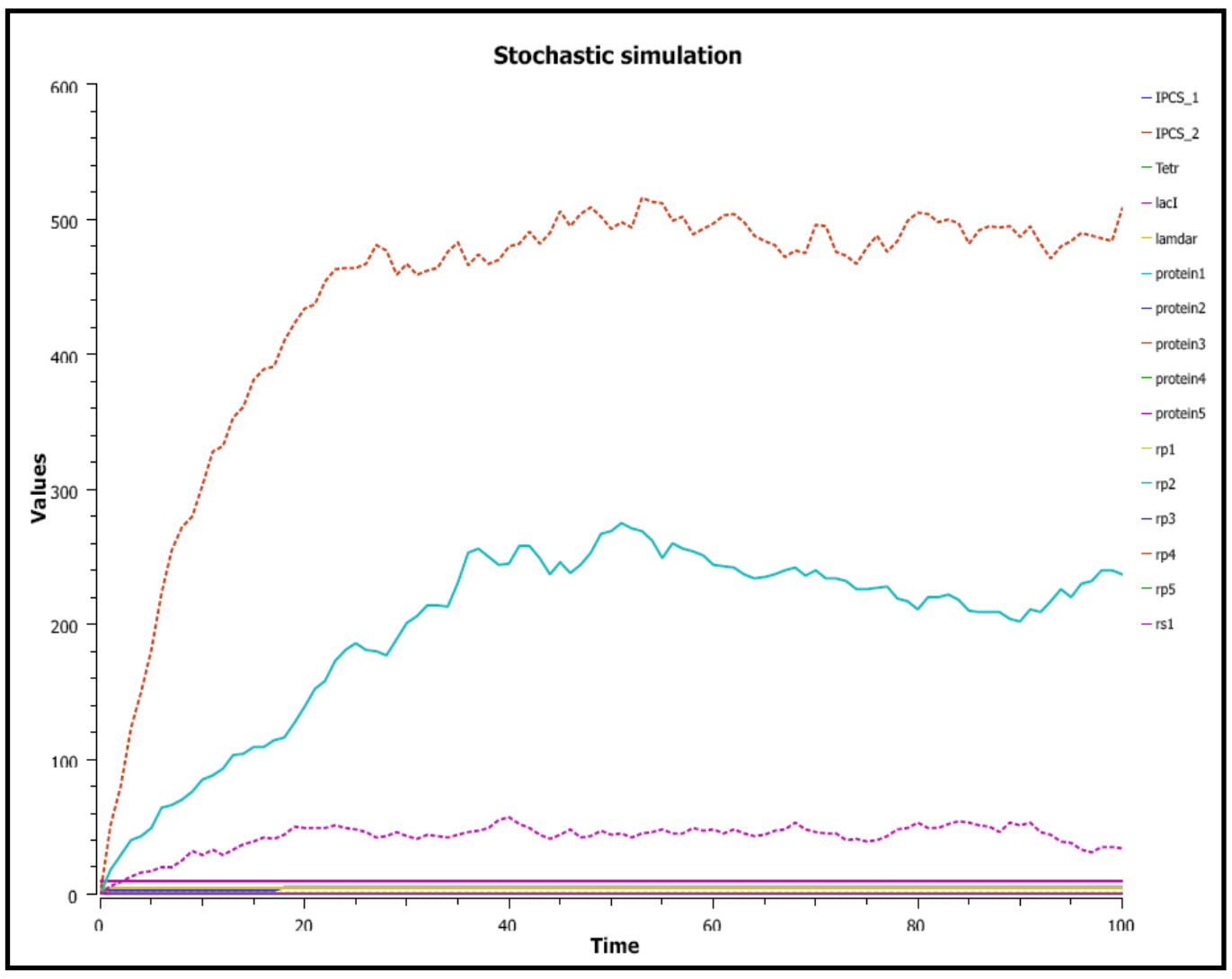

g.) Kd =1.07

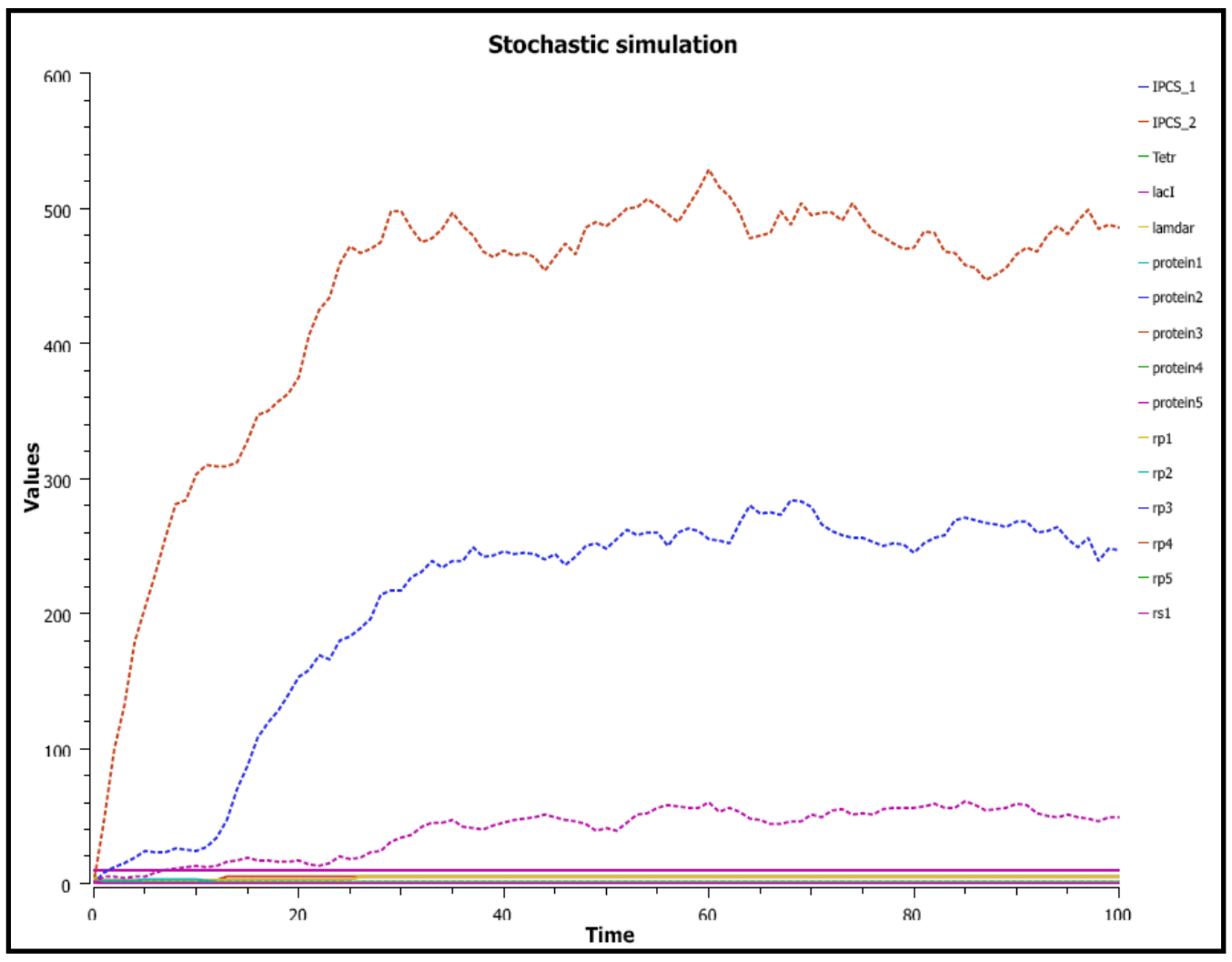

h.) Kd =1.08

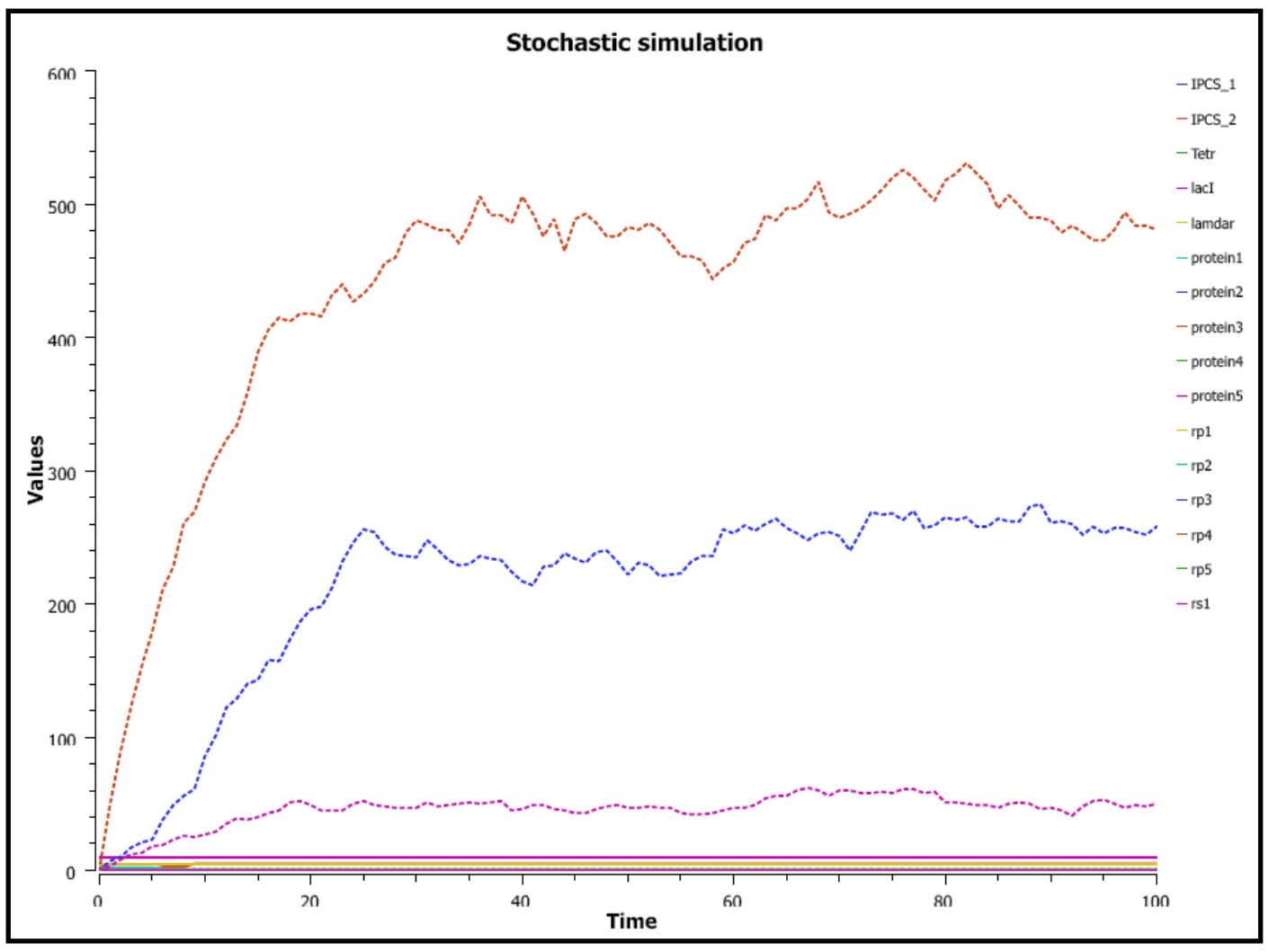

i.) Kd = 1.09

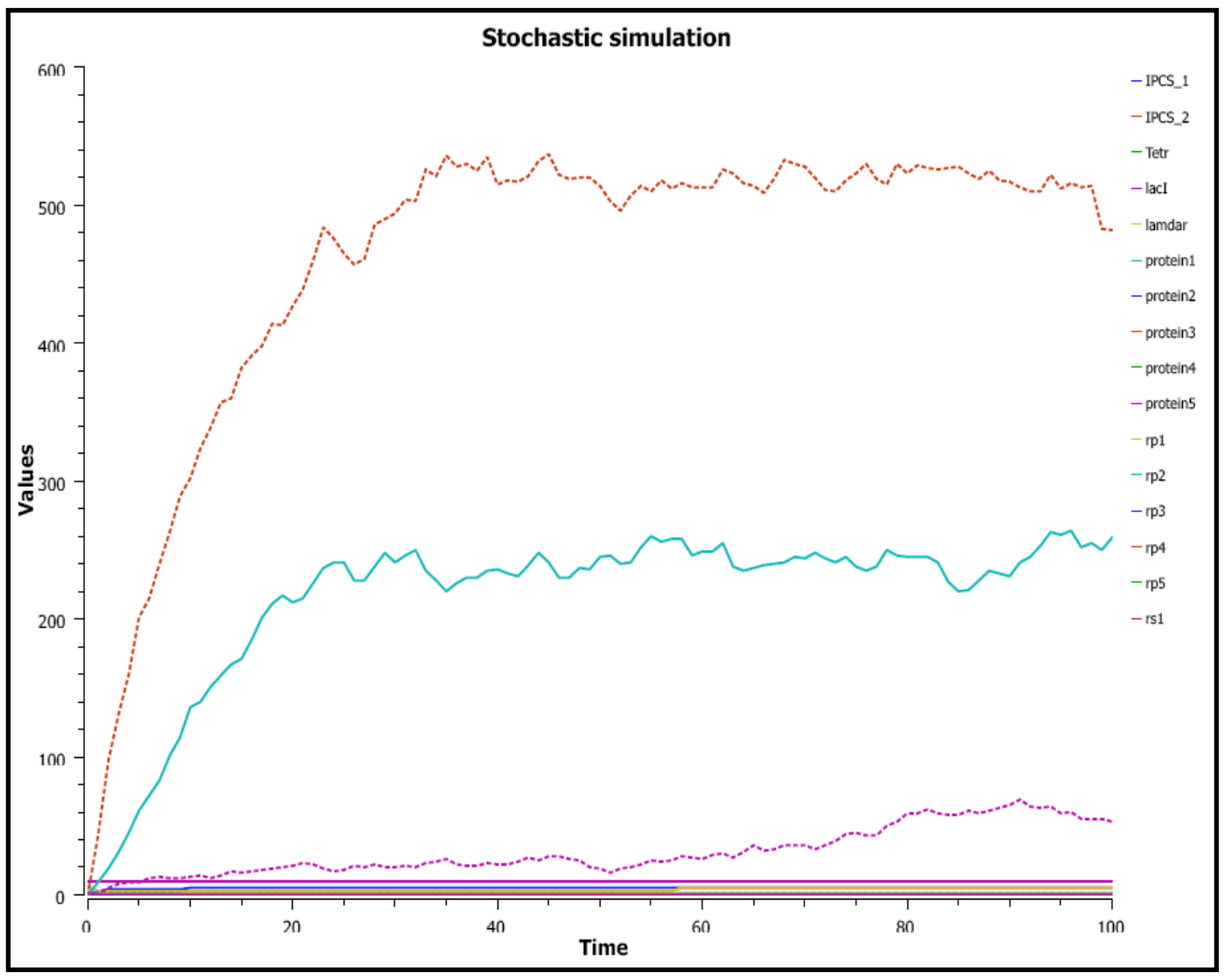

j.) Kd = 2

**Time series data**

```
# Time      IPCS1protein      IPCS2protein      IPCS_1      IPCS_2
      Tetr  Tetrepressor      lacI  lactoserepressor  lamdar
      lamdarepressor    rp1   rp2   rp3   rp4   rp5   rs1
      Compartments[DefaultCompartment]    Values[rp1_strength]
      Values[rp2_strength]    Values[rp3_strength]    Values[rp4_strength]
      Values[rp5_strength]    Values[IPCS1protein_degradation_rate]
      Values[tr1_Kd]    Values[tr1_h]     Values[tr2_Kd]    Values[tr2_h]
      Values[tr3_Kd]    Values[tr3_h]     Values[tr4_Kd]    Values[tr4_h]
      Values[tr5_Kd]    Values[tr5_h]     Values[tr6_Kd]    Values[tr6_h]
      Values[pp1_translation_rate]  Values[pp2_translation_rate]
      Values[pp3_translation_rate]  Values[Tetrepressor_degradation_rate]
      Values[pp4_translation_rate]  Values[pp5_translation_rate]
      Values[IPCS2protein_degradation_rate]
      Values[lamdarepressor_degradation_rate]
      Values[lactoserepressor_degradation_rate]
0     0.08  0.05  4.98753     4.9682      4.98753     0.05  10    0.05
      4.98753     0.05  0.997506    0.993641    0.997506    0.997506
      0.997506    0.997506    1     5     5     5     5     5     0.1   1
      2     1     2     1     2     1     2     1     2     1     2     1
      1     1     0.1   1     1     0.1   0.1   0.1
0.01  0.32167     0.289042    4.61448     4.53115     3.83974     0.0952181
      10    0.5497      4.95507     0.0996474   0.922897    0.906231
      0.990168    0.767949    0.991015    0.767949    1     5     5     5
      5     5     0.1   1     2     1     2     1     2     1     2     1
      2     1     2     1     1     1     0.1   1     1     0.1   0.1   0.1

0.02  0.536912    0.499087    4.00292     3.88116     2.38073     0.125819
      10    1.0489      4.92208     0.148903    0.800585    0.776232
      0.978309    0.476147    0.984416    0.476147    1     5     5     5
      5     5     0.1   1     2     1     2     1     2     1     2     1
      2     1     2     1     1     1     0.1   1     1     0.1   0.1   0.1

0.03  0.721631    0.67724     3.42782     3.28785     1.47272     0.144517
      10    1.5476      4.89771     0.197821    0.685563    0.65757
      0.962341    0.294544    0.979542    0.294544    1     5     5     5
      5     5     0.1   1     2     1     2     1     2     1     2     1
      2     1     2     1     1     1     0.1   1     1     0.1   0.1   0.1

0.04  0.880164    0.828611    2.96455     2.8174      0.96426     0.156311
      10    2.04581     4.88075     0.246486    0.59291     0.563479
      0.942724    0.192852    0.97615     0.192852    1     5     5     5
      5     5     0.1   1     2     1     2     1     2     1     2     1
      2     1     2     1     1     1     0.1   1     1     0.1   0.1   0.1

0.05  1.01805     0.959139    2.60424     2.45527     0.669393    0.164193
      10    2.54351     4.86874     0.29496     0.520848    0.491054
      0.919962    0.133879    0.973748    0.133879    1     5     5     5
      5     5     0.1   1     2     1     2     1     2     1     2     1
      2     1     2     1     1     1     0.1   1     1     0.1   0.1   0.1

0.06  1.13987     1.07358     2.3228      2.17457     0.487997    0.169744
      10    3.04072     4.85997     0.343282    0.46456     0.434915
      0.89458     0.0975993   0.971994    0.0975993   1     5     5     5
      5     5     0.1   1     2     1     2     1     2     1     2     1
      2     1     2     1     1     1     0.1   1     1     0.1   0.1   0.1

0.07  1.24902     1.17544     2.09938     1.95308     0.370003    0.173823
      10    3.53743     4.85336     0.39148     0.419877    0.390616
      0.86711     0.0740006   0.970672    0.0740006   1     5     5     5
```

```
     5      5      0.1   1      2      1      2      1      2      1      2      1
     2      1      2     1      1      1      0.1    1      1      0.1    0.1    0.1

0.08 1.34802      1.26726      1.91869      1.77483      0.289515     0.176921
     10     4.03364      4.84825      0.439571     0.383738     0.354967
     0.838067     0.0579029    0.969649     0.0579029    1      5      5      5
     5      5      0.1   1      2      1      2      1      2      1      2      1
     2      1      2     1      1      1      0.1    1      1      0.1    0.1    0.1

0.09 1.43873      1.35092      1.76992      1.6287       0.232395     0.179337
     10     4.52936      4.8442       0.487569     0.353984     0.325739
     0.807935     0.0464789    0.968841     0.0464789    1      5      5      5
     5      5      0.1   1      2      1      2      1      2      1      2      1
     2      1      2     1      1      1      0.1    1      1      0.1    0.1    0.1

0.1  1.52254      1.42783      1.64544      1.50687      0.190502     0.18126
     10     5.02459      4.84095      0.535483     0.329088     0.301374
     0.777157     0.0381004    0.96819      0.0381004    1      5      5      5
     5      5      0.1   1      2      1      2      1      2      1      2      1
     2      1      2     1      1      1      0.1    1      1      0.1    0.1    0.1

0.11 1.60054      1.49907      1.53979      1.40381      0.158918     0.182818
     10     5.51931      4.83829      0.583319     0.307958     0.280762
     0.746123     0.0317835    0.967658     0.0317835    1      5      5      5
     5      5      0.1   1      2      1      2      1      2      1      2      1
     2      1      2     1      1      1      0.1    1      1      0.1    0.1    0.1

0.12 1.67357      1.56546      1.44899      1.3155       0.134543     0.184097
     10     6.01355      4.8361       0.631083     0.289799     0.2631
     0.715171     0.0269087    0.967219     0.0269087    1      5      5      5
     5      5      0.1   1      2      1      2      1      2      1      2      1
     2      1      2     1      1      1      0.1    1      1      0.1    0.1    0.1

0.13 1.74229      1.62768      1.37011      1.23898      0.115354     0.185158
     10     6.50729      4.83426      0.67878      0.274021     0.247795
     0.684583     0.0230708    0.966853     0.0230708    1      5      5      5
     5      5      0.1   1      2      1      2      1      2      1      2      1
     2      1      2     1      1      1      0.1    1      1      0.1    0.1    0.1

0.14 1.80726      1.68626      1.3009       1.17201      0.0999851    0.186047
     10     7.00053      4.83272      0.726412     0.26018      0.234402
     0.65459      0.019997     0.966545     0.019997     1      5      5      5
     5      5      0.1   1      2      1      2      1      2      1      2      1
     2      1      2     1      1      1      0.1    1      1      0.1    0.1    0.1

0.15 1.8689       1.74164      1.23967      1.11289      0.0874901    0.186795
     10     7.49329      4.83142      0.773982     0.247935     0.222578
     0.625372     0.017498     0.966284     0.017498     1      5      5      5
     5      5      0.1   1      2      1      2      1      2      1      2      1
     2      1      2     1      1      1      0.1    1      1      0.1    0.1    0.1

0.16 1.9276       1.79418      1.18509      1.0603       0.0771975    0.18743
     10     7.98555      4.83031      0.821493     0.237018     0.212061
     0.597068     0.0154395    0.966062     0.0154395    1      5      5      5
     5      5      0.1   1      2      1      2      1      2      1      2      1
     2      1      2     1      1      1      0.1    1      1      0.1    0.1    0.1

0.17 1.98365      1.84418      1.13611      1.0132       0.0686202    0.18797
     10     8.47731      4.82937      0.868946     0.227222     0.20264
     0.569779     0.013724     0.965873     0.013724     1      5      5      5
     5      5      0.1   1      2      1      2      1      2      1      2      1
```

```
     2     1     2     1     1     1     0.1   1     1     0.1   0.1   0.1

0.18 2.03732     1.89189     1.09188     0.970746    0.0613982   0.188431
     10    8.96859     4.82856     0.916343    0.218377    0.194149
     0.543571    0.0122796   0.965711    0.0122796   1     5     5     5
     5     5     0.1   1     2     1     2     1     2     1     2     1
     2     1     2     1     1     1     0.1   1     1     0.1   0.1   0.1

0.19 2.08883     1.93753     1.05174     0.932275    0.0552609   0.188825
     10    9.45938     4.82786     0.963685    0.210347    0.186455
     0.518487    0.0110522   0.965573    0.0110522   1     5     5     5
     5     5     0.1   1     2     1     2     1     2     1     2     1
     2     1     2     1     1     1     0.1   1     1     0.1   0.1   0.1

0.2  2.13838     1.9813      1.01512     0.897238    0.050002    0.189161
     10    9.94967     4.82727     1.01097     0.203023    0.179448
     0.494543    0.0100004   0.965454    0.0100004   1     5     5     5
     5     5     0.1   1     2     1     2     1     2     1     2     1
     2     1     2     1     1     1     0.1   1     1     0.1   0.1   0.1

0.21 2.18612     2.02335     0.981561    0.865184    0.0454617   0.189449
     10    10.4395     4.82676     1.05821     0.196312    0.173037
     0.471741    0.00909234  0.965353    0.00909234  1     5     5     5
     5     5     0.1   1     2     1     2     1     2     1     2     1
     2     1     2     1     1     1     0.1   1     1     0.1   0.1   0.1

0.22 2.23221     2.06381     0.950691    0.835737    0.0415149   0.189694
     10    10.9288     4.82633     1.10539     0.190138    0.167147
     0.450067    0.00830299  0.965266    0.00830299  1     5     5     5
     5     5     0.1   1     2     1     2     1     2     1     2     1
     2     1     2     1     1     1     0.1   1     1     0.1   0.1   0.1

0.23 2.27676     2.10283     0.922187    0.808584    0.0380627   0.189901
     10    11.4176     4.82596     1.15252     0.184437    0.161717
     0.429495    0.00761255  0.965193    0.00761255  1     5     5     5
     5     5     0.1   1     2     1     2     1     2     1     2     1
     2     1     2     1     1     1     0.1   1     1     0.1   0.1   0.1

0.24 2.31991     2.1405      0.895777    0.783458    0.0350258   0.190077
     10    11.906      4.82565     1.19961     0.179155    0.156692
     0.409995    0.00700516  0.965131    0.00700516  1     5     5     5
     5     5     0.1   1     2     1     2     1     2     1     2     1
     2     1     2     1     1     1     0.1   1     1     0.1   0.1   0.1

0.25 2.36173     2.17693     0.871231    0.760135    0.0323402   0.190223
     10    12.3938     4.82539     1.24664     0.174246    0.152027
     0.391526    0.00646803  0.965079    0.00646803  1     5     5     5
     5     5     0.1   1     2     1     2     1     2     1     2     1
     2     1     2     1     1     1     0.1   1     1     0.1   0.1   0.1

0.26 2.40233     2.21219     0.848353    0.738421    0.0299536   0.190344
     10    12.8812     4.82518     1.29362     0.169671    0.147684
     0.374048    0.00599073  0.965036    0.00599073  1     5     5     5
     5     5     0.1   1     2     1     2     1     2     1     2     1
     2     1     2     1     1     1     0.1   1     1     0.1   0.1   0.1

0.27 2.44179     2.24637     0.82697     0.71815     0.0278234   0.190442
     10    13.368      4.82501     1.34056     0.165394    0.14363
     0.357515    0.00556469  0.965001    0.00556469  1     5     5     5
     5     5     0.1   1     2     1     2     1     2     1     2     1
```

```
     2     1     2     1     1     1     0.1   1     1     0.1   0.1   0.1

0.28 2.48017     2.27953     0.806937    0.699179    0.0259141   0.19052
     10    13.8544     4.82487     1.38744     0.161387    0.139836
     0.341881    0.00518281  0.964973    0.00518281  1     5     5     5
     5     5     0.1   1     2     1     2     1     2     1     2     1
     2     1     2     1     1     1     0.1   1     1     0.1   0.1   0.1

0.29 2.51754     2.31175     0.788125    0.681383    0.024196    0.19058
     10    14.3403     4.82476     1.43428     0.157625    0.136277
     0.327102    0.00483921  0.964952    0.00483921  1     5     5     5
     5     5     0.1   1     2     1     2     1     2     1     2     1
     2     1     2     1     1     1     0.1   1     1     0.1   0.1   0.1

0.3  2.55397     2.34306     0.770422    0.664651    0.0226446   0.190624
     10    14.8258     4.82468     1.48107     0.154084    0.13293
     0.31313     0.00452892  0.964937    0.00452892  1     5     5     5
     5     5     0.1   1     2     1     2     1     2     1     2     1
     2     1     2     1     1     1     0.1   1     1     0.1   0.1   0.1

0.31 2.58949     2.37354     0.753728    0.648889    0.0212389   0.190652
     10    15.3107     4.82463     1.52781     0.150746    0.129778
     0.299922    0.00424778  0.964927    0.00424778  1     5     5     5
     5     5     0.1   1     2     1     2     1     2     1     2     1
     2     1     2     1     1     1     0.1   1     1     0.1   0.1   0.1

0.32 2.62417     2.40322     0.737956    0.634011    0.0199612   0.190667
     10    15.7951     4.82461     1.5745      0.147591    0.126802
     0.287434    0.00399223  0.964921    0.00399223  1     5     5     5
     5     5     0.1   1     2     1     2     1     2     1     2     1
     2     1     2     1     1     1     0.1   1     1     0.1   0.1   0.1

0.33 2.65805     2.43215     0.723029    0.619943    0.0187964   0.19067
     10    16.2791     4.8246      1.62115     0.144606    0.123989
     0.275624    0.00375927  0.96492     0.00375927  1     5     5     5
     5     5     0.1   1     2     1     2     1     2     1     2     1
     2     1     2     1     1     1     0.1   1     1     0.1   0.1   0.1

0.34 2.69117     2.46036     0.708879    0.606618    0.0177315   0.190662
     10    16.7626     4.82461     1.66775     0.141776    0.121324
     0.264452    0.0035463   0.964923    0.0035463   1     5     5     5
     5     5     0.1   1     2     1     2     1     2     1     2     1
     2     1     2     1     1     1     0.1   1     1     0.1   0.1   0.1

0.35 2.72357     2.4879      0.695444    0.593976    0.0167555   0.190644
     10    17.2456     4.82465     1.71431     0.139089    0.118795
     0.253881    0.0033511   0.964929    0.0033511   1     5     5     5
     5     5     0.1   1     2     1     2     1     2     1     2     1
     2     1     2     1     1     1     0.1   1     1     0.1   0.1   0.1

0.36 2.75528     2.5148      0.682668    0.581965    0.0158587   0.190616
     10    17.7281     4.8247      1.76082     0.136534    0.116393
     0.243874    0.00317173  0.964939    0.00317173  1     5     5     5
     5     5     0.1   1     2     1     2     1     2     1     2     1
     2     1     2     1     1     1     0.1   1     1     0.1   0.1   0.1

0.37 2.78634     2.54108     0.670504    0.570536    0.0150327   0.19058
     10    18.2101     4.82476     1.80728     0.134101    0.114107
     0.234397    0.00300654  0.964952    0.00300654  1     5     5     5
     5     5     0.1   1     2     1     2     1     2     1     2     1
```

```
     2     1     2     1     1     1     0.1   1     1     0.1   0.1   0.1

0.38 2.81677     2.56678     0.658905    0.559648    0.0142703   0.190536
     10    18.6917     4.82484     1.8537      0.131781    0.11193
     0.225418    0.00285406  0.964968    0.00285406  1     5     5     5
     5     5     0.1   1     2     1     2     1     2     1     2     1
     2     1     2     1     1     1     0.1   1     1     0.1   0.1   0.1

0.39 2.8466      2.59192     0.647833    0.54926     0.0135651   0.190485
     10    19.1727     4.82493     1.90007     0.129567    0.109852
     0.216907    0.00271301  0.964986    0.00271301  1     5     5     5
     5     5     0.1   1     2     1     2     1     2     1     2     1
     2     1     2     1     1     1     0.1   1     1     0.1   0.1   0.1

0.4  2.87587     2.61653     0.63725     0.539338    0.0129115   0.190427
     10    19.6533     4.82503     1.9464      0.12745     0.107868
     0.208835    0.00258229  0.965007    0.00258229  1     5     5     5
     5     5     0.1   1     2     1     2     1     2     1     2     1
     2     1     2     1     1     1     0.1   1     1     0.1   0.1   0.1

0.41 2.90459     2.64063     0.627124    0.529851    0.0123045   0.190362
     10    20.1334     4.82515     1.99268     0.125425    0.10597
     0.201176    0.0024609   0.965029    0.0024609   1     5     5     5
     5     5     0.1   1     2     1     2     1     2     1     2     1
     2     1     2     1     1     1     0.1   1     1     0.1   0.1   0.1

0.42 2.93278     2.66424     0.617424    0.520769    0.0117399   0.190292
     10    20.6131     4.82527     2.03892     0.123485    0.104154
     0.193905    0.00234798  0.965054    0.00234798  1     5     5     5
     5     5     0.1   1     2     1     2     1     2     1     2     1
     2     1     2     1     1     1     0.1   1     1     0.1   0.1   0.1

0.43 2.96047     2.68738     0.608123    0.512065    0.0112138   0.190217
     10    21.0922     4.82541     2.08511     0.121625    0.102413
     0.186997    0.00224275  0.965081    0.00224275  1     5     5     5
     5     5     0.1   1     2     1     2     1     2     1     2     1
     2     1     2     1     1     1     0.1   1     1     0.1   0.1   0.1

0.44 2.98768     2.71008     0.599197    0.503717    0.0107227   0.190136
     10    21.5709     4.82555     2.13125     0.119839    0.100743
     0.180432    0.00214453  0.96511     0.00214453  1     5     5     5
     5     5     0.1   1     2     1     2     1     2     1     2     1
     2     1     2     1     1     1     0.1   1     1     0.1   0.1   0.1

0.45 3.01442     2.73234     0.590621    0.495701    0.0102635   0.190051
     10    22.0491     4.8257      2.17736     0.118124    0.0991402
     0.174189    0.00205271  0.96514     0.00205271  1     5     5     5
     5     5     0.1   1     2     1     2     1     2     1     2     1
     2     1     2     1     1     1     0.1   1     1     0.1   0.1   0.1

0.46 3.04072     2.75419     0.582375    0.487998    0.0098337   0.189961
     10    22.5268     4.82586     2.22341     0.116475    0.0975995
     0.168249    0.00196674  0.965171    0.00196674  1     5     5     5
     5     5     0.1   1     2     1     2     1     2     1     2     1
     2     1     2     1     1     1     0.1   1     1     0.1   0.1   0.1

0.47 3.06658     2.77563     0.574439    0.480588    0.00943068  0.189868
     10    23.004      4.82602     2.26943     0.114888    0.0961177
     0.162594    0.00188614  0.965205    0.00188614  1     5     5     5
     5     5     0.1   1     2     1     2     1     2     1     2     1
```

```
     2     1     2     1     1     1     0.1   1     1     0.1   0.1   0.1

0.48 3.09203     2.7967      0.566796    0.473456    0.00905229  0.18977
     10    23.4808     4.8262      2.31539     0.113359    0.0946911
     0.157207    0.00181046  0.965239    0.00181046  1     5     5     5
     5     5     0.1   1     2     1     2     1     2     1     2     1
     2     1     2     1     1     1     0.1   1     1     0.1   0.1   0.1

0.49 3.11708     2.81739     0.559429    0.466584    0.00869656  0.189669
     10    23.957      4.82637     2.36132     0.111886    0.0933167
     0.152072    0.00173931  0.965275    0.00173931  1     5     5     5
     5     5     0.1   1     2     1     2     1     2     1     2     1
     2     1     2     1     1     1     0.1   1     1     0.1   0.1   0.1

0.5  3.14174     2.83773     0.552322    0.459958    0.00836171  0.189565
     10    24.4328     4.82656     2.4072      0.110464    0.0919916
     0.147175    0.00167234  0.965312    0.00167234  1     5     5     5
     5     5     0.1   1     2     1     2     1     2     1     2     1
     2     1     2     1     1     1     0.1   1     1     0.1   0.1   0.1

0.51 3.16603     2.85771     0.545463    0.453565    0.00804612  0.189457
     10    24.9082     4.82675     2.45304     0.109093    0.0907131
     0.142503    0.00160922  0.96535     0.00160922  1     5     5     5
     5     5     0.1   1     2     1     2     1     2     1     2     1
     2     1     2     1     1     1     0.1   1     1     0.1   0.1   0.1

0.52 3.18996     2.87737     0.538836    0.447393    0.00774835  0.189347
     10    25.383      4.82694     2.49883     0.107767    0.0894786
     0.138043    0.00154967  0.965389    0.00154967  1     5     5     5
     5     5     0.1   1     2     1     2     1     2     1     2     1
     2     1     2     1     1     1     0.1   1     1     0.1   0.1   0.1

0.53 3.21354     2.8967      0.532431    0.441429    0.00746708  0.189234
     10    25.8574     4.82714     2.54458     0.106486    0.0882858
     0.133782    0.00149342  0.965429    0.00149342  1     5     5     5
     5     5     0.1   1     2     1     2     1     2     1     2     1
     2     1     2     1     1     1     0.1   1     1     0.1   0.1   0.1

0.54 3.23678     2.91572     0.526236    0.435663    0.0072011   0.189118
     10    26.3313     4.82735     2.59028     0.105247    0.0871327
     0.129709    0.00144022  0.965469    0.00144022  1     5     5     5
     5     5     0.1   1     2     1     2     1     2     1     2     1
     2     1     2     1     1     1     0.1   1     1     0.1   0.1   0.1

0.55 3.25969     2.93444     0.52024     0.430085    0.00694932  0.189 10
     26.8048     4.82756     2.63594     0.104048    0.086017    0.125815
     0.00138986  0.965511    0.00138986  1     5     5     5     5     5
     0.1   1     2     1     2     1     2     1     2     1     2     1
     2     1     1     1     0.1   1     1     0.1   0.1   0.1
0.56 3.28229     2.95287     0.514434    0.424685    0.00671075  0.188879
     10    27.2777     4.82777     2.68156     0.102887    0.0849371
     0.122089    0.00134215  0.965554    0.00134215  1     5     5     5
     5     5     0.1   1     2     1     2     1     2     1     2     1
     2     1     2     1     1     1     0.1   1     1     0.1   0.1   0.1

0.57 3.30458     2.97101     0.508809    0.419455    0.00648447  0.188756
     10    27.7502     4.82798     2.72714     0.101762    0.0838911
     0.118522    0.00129689  0.965597    0.00129689  1     5     5     5
     5     5     0.1   1     2     1     2     1     2     1     2     1
     2     1     2     1     1     1     0.1   1     1     0.1   0.1   0.1
```

```
0.58  3.32656     2.98887     0.503355    0.414387    0.00626965  0.188631
      10    28.2222     4.8282      2.77267     0.100671    0.0828773
      0.115106    0.00125393  0.965641    0.00125393  1     5     5     5
      5     5     0.1   1     2     1     2     1     2     1     2     1
      2     1     2     1     1     1     0.1   1     1     0.1   0.1   0.1

0.59  3.34826     3.00647     0.498064    0.409472    0.00606552  0.188504
      10    28.6938     4.82843     2.81815     0.0996129   0.0818944
      0.111832    0.0012131   0.965685    0.0012131   1     5     5     5
      5     5     0.1   1     2     1     2     1     2     1     2     1
      2     1     2     1     1     1     0.1   1     1     0.1   0.1   0.1

0.6   3.36968     3.02381     0.492931    0.404704    0.00587139  0.188376
      10    29.1648     4.82865     2.8636      0.0985861   0.0809407
      0.108693    0.00117428  0.965731    0.00117428  1     5     5     5
      5     5     0.1   1     2     1     2     1     2     1     2     1
      2     1     2     1     1     1     0.1   1     1     0.1   0.1   0.1

0.61  3.39082     3.04089     0.487946    0.400075    0.00568661  0.188245
      10    29.6354     4.82888     2.909 0.0975891   0.0800151   0.105683
      0.00113732  0.965777    0.00113732  1     5     5     5     5     5
      0.1   1     2     1     2     1     2     1     2     1     2     1
      2     1     1     1     0.1   1     1     0.1   0.1   0.1
0.62  3.41169     3.05774     0.483103    0.395581    0.00551059  0.188113
      10    30.1056     4.82911     2.95436     0.0966207   0.0791162
      0.102794    0.00110212  0.965823    0.00110212  1     5     5     5
      5     5     0.1   1     2     1     2     1     2     1     2     1
      2     1     2     1     1     1     0.1   1     1     0.1   0.1   0.1

0.63  3.43231     3.07434     0.478397    0.391214    0.00534277  0.187979
      10    30.5752     4.82935     2.99967     0.0956795   0.0782428
      0.10002     0.00106855  0.96587     0.00106855  1     5     5     5
      5     5     0.1   1     2     1     2     1     2     1     2     1
      2     1     2     1     1     1     0.1   1     1     0.1   0.1   0.1

0.64  3.45267     3.09071     0.473822    0.386969    0.00518266  0.187844
      10    31.0444     4.82959     3.04495     0.0947643   0.0773939
      0.0973549   0.00103653  0.965917    0.00103653  1     5     5     5
      5     5     0.1   1     2     1     2     1     2     1     2     1
      2     1     2     1     1     1     0.1   1     1     0.1   0.1   0.1

0.65  3.47279     3.10686     0.469371    0.382841    0.00502979  0.187707
      10    31.5131     4.82983     3.09018     0.0938741   0.0765683
      0.094794    0.00100596  0.965965    0.00100596  1     5     5     5
      5     5     0.1   1     2     1     2     1     2     1     2     1
      2     1     2     1     1     1     0.1   1     1     0.1   0.1   0.1

0.66  3.49266     3.12278     0.465039    0.378825    0.00488373  0.187569
      10    31.9814     4.83007     3.13536     0.0930079   0.0757651
      0.0923319   0.000976745 0.966014    0.000976745 1     5     5     5
      5     5     0.1   1     2     1     2     1     2     1     2     1
      2     1     2     1     1     1     0.1   1     1     0.1   0.1   0.1

0.67  3.51231     3.1385      0.460823    0.374917    0.00474407  0.18743
      10    32.4492     4.83031     3.18051     0.0921646   0.0749834
      0.0899635   0.000948813 0.966062    0.000948813 1     5     5     5
      5     5     0.1   1     2     1     2     1     2     1     2     1
      2     1     2     1     1     1     0.1   1     1     0.1   0.1   0.1

0.68  3.53172     3.154 0.456716    0.371111    0.00461044  0.187289    10
      32.9165     4.83056     3.22561     0.0913433   0.0742222   0.0876843
```

```
     0.000922089 0.966112    0.000922089 1     5     5     5     5     5
     0.1   1     2     1     2     1     2     1     2     1     2     1
     2     1     1     1     0.1   1     1     0.1   0.1   0.1
0.69 3.55092     3.1693      0.452715    0.367404    0.00448251  0.187147
     10    33.3833     4.83081     3.27067     0.0905431   0.0734808
     0.0854901   0.000896502 0.966161    0.000896502 1     5     5     5
     5     5     0.1   1     2     1     2     1     2     1     2     1
     2     1     2     1     1     1     0.1   1     1     0.1   0.1   0.1

0.7  3.56989     3.1844      0.448816    0.363792    0.00435995  0.187004
     10    33.8497     4.83106     3.31568     0.0897632   0.0727583
     0.0833767   0.000871989 0.966211    0.000871989 1     5     5     5
     5     5     0.1   1     2     1     2     1     2     1     2     1
     2     1     2     1     1     1     0.1   1     1     0.1   0.1   0.1

0.71 3.58866     3.19931     0.445014    0.36027     0.00424246  0.18686
     10    34.3156     4.83131     3.36066     0.0890028   0.072054
     0.0813404   0.000848491 0.966261    0.000848491 1     5     5     5
     5     5     0.1   1     2     1     2     1     2     1     2     1
     2     1     2     1     1     1     0.1   1     1     0.1   0.1   0.1

0.72 3.60722     3.21403     0.441306    0.356836    0.00412976  0.186715
     10    34.7811     4.83156     3.40559     0.0882612   0.0713673
     0.0793775   0.000825952 0.966312    0.000825952 1     5     5     5
     5     5     0.1   1     2     1     2     1     2     1     2     1
     2     1     2     1     1     1     0.1   1     1     0.1   0.1   0.1

0.73 3.62558     3.22857     0.437688    0.353487    0.0040216   0.18657
     10    35.2461     4.83181     3.45048     0.0875376   0.0706973
     0.0774846   0.000804321 0.966363    0.000804321 1     5     5     5
     5     5     0.1   1     2     1     2     1     2     1     2     1
     2     1     2     1     1     1     0.1   1     1     0.1   0.1   0.1

0.74 3.64374     3.24293     0.434157    0.350218    0.00391774  0.186423
     10    35.7106     4.83207     3.49532     0.0868314   0.0700436
     0.0756585   0.000783548 0.966414    0.000783548 1     5     5     5
     5     5     0.1   1     2     1     2     1     2     1     2     1
     2     1     2     1     1     1     0.1   1     1     0.1   0.1   0.1

0.75 3.66171     3.25711     0.43071     0.347027    0.00381795  0.186275
     10    36.1746     4.83233     3.54013     0.086142    0.0694054
     0.0738962   0.000763589 0.966465    0.000763589 1     5     5     5
     5     5     0.1   1     2     1     2     1     2     1     2     1
     2     1     2     1     1     1     0.1   1     1     0.1   0.1   0.1

0.76 3.67949     3.27112     0.427343    0.343912    0.00372201  0.186127
     10    36.6382     4.83258     3.58489     0.0854687   0.0687823
     0.0721947   0.000744402 0.966517    0.000744402 1     5     5     5
     5     5     0.1   1     2     1     2     1     2     1     2     1
     2     1     2     1     1     1     0.1   1     1     0.1   0.1   0.1

0.77 3.69708     3.28496     0.424054    0.340868    0.00362973  0.185977
     10    37.1014     4.83284     3.62961     0.0848109   0.0681737
     0.0705514   0.000725947 0.966569    0.000725947 1     5     5     5
     5     5     0.1   1     2     1     2     1     2     1     2     1
     2     1     2     1     1     1     0.1   1     1     0.1   0.1   0.1

0.78 3.7145      3.29863     0.420841    0.337895    0.00354093  0.185827
     10    37.564      4.8331      3.67429     0.0841681   0.067579
     0.0689637   0.000708187 0.966621    0.000708187 1     5     5     5
     5     5     0.1   1     2     1     2     1     2     1     2     1
```

```
     2     1     2     1     1     1     0.1   1     1     0.1   0.1   0.1

0.79 3.73174     3.31215     0.417699    0.334989    0.00345544  0.185676
     10    38.0262     4.83337     3.71892     0.0835399   0.0669977
     0.0674291   0.000691087 0.966673    0.000691087 1     5     5     5
     5     5     0.1   1     2     1     2     1     2     1     2     1
     2     1     2     1     1     1     0.1   1     1     0.1   0.1   0.1

0.8  3.74881     3.32551     0.414628    0.332148    0.00337308  0.185525
     10    38.488      4.83363     3.76352     0.0829256   0.0664295
     0.0659454   0.000674616 0.966726    0.000674616 1     5     5     5
     5     5     0.1   1     2     1     2     1     2     1     2     1
     2     1     2     1     1     1     0.1   1     1     0.1   0.1   0.1

0.81 3.76571     3.33871     0.411624    0.329369    0.00329371  0.185373
     10    38.9493     4.83389     3.80807     0.0823248   0.0658738
     0.0645104   0.000658742 0.966778    0.000658742 1     5     5     5
     5     5     0.1   1     2     1     2     1     2     1     2     1
     2     1     2     1     1     1     0.1   1     1     0.1   0.1   0.1

0.82 3.78244     3.35177     0.408685    0.326651    0.00321718  0.18522
     10    39.4101     4.83416     3.85258     0.081737    0.0653303
     0.0631219   0.000643437 0.966831    0.000643437 1     5     5     5
     5     5     0.1   1     2     1     2     1     2     1     2     1
     2     1     2     1     1     1     0.1   1     1     0.1   0.1   0.1

0.83 3.79901     3.36468     0.40581     0.323992    0.00314336  0.185067
     10    39.8704     4.83442     3.89705     0.0811619   0.0647984
     0.0617781   0.000628673 0.966884    0.000628673 1     5     5     5
     5     5     0.1   1     2     1     2     1     2     1     2     1
     2     1     2     1     1     1     0.1   1     1     0.1   0.1   0.1

0.84 3.81542     3.37744     0.402995    0.321389    0.00307213  0.184913
     10    40.3303     4.83469     3.94147     0.080599    0.0642779
     0.060477    0.000614426 0.966938    0.000614426 1     5     5     5
     5     5     0.1   1     2     1     2     1     2     1     2     1
     2     1     2     1     1     1     0.1   1     1     0.1   0.1   0.1

0.85 3.83168     3.39006     0.40024     0.318842    0.00300335  0.184758
     10    40.7898     4.83496     3.98586     0.0800479   0.0637683
     0.059217    0.00060067  0.966991    0.00060067  1     5     5     5
     5     5     0.1   1     2     1     2     1     2     1     2     1
     2     1     2     1     1     1     0.1   1     1     0.1   0.1   0.1

0.86 3.84778     3.40254     0.397541    0.316347    0.00293693  0.184603
     10    41.2488     4.83522     4.0302      0.0795082   0.0632693
     0.0579962   0.000587385 0.967045    0.000587385 1     5     5     5
     5     5     0.1   1     2     1     2     1     2     1     2     1
     2     1     2     1     1     1     0.1   1     1     0.1   0.1   0.1

0.87 3.86374     3.41489     0.394898    0.313903    0.00287274  0.184448
     10    41.7073     4.83549     4.0745      0.0789797   0.0627806
     0.0568132   0.000574548 0.967098    0.000574548 1     5     5     5
     5     5     0.1   1     2     1     2     1     2     1     2     1
     2     1     2     1     1     1     0.1   1     1     0.1   0.1   0.1

0.88 3.87955     3.42711     0.392309    0.311509    0.0028107   0.184292
     10    42.1653     4.83576     4.11876     0.0784618   0.0623019
     0.0556663   0.00056214  0.967152    0.00056214  1     5     5     5
     5     5     0.1   1     2     1     2     1     2     1     2     1
```

```
     2     1     2     1     1     1     0.1   1     1     0.1   0.1   0.1

0.89 3.89521     3.43919     0.389772    0.309163    0.0027507   0.184136
     10    42.623      4.83603     4.16298     0.0779543   0.0618327
     0.0545542   0.000550141 0.967206    0.000550141 1     5     5     5
     5     5     0.1   1     2     1     2     1     2     1     2     1
     2     1     2     1     1     1     0.1   1     1     0.1   0.1   0.1

0.9  3.91073     3.45115     0.387285    0.306864    0.00269267  0.183979
     10    43.0801     4.8363      4.20716     0.0774569   0.0613729
     0.0534755   0.000538533 0.96726     0.000538533 1     5     5     5
     5     5     0.1   1     2     1     2     1     2     1     2     1
     2     1     2     1     1     1     0.1   1     1     0.1   0.1   0.1

0.91 3.92612     3.46297     0.384846    0.30461     0.0026365   0.183821
     10    43.5368     4.83657     4.25129     0.0769693   0.0609221
     0.0524288   0.0005273   0.967314    0.0005273   1     5     5     5
     5     5     0.1   1     2     1     2     1     2     1     2     1
     2     1     2     1     1     1     0.1   1     1     0.1   0.1   0.1

0.92 3.94137     3.47468     0.382456    0.3024      0.00258213  0.183664
     10    43.993      4.83684     4.29538     0.0764912   0.06048
     0.051413    0.000516426 0.967368    0.000516426 1     5     5     5
     5     5     0.1   1     2     1     2     1     2     1     2     1
     2     1     2     1     1     1     0.1   1     1     0.1   0.1   0.1

0.93 3.95648     3.48627     0.380111    0.300233    0.00252947  0.183506
     10    44.4488     4.83711     4.33944     0.0760222   0.0600465
     0.0504268   0.000505895 0.967423    0.000505895 1     5     5     5
     5     5     0.1   1     2     1     2     1     2     1     2     1
     2     1     2     1     1     1     0.1   1     1     0.1   0.1   0.1

0.94 3.97147     3.49773     0.377811    0.298106    0.00247846  0.183347
     10    44.9041     4.83739     4.38345     0.0755622   0.0596212
     0.0494692   0.000495692 0.967477    0.000495692 1     5     5     5
     5     5     0.1   1     2     1     2     1     2     1     2     1
     2     1     2     1     1     1     0.1   1     1     0.1   0.1   0.1

0.95 3.98632     3.50908     0.375554    0.29602     0.00242903  0.183189
     10    45.359      4.83766     4.42742     0.0751109   0.059204
     0.0485389   0.000485805 0.967531    0.000485805 1     5     5     5
     5     5     0.1   1     2     1     2     1     2     1     2     1
     2     1     2     1     1     1     0.1   1     1     0.1   0.1   0.1

0.96 4.00105     3.52032     0.37334     0.293972    0.0023811   0.18303
     10    45.8134     4.83793     4.47135     0.0746679   0.0587945
     0.0476351   0.00047622  0.967586    0.00047622  1     5     5     5
     5     5     0.1   1     2     1     2     1     2     1     2     1
     2     1     2     1     1     1     0.1   1     1     0.1   0.1   0.1

0.97 4.01565     3.53144     0.371166    0.291963    0.00233463  0.18287
     10    46.2674     4.8382      4.51523     0.0742332   0.0583925
     0.0467567   0.000466925 0.967641    0.000466925 1     5     5     5
     5     5     0.1   1     2     1     2     1     2     1     2     1
     2     1     2     1     1     1     0.1   1     1     0.1   0.1   0.1

0.98 4.03014     3.54245     0.369032    0.28999     0.00228954  0.182711
     10    46.7209     4.83848     4.55908     0.0738064   0.0579979
     0.0459027   0.000457909 0.967695    0.000457909 1     5     5     5
     5     5     0.1   1     2     1     2     1     2     1     2     1
```

```
      2     1     2     1     1     1     0.1   1     1     0.1   0.1   0.1

0.99  4.0445      3.55335     0.366937    0.288052    0.0022458   0.182551
      10    47.1739     4.83875     4.60288     0.0733874   0.0576104
      0.0450724   0.00044916  0.96775     0.00044916  1     5     5     5
      5     5     0.1   1     2     1     2     1     2     1     2     1
      2     1     2     1     1     1     0.1   1     1     0.1   0.1   0.1

1     4.05874     3.56415     0.364879    0.28615     0.00220334  0.18239
      10    47.6265     4.83902     4.64665     0.0729759   0.0572299
      0.0442647   0.000440667 0.967805    0.000440667 1     5     5     5
      5     5     0.1   1     2     1     2     1     2     1     2     1
      2     1     2     1     1     1     0.1   1     1     0.1   0.1   0.1
```

**GRENITS commands**

Input file for GRENITS

```
      v1    v2    v3    v4    v5    v6    v7    v8    v9    v10   v11   v12
      v13   v14   v15   v16   v17   v18   v19   v20   v21   v22   v23   v24
      v25   v26   v27   v28   v29   v30   v31   v32   v33   v34   v35   v36
      v37   v38   v39   v40   v41   v42   v43   v44   v45   v46   v47   v48
      v49   v50   v51   v52   v53   v54   v55   v56   v57   v58   v59   v60
      v61   v62   v63   v64   v65   v66   v67   v68   v69   v70   v71   v72
      v73   v74   v75   v76   v77   v78   v79   v80   v81   v82   v83   v84
      v85   v86   v87   v88   v89   v90   v91   v92   v93   v94   v95   v96
      v97   v98   v99   v100  v101
IPCS_1      4.98753     4.61448     4.00292     3.42782     2.96455
      2.60424     2.3228      2.09938     1.91869     1.76992     1.64544
      1.53979     1.44899     1.37011     1.3009      1.23967     1.18509
      1.13611     1.09188     1.05174     1.01512     0.981561    0.950691
      0.922187    0.895777    0.871231    0.848353    0.82697     0.806937
      0.788125    0.770422    0.753728    0.737956    0.723029    0.708879
      0.695444    0.682668    0.670504    0.658905    0.647833    0.63725
      0.627124    0.617424    0.608123    0.599197    0.590621    0.582375
      0.574439    0.566796    0.559429    0.552322    0.545463    0.538836
      0.532431    0.526236    0.52024     0.514434    0.508809    0.503355
      0.498064    0.492931    0.487946    0.483103    0.478397    0.473822
      0.469371    0.465039    0.460823    0.456716    0.452715    0.448816
      0.445014    0.441306    0.437688    0.434157    0.43071     0.427343
      0.424054    0.420841    0.417699    0.414628    0.411624    0.408685
      0.40581     0.402995    0.40024     0.397541    0.394898    0.392309
      0.389772    0.387285    0.384846    0.382456    0.380111    0.377811
      0.375554    0.37334     0.371166    0.369032    0.366937    0.364879
IPCS_2      4.9682      4.53115     3.88116     3.28785     2.8174
      2.45527     2.17457     1.95308     1.77483     1.6287      1.50687
      1.40381     1.3155      1.23898     1.17201     1.11289     1.0603
      1.0132      0.970746    0.932275    0.897238    0.865184    0.835737
      0.808584    0.783458    0.760135    0.738421    0.71815     0.699179
      0.681383    0.664651    0.648889    0.634011    0.619943    0.606618
      0.593976    0.581965    0.570536    0.559648    0.54926     0.539338
      0.529851    0.520769    0.512065    0.503717    0.495701    0.487998
      0.480588    0.473456    0.466584    0.459958    0.453565    0.447393
      0.441429    0.435663    0.430085    0.424685    0.419455    0.414387
      0.409472    0.404704    0.400075    0.395581    0.391214    0.386969
      0.382841    0.378825    0.374917    0.371111    0.367404    0.363792
      0.36027     0.356836    0.353487    0.350218    0.347027    0.343912
      0.340868    0.337895    0.334989    0.332148    0.329369    0.326651
      0.323992    0.321389    0.318842    0.316347    0.313903    0.311509
```

```
     0.309163    0.306864    0.30461     0.3024      0.300233    0.298106
     0.29602     0.293972    0.291963    0.28999     0.288052    0.28615
Tetr 4.98753     3.83974     2.38073     1.47272     0.96426     0.669393
     0.487997    0.370003    0.289515    0.232395    0.190502    0.158918
     0.134543    0.115354    0.0999851   0.0874901   0.0771975   0.0686202
     0.0613982   0.0552609   0.050002    0.0454617   0.0415149   0.0380627
     0.0350258   0.0323402   0.0299536   0.0278234   0.0259141   0.024196
     0.0226446   0.0212389   0.0199612   0.0187964   0.0177315   0.0167555
     0.0158587   0.0150327   0.0142703   0.0135651   0.0129115   0.0123045
     0.0117399   0.0112138   0.0107227   0.0102635   0.0098337
     0.00943068  0.00905229  0.00869656  0.00836171  0.00804612
     0.00774835  0.00746708  0.0072011   0.00694932  0.00671075
     0.00648447  0.00626965  0.00606552  0.00587139  0.00568661
     0.00551059  0.00534277  0.00518266  0.00502979  0.00488373
     0.00474407  0.00461044  0.00448251  0.00435995  0.00424246
     0.00412976  0.0040216   0.00391774  0.00381795  0.00372201
     0.00362973  0.00354093  0.00345544  0.00337308  0.00329371
     0.00321718  0.00314336  0.00307213  0.00300335  0.00293693
     0.00287274  0.0028107   0.0027507   0.00269267  0.0026365
     0.00258213  0.00252947  0.00247846  0.00242903  0.0023811
     0.00233463  0.00228954  0.0022458   0.00220334
lacI 10    10    10    10    10    10    10    10    10    10    10    10
     10    10    10    10    10    10    10    10    10    10    10    10
     10    10    10    10    10    10    10    10    10    10    10    10
     10    10    10    10    10    10    10    10    10    10    10    10
     10    10    10    10    10    10    10    10    10    10    10    10
     10    10    10    10    10    10    10    10    10    10    10    10
     10    10    10    10    10    10    10    10    10    10    10    10
     10    10    10    10    10    10    10    10    10    10    10    10
     10    10    10    10    10
lamdar     4.98753     4.95507     4.92208     4.89771     4.88075
     4.86874     4.85997     4.85336     4.84825     4.8442      4.84095
     4.83829     4.8361      4.83426     4.83272     4.83142     4.83031
     4.82937     4.82856     4.82786     4.82727     4.82676     4.82633
     4.82596     4.82565     4.82539     4.82518     4.82501     4.82487
     4.82476     4.82468     4.82463     4.82461     4.8246      4.82461
     4.82465     4.8247      4.82476     4.82484     4.82493     4.82503
     4.82515     4.82527     4.82541     4.82555     4.8257      4.82586
     4.82602     4.8262      4.82637     4.82656     4.82675     4.82694
     4.82714     4.82735     4.82756     4.82777     4.82798     4.8282
     4.82843     4.82865     4.82888     4.82911     4.82935     4.82959
     4.82983     4.83007     4.83031     4.83056     4.83081     4.83106
     4.83131     4.83156     4.83181     4.83207     4.83233     4.83258
     4.83284     4.8331      4.83337     4.83363     4.83389     4.83416
     4.83442     4.83469     4.83496     4.83522     4.83549     4.83576
     4.83603     4.8363      4.83657     4.83684     4.83711     4.83739
     4.83766     4.83793     4.8382      4.83848     4.83875     4.83902
```

Commands

```
>library(GRENITS)
> data(SWITCH_ODE)

>dim(SWITCH_ODE)

>output.folder <- paste(tempdir(), "/Example_LinearNetswitch2", sep="")

> prob.file <- paste(output.folder, "/NetworkProbability_Matrix.txt", sep = "")

> LinearNet(output.folder, SWITCH_ODE)
```

Started MCMC chain 1 =============

MCMC chain 1 finished!

Started MCMC chain 2 =============

MCMC chain 2 finished!

> analyse.output(output.folder)

> prob.file <- paste(output.folder, "/NetworkProbability_Matrix.txt", sep = "")

> prob.mat <- read.table(prob.file)

> print(prob.mat)

> data(ipcs_ode)

> plot.ts( t(ipcs_ode), plot.type = "single",col = 1:5,xlim = c(0,65),main = "Circadian Clock Network \n ODE simulated data",xlab = "Time (s)",ylab ="Expression")

> legend("topright",rownames(ipcs_ode),lty = 1,col = 1:5)

> library(network)

> inferred.net <-1*(prob.mat >0.0

> print(inferred.net)

> inferred.net <-network(inferred.net)

> par(mfrow =c(1,2), cex=1.76,cex.lab =1.3,cex.main=1.4)

> prob.vec<-sort(as.vector(as.matrix(prob.mat)), T)

> prob.vec <-prob.vec[4:0 - length(prob.vec)]

> plot(x =prob.vec,y =1:length(prob.vec),xlim=c(0,1),main="connection included vs Threshold",xlab = "probability threshold",ylab="connection included")

> lines(c(0.1,0.1),c(0,30), col="red" ,lty=2,lwd =2)

> plot(inferred.net,label=network.vertex.names(inferred.net), main="IPCS switch inferred Network",mode ="circle",vertex.cex=7,arrowhead.cex=2,vertex.col="green")

MCMC default parameter

"x"

"" "LinearNet"

"samples" "1e+05"

"burn.in" "10000"

```
"thin" "10"

"c" "0.5"

"d" "0.5"

"sigma.s" "2"

"a" "2"

"b" "0.01"

"sigma.mu" "2"
```

BoolNet commands

```
> bin<-binarizeTimeSeries(ipcs_1,method="edgeDetector")

> print(bin)

$binarizedMeasurements

       X1 X2 X3 X4 X5 X6 X7 X8 X9 X10 X11

ipcs_1  1  1  1  0  0  0  0  0  0   0   0

ipcs_2  1  1  1  0  0  0  0  0  0   0   0

tetr    1  1  1  0  0  0  0  0  0   0   0

laci    0  0  0  0  0  0  0  0  0   0   0

lamdar  1  0  0  0  0  0  0  0  0   0   0


$thresholds

   ipcs_1     ipcs_2       tetr       laci     lamdar

-0.0533765 -0.1122500 -1.4730250         NA  0.6914085


> reconstructed<-
reconstructNetwork(bin$binarizedMeasurements,method="bestfit",maxK=4)

> print(reconstructed)

Probabilistic Boolean network with 5 genes


Involved genes:
```

ipcs_1 ipcs_2 tetr laci lamdar

Transition functions:

Alternative transition functions for gene ipcs_1:

ipcs_1 = <f(lamdar){01}> (probability: 0.25, error: 1)

ipcs_1 = <f(tetr){01}> (probability: 0.25, error: 1)

ipcs_1 = <f(ipcs_2){01}> (probability: 0.25, error: 1)

ipcs_1 = <f(ipcs_1){01}> (probability: 0.25, error: 1)

Alternative transition functions for gene ipcs_2:

ipcs_2 = <f(lamdar){01}> (probability: 0.25, error: 1)

ipcs_2 = <f(tetr){01}> (probability: 0.25, error: 1)

ipcs_2 = <f(ipcs_2){01}> (probability: 0.25, error: 1)

ipcs_2 = <f(ipcs_1){01}> (probability: 0.25, error: 1)

Alternative transition functions for gene tetr:

tetr = <f(lamdar){01}> (probability: 0.25, error: 1)

tetr = <f(tetr){01}> (probability: 0.25, error: 1)

tetr = <f(ipcs_2){01}> (probability: 0.25, error: 1)

tetr = <f(ipcs_1){01}> (probability: 0.25, error: 1)

Alternative transition functions for gene laci:

laci = 0 (probability: 1, error: 0)

Alternative transition functions for gene lamdar:

lamdar = 0 (probability: 1, error: 0)

Knocked-out and over-expressed genes:

laci = 0

lamdar = 0

> example_PBN<-reconstructed

> print(example_PBN)

Probabilistic Boolean network with 5 genes

Involved genes:

ipcs_1 ipcs_2 tetr laci lamdar

Transition functions:

Alternative transition functions for gene ipcs_1:

ipcs_1 = <f(lamdar){01}> (probability: 0.25, error: 1)

ipcs_1 = <f(tetr){01}> (probability: 0.25, error: 1)

ipcs_1 = <f(ipcs_2){01}> (probability: 0.25, error: 1)

ipcs_1 = <f(ipcs_1){01}> (probability: 0.25, error: 1)

Alternative transition functions for gene ipcs_2:

ipcs_2 = <f(lamdar){01}> (probability: 0.25, error: 1)

ipcs_2 = <f(tetr){01}> (probability: 0.25, error: 1)

ipcs_2 = <f(ipcs_2){01}> (probability: 0.25, error: 1)

ipcs_2 = <f(ipcs_1){01}> (probability: 0.25, error: 1)

Alternative transition functions for gene tetr:

tetr = <f(lamdar){01}> (probability: 0.25, error: 1)

tetr = <f(tetr){01}> (probability: 0.25, error: 1)

tetr = <f(ipcs_2){01}> (probability: 0.25, error: 1)

tetr = <f(ipcs_1){01}> (probability: 0.25, error: 1)

Alternative transition functions for gene laci:

laci = 0 (probability: 1, error: 0)

Alternative transition functions for gene lamdar:

lamdar = 0 (probability: 1, error: 0)

Knocked-out and over-expressed genes:

laci = 0

lamdar = 0

```
> net<-chooseNetwork(example_PBN,rep(1,length(example_PBN$genes)))
> attr<-getAttractors(net)
> attr
```

Attractor 1 is a simple attractor consisting of 1 state(s) and has a basin of 12 state(s):

```
 |--<----|
 V       |
 10011   |
 |       |
 V       |
 |-->----|
```

Genes are encoded in the following order: Gene 1 Gene 2 Gene 3 Gene 4 Gene 5

Attractor 2 is a simple attractor consisting of 3 state(s) and has a basin of 17 state(s):

```
 |--<----|
 V       |
 00000   |
 |       |
 10110   |
 |       |
 10001   |
 |       |
 V       |
 |-->----|
> plotAttractors(attr,subset=1)
$`1`
$`2`
NULL
> plotAttractors(attr,subset=2)
$`1`
NULL
$`2`
Genes are encoded in the following order: Gene 1 Gene 2 Gene 3 Gene 4 Gene 5
> par(mfrow=c(2,length(attractors$attractors)))
> tt <-getTransitionTable(attr)
> net <-generateRandomNKNetwork(n=5,k=5)
```

> print(net)

Boolean network with 5 genes

Involved genes:

Gene 1 Gene 2 Gene 3 Gene 4 Gene 5

Transition functions:

Gene 1 = <f(Gene 3,Gene 4,Gene 5,Gene 1,Gene 2){01100000011111100010101000010111}>

Gene 2 = <f(Gene 3,Gene 4,Gene 2,Gene 5,Gene 1){01001010011100011110011000011110}>

Gene 3 = <f(Gene 2,Gene 5,Gene 3,Gene 4,Gene 1){00101100101100101101111100101011}>

Gene 4 = <f(Gene 4,Gene 2,Gene 5,Gene 3,Gene 1){00011000010110111111101000001001}>

Gene 5 = <f(Gene 1,Gene 4,Gene 3,Gene 2,Gene 5){10000011101101010011100001111011}>

>path <-getPathToAttractor(net,rep(0,5))

> path

> toPajek(attr,file="net.net")

> toPajek(attr,file="net.net",includeLabels=TRUE)

Network file format

This section provides a full language description for the network _le format of BoolNet. The language is described in Extended Backus-Naur Form (EBNF).

Rule = GeneName Separator Boolean Expression [Separator Probability];

Boolean Expression = GeneName

| "!" Boolean Expression

| "(" Boolean Expression ")"

| Boolean Expression " & " Boolean Expression

| Boolean Expression " | " Boolean Expression;

GeneName =? A gene name from the list of involved genes?

Separator = ",";

Probability =? A floating-point number?